\documentclass{article}

\usepackage{epubtk}
\usepackage{graphicx}
\usepackage[square,comma,sort]{natbib}

\begin{document}

\title{Probes and Tests of Strong-Field Gravity with Observations
in the Electromagnetic Spectrum}

\author{%
\epubtkAuthorData{Dimitrios Psaltis}{%
Physics and Astronomy Departments\\
University of Arizona}{%
dpsaltis@physics.arizona.edu}{%
http//www.physics.arizona.edu/\~dpsaltis}%
}

\date{}
\maketitle

\begin{abstract}
Neutron stars and black holes are the astrophysical systems with the
strongest gravitational fields in the universe. In this article, I
review the prospect of probing with observations of such compact
objects some of the most intriguing General Relativistic predictions
in the strong-field regime: the absence of stable circular orbits near
a compact object and the presence of event horizons around black-hole
singularities. I discuss the need for a theoretical framework within
which future experiments will provide detailed, quantitative tests of
gravity theories. Finally, I summarize the constraints imposed by
current observations of neutron stars on potential deviations from
General Relativity.
\end{abstract}

\epubtkKeywords{}

\newpage

\section{Introduction}
\label{sec:int}

Over the past 90 years, the basic ingredients of General Relativity
have been tested in many different ways and in many different
settings. From the solar eclipse expedition of 1917 to the modern
observations of double neutron stars, General Relativity has passed all
tests with flying colors~\cite{Will06}. Yet, our inability to devise a
renormalizable quantum gravity theory as well as the mathematical
singularities found in many solutions of Einstein's equations suggest
that we should look harder for gravitational phenomena not described
by General Relativity.

The search for such deviations has been very fruitful in the regime of
very weak fields. Observations of high-redshift
supernovae~\cite{Perl97, Riess98} and of the cosmic microwave
background with WMAP~\cite{Spergel03} have measured a non-zero
cosmological constant (or a slowly rolling field that behaves as such
at late times). This discovery can be incorporated within the
framework of General Relativity, if interpreted simply as a constant
in the Einstein--Hilbert action. It nevertheless brought to the surface
a major problem in trying to connect gravity to basic ideas of quantum
vacuum fluctuations~\cite{Weinberg89,Carroll01}.

In the strong-field regime, on the other hand, which is relevant for
the evolution of the very early universe and for determining the
properties of black holes and neutron stars, little progress has been
made in testing the predictions of general
relativity~\cite{Stairs03}. There are two reasons that have been
responsible for this lag. First, phenomena that occur in strong
gravitational fields are complex and often explosive, making it very
difficult to find observable properties that depend cleanly on the
gravitational field and that allow for quantitative tests of gravity
theories. Second, there exists no general theoretical framework within
which to quantify deviations from general relativistic predictions in
the strong-field regime.

During the current decade, technological advances and increased
theoretical activity have led to developments that promise to make
strong-field gravity tests a routine in the near future. The first
generation of earth-based gravitational wave observatories (such as
LIGO~\cite{LIGO}, GEO600~\cite{GEO}, TAMA300~\cite{TAMA}, and
VIRGO~\cite{VIRGO}) as well as the Beyond Einstein Missions (such as
Constellation-X, LISA, and the Black Hole Imager~\cite{beyondEins})
will offer an unprecedented look into the near fields of black holes
and neutron stars. Moreover, recent ideas on quantum
gravity~\cite{Burgess04}, brane-world gravity~\cite{Maartens04}, or
other Lagrangian extensions of general
relativity~\cite{Woodard06,Sotiriou08} will provide the means with
which the experimental results will be interpreted.

In this article, I review the theoretical and experimental prospects
of testing strong-field General Relativity with observations in the
electromagnetic spectrum. In the first few sections, I discuss the
motivation for performing such tests and then describe the
astrophysical settings in which strong-field effects can be
measured. In Section~\ref{sec:need}, I elaborate on the need for a
theoretical framework within which strong-field gravity tests can be
performed and in Section~\ref{sec:tests} I review the current
quantitative tests of General Relativity in the strong-field regime
that use neutron stars. Finally, in Section~\ref{section:beyond} I
discuss the prospect of probing and testing strong gravitational
fields with upcoming experiments and observatories.

\newpage

\section{The motivation for strong-field tests}
\label{sec:motiv}

Most physical scientists would agree that there is very little need to
motivate testing one of the fundamental theories of physics in a
regime that experiments have probed only marginally, so far. However,
in the particular case of testing the strong-field predictions of
General Relativity, there exist at least three arguments that provide
additional strong support to such an endeavor. First, there is no
fundamental reason to choose Einstein's equations over other
alternatives. Second, gravitational tests to date seldom probe strong
gravitational fields. Finally, it is known that General Relativity
breaks down at the strong-field regime. I will now elaborate on each
of these arguments.

\smallskip

\noindent $\bullet$ {\em There is no fundamental reason to choose
Einstein's equations over other alternatives.~--\/} All theories of
gravity, including Newton's theory and General Relativity, have two
distinct ingredients. The first describes how matter moves in the
presence of a gravitational field. The second describes how the
gravitational field is generated in the presence of matter. For
Newtonian dynamics, the first ingredient is Newton's second law
together with the assertion that the gravitational and inertial masses
of an object are the same; the second ingredient is Poisson's
equation. For General Relativity, the first ingredient arises from the
equivalence principle, whereas the second is Einstein's field
equation.

The equivalence principle, in its various formulations, dictates the
geometric aspect of the theory~\cite{Will06}: it is impossible to tell
the difference between a reference frame at rest and one free-falling
in a gravitational field, by performing local, non-gravitational (for
the Einstein Equivalence Principle) or even gravitational (for the
Strong Equivalence Principle) experiments.  Moreover, the equivalence
principle encompasses the Lorentz symmetry, as well as our belief that
there is no preferred frame and position anywhere in the
spacetime. Because of its central importance in any gravity theory,
there have been many attempts during the last century at testing the
validity of the equivalence principle. These were performed mostly in
the weak-field regime and have resulted in upper limits on possible
violations of this principle that are as stringent as one part in
$10^{12}$~\cite{Will06}.

Contrary to the case of the equivalence principle, there are no
compelling arguments one can make that lead uniquely to Einstein's
field equation. In fact, Einstein reached the field equation, more or
less, by reverse engineering (see the informative discussion
in~\cite{Misner73, Pais82}) and, soon afterwards, Hilbert
constructed a Lagrangian action that leads to the same equation. The
Einstein--Hilbert action is directly proportional to the Ricci scalar,
$R$,
\begin{equation}
           S= \frac{c^4}{16\pi G}\int d^4 x  \, \sqrt{-g} \, (R-2\Lambda)\;,
\label{EHaction}
\end{equation}
where $g\equiv\det\vert g_{\mu\nu}\vert$, $g_{\mu\nu}$ is the
spacetime metric, $c$ is the speed of light, $G$ is the gravitational
constant, and $\Lambda$ is the cosmological constant.  While such a
theory is entirely self-consistent at the classical level, it may
represent only an approximation that is valid at the scales of
curvature that are found in terrestrial, solar, and stellar-system
tests.

Indeed, a self-consistent theory of gravity can also be constructed
for any other action that obeys the following four simple
requirements~\cite{Misner73}. It has to: {\em (i)\/} reproduce the
Minkowski spacetime in the absence of matter and of the cosmological
constant, {\em (ii)\/} be constructed from only the Riemann curvature
tensor and the metric, {\em (iii)\/} follow the symmetries and
conservation laws of the stress-energy tensor of matter, and {\em
(iv)\/} reproduce Poisson's equation in the Newtonian limit. Of all
the possibilities that meet these requirements, the field equations
that are derived from the Einstein--Hilbert action are the only ones
that are also linear in the Riemann tensor. Albeit simple and elegant,
a more general classical action of the form~\cite{Sotiriou08}
\begin{equation}
           S =\frac{c^4}{16\pi G}\
                 \int d^4 x  \, \sqrt{-g} \,
            f(R)\;,
\label{EHaction2}
\end{equation}
also obeys the same requirements.  Indeed, the
action~(\ref{EHaction2}) results in a field equation that allows for
the Minkowski solution in the absence of matter, is constructed only
from the Riemann tensor, obeys the usual symmetries and conservation
laws~\cite{Wald84}, and can be made to produce negligible corrections
at the small curvatures probed by weak-field gravitational
experiments.  On the other hand, the predictions of the theory may be
significantly different at the strong curvatures probed by
gravitational tests involving compact objects.

The single, rank-2 tensor field $g_{\mu\nu}$ (i.e., the metric) of the
Einstein--Hilbert action may also not be adequate to describe
completely the gravitational force (although, if additional fields are
introduced, then the strong equivalence principle is violated, with
important implications for the frame- and time- dependence of
gravitational experiments). In fact, a variant of such theories with
an additional scalar field, the Brans--Dicke theory~\cite{Brans61},
has been the most widely used alternative to General Relativity to be
tested against experiments.  Today, scalar-tensor theories are among
the prime candidates for explaining the acceleration of the universe
at late times (the ``dark energy''~\cite{Peebles03}). Depending on the
coupling between the metric, the scalar field, and matter, the
relative contribution of such additional fields may become significant
only at the high curvatures found in early universe or in the vicinity
of compact objects.

Although the above discussion has considered only the classical action
of the gravitational field in a phenomenological manner, it is
important to also note that corrections to the Einstein--Hilbert action
occur naturally in quantum gravity theories and in string theory. For
example, if we choose to interpret the metric $g_{\mu\nu}$ as a
quantum field, we can take Equation~(\ref{EHaction}) as a quantum
field-theoretic action defined at an ultraviolet scale (such as the
Planck scale), and proceed to perform quantum-mechanical calculations
in the usual way~\cite{Donoghue94}.  However, radiative corrections
will induce an infinite series of counterterms as we flow to lower
energies and such counterterms will not be reabsorbed into the
original Lagrangian by adjusting its bare parameters.  Instead, such
terms will appear as new, higher-derivative correction terms in the
Einstein--Hilbert action~(\ref{EHaction}). 

Finally, it is worth emphasizing that the previous discussion focuses
on Lagrangian gravity in a four-dimensional spacetime.  In the context
of string theory, General Relativity emerges only as a leading
approximation. String theory also predicts an infinite set of
non-linear terms in the scalar curvature, all suppressed by powers of
the Planck scale. Moreover, the low-energy effective action of string
theory contains additional scalar (dilatonic) and vector gravitational
fields~\cite{Green88}. Motivated by ideas of string theory, brane-world
gravity~\cite{Maartens04,DGP00a,DGP00b,DGP00c} also provides a
self-consistent theory that is consistent with all current tests of
gravity.

All the above strongly support the notion that the field equation that
arises from the Einstein--Hilbert action may be appropriate only at the
scales that have been probed by current gravitational tests. But how deep
have we looked?

\smallskip

\noindent $\bullet$ {\em Gravitational tests to date seldom probe
strong gravitational fields.~--\/} All historical tests of general
relativity have been performed in our solar system. The strongest
gravitational field they can, therefore, probe is that at the surface
of the Sun, which corresponds to a gravitational redshift of
\begin{equation}
z_\odot\simeq \frac{GM_\odot}{R_\odot c^2}\simeq 2\times 10^{-6}\;,
\end{equation}
and to a spacetime curvature of
\begin{equation}
\frac{GM_\odot}{R_\odot^3 c^2}\simeq 4\times 10^{-28}~\mbox{cm}^{-2}\;.
\end{equation}
Coincidentally, the gravitational fields that have been probed in
tests using double neutron stars are of the same magnitude, since the
masses and separation of the two neutron stars in the systems under
consideration are comparable to the mass and radius of the Sun,
respectively\footnote{Note, however, that some of the phenomena
observed in double neutron stars depend on the coupling of matter and
gravity in the strong-field regime~\cite{Damour93}.}. These are
substantially weaker fields than those found in the vicinities of
neutron stars and stellar-mass black holes, which correspond to a
redshift of $\sim 1$ and a spacetime curvature of $\simeq
2\times 10^{-13}$~cm$^{-2}$.

It is instructive to compare the degree to which current tests verify
the predictions of General Relativity to the increase in the strength
of the gravitational field going from the solar system to the vicinity
of a compact object. Current constraints on the deviation of the PPN
parameters from the General Relativistic predictions are of order
$\simeq 10^{-5}$~\cite{Will06}. It is conceivable, therefore, that
deviations consistent with these constraints can grow and become of order
unity when the redshift of the gravitational field probed is increased
by six orders of magnitude and the spacetime curvature by fifteen!
 
Is it possible, however, that General Relativity still describes
accurately phenomena that occur in the strong gravitational fields
found in the vicinity of stellar-mass black holes and neutron stars?

\smallskip

\noindent $\bullet$ {\em General Relativity breaks down at the
strong-field regime.~--\/} Our current understanding of the physical
world leaves very little doubt that the theory of General Relativity
itself breaks down at the limit of very strong gravitational
fields. Considering the theory simply as a classical, geometric
description of the spacetime leads to predictions of infinite matter
densities and curvatures in two different settings. Integrating
forward in time the Oppenheimer--Snyder equations, which describe the
collapse of a cloud of dust~\cite{Oppenheimer39}, leads to the
formation of a black hole with a singularity at its
center. Integrating backwards in time the Friedmann equation, which
describes the evolution of a uniform and isotropic universe, always
results in a singularity at the beginning of time, the big
bang. Clearly, the outcome in both of these settings is unphysical.

It is widely believed that quantum gravity prohibits these unphysical
situations that occur at the limit of infinitely strong gravitational
fields.  Even though none of the observable astrophysical objects
offer the possibility of testing gravity at the Planck scale, they
will nevertheless allow placing constraints on deviations from general
relativity that are as large as $\sim 10$ orders of magnitude more
stringent compared to all other current tests. This is the best result
we can expect in the near future to come out of the detection of
gravitational waves and the observation of the innermost regions of
neutron stars and black holes with NASA's Beyond Einstein missions. If
the history of the recent detection of a minute yet non-zero
cosmological constant is any measure of our inability to predict even
the order of magnitude of gravitational effects that we have not
directly probed, then we might be up for a pleasant surprise!

\newpage

\section{Astrophysical and Cosmological Settings of Strong Gravitational
Fields}
\label{sec:strong}

\subsection{When is a gravitational field strong?}
\label{subsec:definition}

Looking at the Schwarzschild spacetime, it is natural to measure the
``strength'' of the gravitational field at a distance $r$ away from an
object of mass $M$ by the parameter
\begin{equation}
\epsilon\equiv \frac{GM}{rc^2}\;,
\end{equation}
which is proportional to the Newtonian gravitational potential and is
also directly related to the redshift.  Infinitesimal gravitational
fields correspond to the limit $\epsilon\rightarrow 0$, which leads to
the Minkowski spacetime of special relativity.  Weak gravitational
fields correspond to $\epsilon\ll 1$, which leads to Newtonian
gravity. Finally, the strongest gravitational fields accessible to an
observer are characterized by $\epsilon\rightarrow 1$, at which point
the black-hole horizon of an object of mass $M$ is approached. (Note
that formally the horizon of a Schwarzschild black hole occurs at
$\epsilon=2$; I drop here the factor of 2, as I am mostly interested
in dimensional arguments).

Albeit useful in defining post-Newtonian expansions, the parameter
$\epsilon$ is not fundamental in characterizing a gravitational field
in Einstein's theory. Indeed, the geodesic equation and the Einstein
field equation (or equivalently, the Einstein--Hilbert
action~[\ref{EHaction}]) are written in terms of the Ricci scalar, the
Ricci tensor, and the Riemann tensor, all of which measure the
curvature of the field and not its potential. As a result, when we
consider deviations from General Relativity that arise by adding
linearly terms to the Einstein--Hilbert action, the critical strength
of the gravitational field beyond which these additional terms become 
important is typically given in terms of the spacetime curvature.

For example, in the presence of a cosmological
constant, the metric of a spherically symmetric object becomes
\begin{equation}
ds^2=-\left(1-\frac{2GM}{rc^2}-\frac{\Lambda r^2}{3}\right)dt^2
     +\left(1-\frac{2GM}{rc^2}-\frac{\Lambda r^2}{3}\right)^{-1}dr^2
     +r^2(d\theta^2+\sin^2\theta d\phi^2)
\end{equation}
and the Newtonian approximation becomes invalid when
\begin{equation}
   \frac{GM}{c^2r^3}\ll \frac{1}{6}\Lambda\;.
\label{eq:curv_lambda}
\end{equation}
In this case, a gravitational field is ``weak'' if the spacetime
curvature is smaller than $\Lambda/3$, independent of the value of the
parameter $\epsilon$. In the opposite extreme, if there are additional
terms in the action of the gravitational field beyond the
Einstein--Hilbert term, such as
\begin{equation}
           S =\frac{c^4}{16\pi G}\
                 \int d^4 x  \, \sqrt{-g} \,
            (R+\alpha R^2)\;,
\label{eq:EH3}
\end{equation}
then the General Relativistic predictions become inaccurate at
strong gravitational fields defined by the condition
\begin{equation}
   \frac{GM}{c^2r^3}\gg \frac{1}{\alpha}\;,
\label{eq:curv_nonlinear}
\end{equation}
even if the parameter $\epsilon$ is much smaller than unity. Note
that, in Equation~(\ref{eq:EH3}), $\alpha$ is an appropriate constant
with units of (length)$^2$ and I have set the Ricci scalar to $R\sim
GM/r^3c^2$ (I use this here as an order of magnitude estimate and do
not consider the fact that, if the distance $r$ is larger than the
radius of the object, then the Ricci scalar in General Relativity
vanishes).

Similar considerations lead to a condition on curvature when we add to
the Einstein--Hilbert action terms that invoke additional scalar,
vector, and tensor fields.  In all these cases, a strong gravitational
field is characterized not by a large gravitational potential (i.e., a
high value of the parameter $\epsilon$) but rather by a large
curvature
\begin{equation}
\xi\equiv \frac{GM}{r^3c^2}\;.
\label{eq:xi}
\end{equation}
Because the condition that the curvature needs to satisfy in order for
a gravitational field to be considered ``strong'' depends on the
particular deviation from General Relativity under study (cf.\
Equations~[\ref{eq:curv_lambda}] and [\ref{eq:curv_nonlinear}]) I will
not normalize the parameter $\xi$ to any particular energy density but
rather leave it, hereafter, as a dimensional quantity.

This is an appropriate parameter with which to measure the strength of
a gravitational field in a geometric theory of gravity, such as
General Relativity, because the curvature is the lowest order quantity
of the gravitational field that cannot be set to zero by a coordinate
transformation. Moreover, because the curvature measures energy
density, a limit on curvature will correspond to an energy scale
beyond which additional gravitational degrees of freedom may become
important.

\subsection{A parameter space for tests of gravity}
\label{subsec:parameters}

The two parameters, $\epsilon$ and $\xi$, define a parameter space on
which we can quantify the strengths of the gravitational fields probed
by different tests of gravity (see Figure~\ref{fig:fields}). Only a
fraction of this parameter space is accessible to experiments. Regions
of the parameter space with potential $\epsilon>1$ correspond to
distances from a gravitating object that are smaller than the horizon
radius and are, therefore, inaccessible to observers. (I neglect here,
for simplicity, the small numerical factor in the horizon radius that
depends on the spin of the black hole.) In Figure~\ref{fig:fields},
this region is outlined by the vertical red line.

\epubtkImage{strong_fields.png}{
\begin{figure}[htbp]
\centerline{\includegraphics[height=18.0cm]{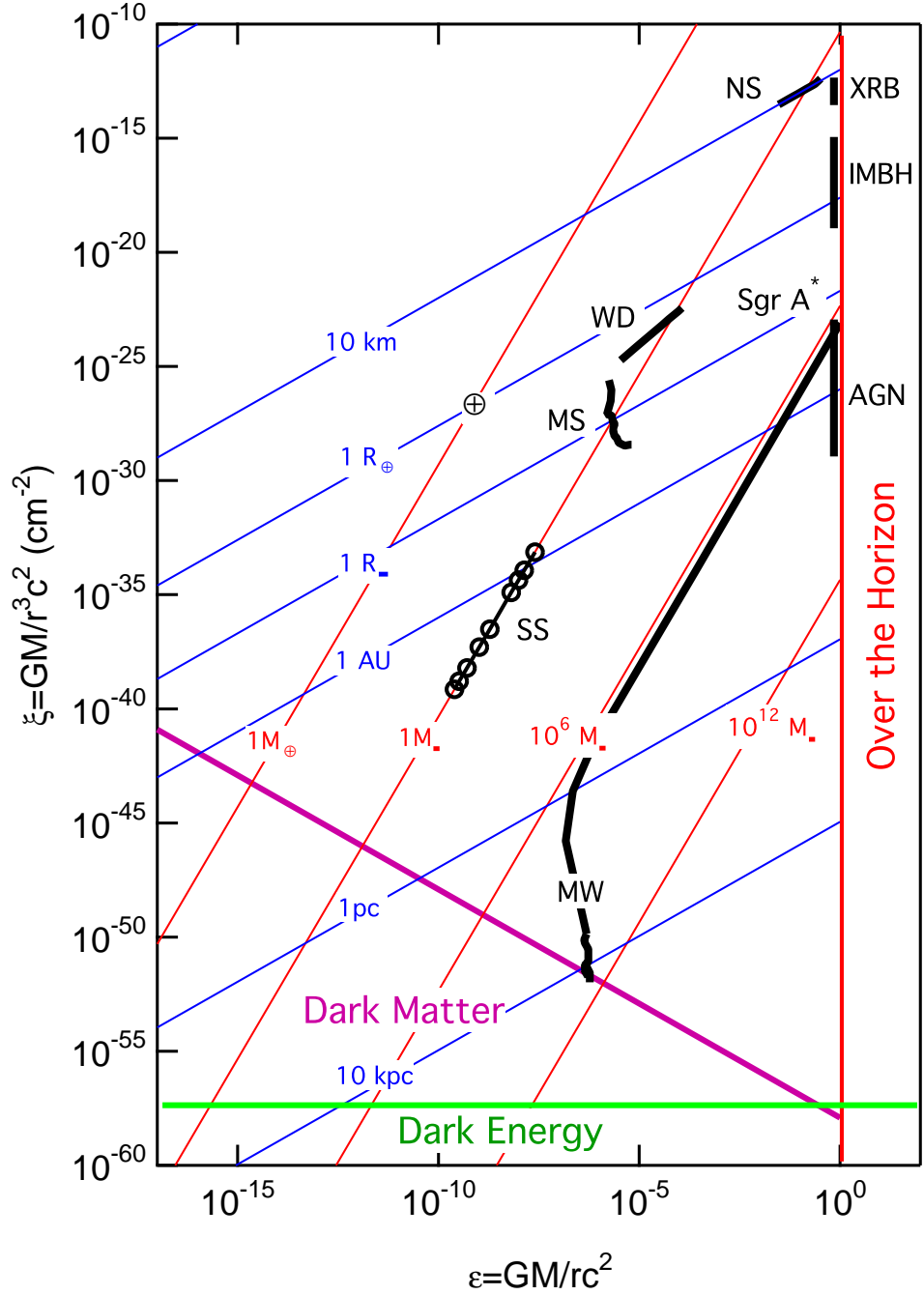}}
\caption{A parameter space for quantifying the strength
of a gravitational field.  The $x$-axis measures the potential
$\epsilon\equiv GM/rc^2$ and the $y$-axis measures the spacetime
curvature $\xi\equiv GM/r^3c^2$ of the gravitational field at a radius
$r$ away from a central object of mass $M$. These two parameters
provide two different quantitative measures of the strength of the
gravitational fields. The various curves, points, and legends are
described in the text.}
\label{fig:fields}
\end{figure}}

Quantifying deviations from General Relativity for part of the
parameter space requires a detailed understanding of the properties of
dark matter and dark energy, which is beyond current capabilities.  In
the limit of very small values of the curvature, the presence of a
non-zero cosmological constant affects the outcome of gravitational
experiments when (see eq.~[\ref{eq:curv_lambda}])
\begin{equation}
\xi\le \frac{3G\Omega_\Lambda H_0^2}{8\pi c^2}\simeq 5\times 
   10^{-58}\left(\frac{\Omega_\Lambda}{0.73}\right)
   \left(\frac{H_0}{73~\mbox{km/s/Mpc}}\right)^2~\mbox{cm}^{-2}\;,
\end{equation}
where $\Omega_\Lambda$ is the current density of dark energy in units
of the critical density and $H_0$ is the current value of the Hubble
constant. Phenomena that probe such low values of curvature (i.e.,
below the horizontal green line in Figure~\ref{fig:fields}) can lead to
quantitative tests of General Relativity only if a specific model of
dark energy (e.g., a cosmological constant) is assumed.

The ability to perform a quantitative test of a gravity theory also
relies on an independent measurement of the mass that generates the
gravitational field. This is not always possible, especially in
various cosmological settings, where gravitational phenomena are used
mostly to infer the presence of dark matter and not to test General
Relativistic predictions. Dark matter is typically required in systems
for which the acceleration drops below the so-called MOND acceleration
scale $a_0\simeq 10^{-8}$~cm~s$^{-2}$~\cite{Milgrom83, Sanders02,
Bekenstein07}. (This is an observed fact, independent of whether the
inability of Newtonian gravity to account for observations is due to
the presence of dark matter or to the breakdown of the theory itself.)
This acceleration scale is also comparable to $a_0\simeq c
H_0$. Systems for which dark matter is necessary to account for their
gravitational fields are characterized by
\begin{equation}
\xi \le \left(\frac{a_0}{c^2}\right)^2 \frac{1}{\epsilon}
\simeq \left(\frac{H_0}{c}\right)^2 \frac{1}{\epsilon}\;.
\end{equation}
This region of the parameter space is outlined by the purple line in
Figure~\ref{fig:fields}. The fact that the three lines that correspond
to the Schwarzschild horizon, the MOND acceleration scale, and the
dark energy all seem to intersect roughly in one point in the
parameter space is directly related to the cosmic coincidence problem,
i.e., that the universe is flat, with comparable amounts of (mostly
dark) matter and dark energy. 

In the opposite limit of very strong gravitational fields, General
Relativity is expected to break down when quantum effects become
impossible to neglect. This is expected to happen if a gravitational
test probes a distance from an object of mass $M$ that is comparable
to the Compton wavelength $\lambda_{\rm C}\equiv h/Mc$, where $h$
is Planck's constant. Quantum effects are, therefore, expected to 
dominate when 
\begin{equation}
\xi\ge \frac{1}{L_{\rm P}^2}\epsilon^2\;,
\end{equation}
where $L_{\rm P}\equiv (Gh/c^3)^{1/2}\simeq 4\times 10^{-33}$~cm is
the Planck length. This part of the parameter space is not shown in
Figure~\ref{fig:fields}, as it is many orders of magnitude away from
the values of the parameters that correspond to astrophysical systems.

Having defined the parameter space and outlined the various limiting
cases, I can now identify the various astrophysical systems that
probe its various regimes. In general, systems of constant central
mass $M$ will follow curves of the form
\begin{equation}
\xi=\frac{c^4}{G^2M^2}\epsilon^3\;,
\end{equation}
whereas probes at a constant distance $r$ away from the central object
will follow curves of the form
\begin{equation}
\xi=\frac{1}{r^2}\epsilon\;.
\end{equation}
Figure~\ref{fig:fields} shows a number of representative contours of
constant mass and distance. 

The strongest gravitational fields around astrophysical systems can be
found in the vicinities of neutron stars (NS in
Figure~\ref{fig:fields}) and black holes in X-ray binaries
(XRB). Large gravitational potentials but smaller curvatures can be
found around the horizons of intermediate mass black holes ($\sim
10^2\mbox{\,--\,}10^4\,M_\odot$; IMBHs) and in active galactic nuclei
($10^6\mbox{\,--\,}10^{10}\,M_\odot$; AGN). Weaker gravitational fields exist near
the surfaces of white dwarfs (WD), main-sequence stars (MS), or at the
distances of the various planets in our solar system (SS). Finally,
even weaker gravitational fields are probed by observations of the
motions of stars in the vicinity of the black hole in the center of
the Milky Way (Sgr~A$^*$), and by studies of the rotational curve of
the Milky Way (MW) and other galaxies. In placing the various systems
on the parameter space shown in Figure~\ref{fig:fields}, I have used a
typical mass-radius relation for neutron stars and white
dwarfs~\cite{Shapiro84}, the calculated mass-radius relation of
main-sequence stars~\cite{Clayton83}, and the inferred mass-radius
profile of the inner region around Sgr~A$^*$~\cite{Schodel02}, which
smoothly approaches the mass profile inferred from the rotation curve
of the Milky Way~\cite{Dehnen98}.

\epubtkImage{tests.png}{
\begin{figure}[t]
\centerline{\includegraphics[width=12.0cm]{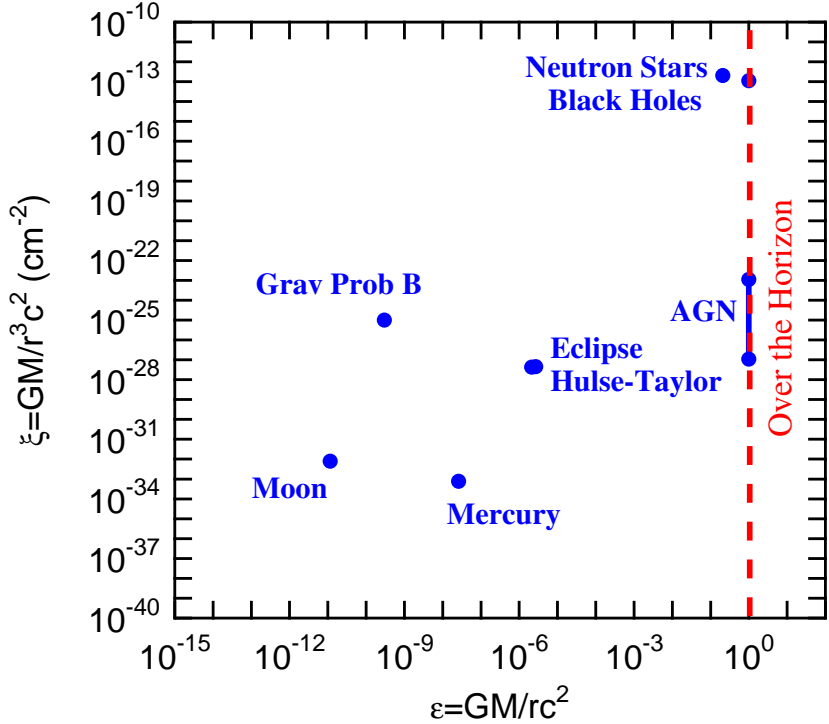}}
\caption{Tests of General Relativity placed on an appropriate
  parameter space.  The long-dashed line represents the event horizon
  of Schwarzschild black holes.}
\label{fig:tests}
\end{figure}}

Current tests of General Relativity with astrophysical objects probe a
wide range of gravitational potentials and curvatures (see
Figure~\ref{fig:tests}).  However, they fall short of probing the most
extreme phenomena that are predicted by the theory to occur in the
vicinities of compact objects. For example, tests during solar
eclipses, with double neutron stars (such as the Hulse--Taylor pulsar),
or with Grav Prob B probe curvatures that are the same as those found
near the horizons of supermassive black holes, but potentials that are
smaller by six to ten orders of magnitude. Moreover, all these tests
probe curvatures that are smaller by thirteen or more orders of
magnitude from those found near the surfaces of neutron stars and the
horizons of stellar-mass black holes. Future experiments, such as the
gravitational wave detectors and the Beyond Einstein missions, will
offer for the first time the opportunity to probe directly such strong
gravitational fields.

The whole range of gravitational fields, from the weakest to the
strongest, can also be found during various epochs of the evolution of
the universe. As a result, observations of cosmological phenomena may
also probe very strong gravitational fields.  The scalar curvature of
a flat universe is given by
\begin{equation}
R=\frac{6}{\alpha^2}\left(\alpha \ddot{\alpha}+\dot{\alpha}^2\right)\;,
\end{equation}
where $\alpha$ is the scale factor. Using the Friedmann equation,
the scalar curvature becomes
\begin{equation}
R=3\left(\frac{\Omega_{\rm m}^0}{\alpha^3}+4\Omega_{\rm d}^0\right)
   \left(\frac{H_0}{c}\right)^2\;,
\label{eq:Rcosmo}
\end{equation}
where $\Omega_{\rm m}^0$ and $\Omega_{\rm d}^0$ are the
(non-relativistic) matter and dark energy densities in the present
universe, respectively, in units of the critical
density. Equation~(\ref{eq:Rcosmo}) shows that, at late times, the
radius of curvature of the universe is comparable to the Hubble
distance.

The evolution of the scalar curvature with redshift for a flat
universe and for the best-fit cosmological parameters obtained by the
WMAP mission~\cite{Spergel03} is shown in Figure~\ref{fig:cosmo}.
Identified on this figure are several characteristic epochs that have
been used in testing General Relativistic predictions: the $z\simeq 1$
epoch of type I supernovae that are used to measure the value of the
cosmological constant~\cite{Perl97, Riess98}; the $z\simeq 1000$ epoch
at which the acoustic peaks of the cosmic microwave background
observed by WMAP are produced; and the period of nucleosynthesis
during which the temperature of the universe was in the range 60~keV
-- 1~MeV~\cite{Santiago97, Carroll02}.  The period of big-bang
nucleosynthesis is the earliest epoch for which quantitative tests
have been performed. The corresponding scalar curvature of the
universe at that time, however, is still small and comparable to the
curvatures of gravitational fields probed by current tests of General
Relativity in the solar system. It was only when the temperature of
the Universe was $\sim 100$~GeV that its curvature was $\simeq
10^{-12}$~cm$^{-2}$, i.e., comparable to that found around a neutron
star or stellar-mass black hole. This is the period of electroweak
baryogenesis, for which no detailed theoretical models or data exist
to date.

\epubtkImage{cosmology.png}{
\begin{figure}[htbp]
\centerline{\includegraphics[width=12.0cm]{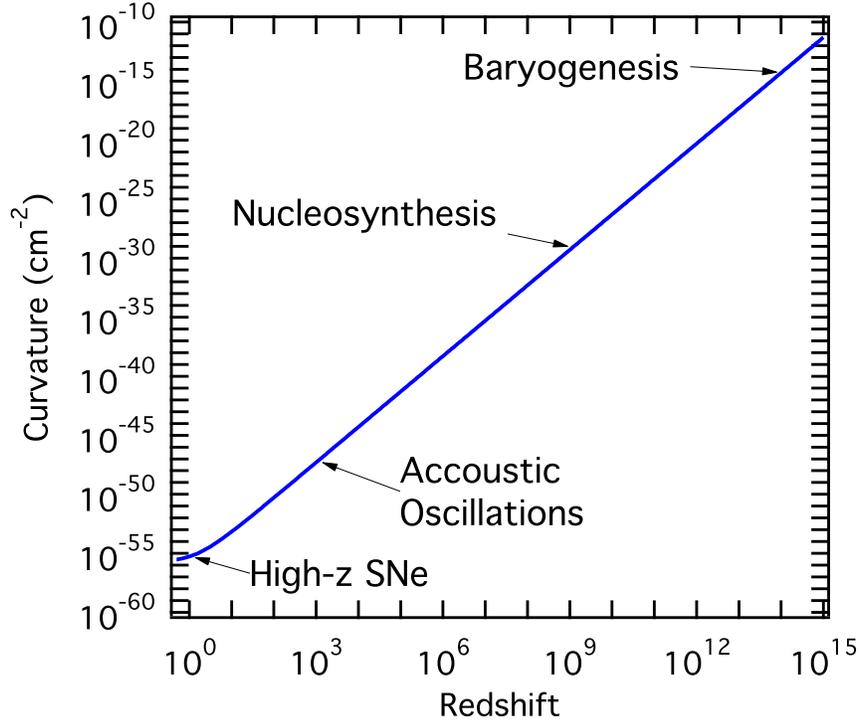}}
\caption{The scalar curvature of our universe, as a function
of redshift. The curve corresponds to a flat universe with the
best-fit values of the cosmological parameters obtained by the WMAP
mission~\cite{Spergel03}. The arrows point to the curvature and
redshift of the universe during various epochs.}
\label{fig:cosmo}
\end{figure}}

\subsection{Probing versus testing strong-field gravity}
\label{subsec:probvstest}

The parameter space shown in Figure~\ref{fig:fields} is useful in
identifying the strength of the gravitational field probed by a
particular test of gravity. However, it is important to emphasize that
probing a gravitational field of a given strength is not necessarily
the same as testing General Relativity in that regime. I discuss
bellow the difference with two examples from scalar-tensor gravity
that illustrate the two opposite extremes.

First, a phenomenon that occurs in a weak gravitational field may
actually be testing the strong-field regime of gravity. In General
Relativity, Birkhoff's theorem states that the external spacetime of a
spherically symmetric object is described by the Schwarzschild metric,
independent of the properties of the object itself. Birkhoff's
theorem, however, does not apply to a variety of gravity theories,
such as scalar-tensor or non-linear (e.g., $R+R^2$) theories. In fact,
in these theories, the spacetime at any point around a spherically
symmetric object depends on the mass distribution that generates the
spacetime, which may itself lie in a strong gravitational field and,
therefore, probe that regime of the theory. For example, in
Brans--Dicke gravity, which is a special case of scalar-tensor
theories, the evolution of the binary orbit in a system with two
neutron stars due to the emission of gravitational waves depends on
the coupling of matter to the scalar field, which occurs in the strong
gravitational field of each neutron
star~\cite{Eardley75, Will89, Damour96}.  As a result, even though the
gravitational field that corresponds to a double-neutron star orbit is
rather weak (see Figure~\ref{fig:tests}), observations of the orbital
decay of the binary actually test General Relativity against
scalar-tensor theories in the strong-field regime~\cite{Damour96}.

In the opposite extreme, phenomena that probe strong gravitational
fields cannot necessarily be used in testing General Relativity in
this regime. Analytical and numerical studies strongly suggest that
the end state of the collapse of a star in Brans--Dicke gravity is a
black hole described by the Kerr spacetime of General
Relativity~\cite{Thorne71, Bekenstein72, Hawking72, Scheel95,
Psaltis07b}. Therefore, the observation of a phenomenon that occurs
even just above the horizon of a black hole cannot be used in testing
General Relativity against Brans--Dicke gravity in the strong-field
regime, because both theories make the exact same prediction for that
phenomenon.

In the following, I will distinguish attempts to probe phenomena that
occur exclusively in the strong-field regime of General Relativity
from those that aim to test the strong-field predictions of the theory
against various alternatives.

\newpage

\section{Probing Strong Gravitational Fields with Astrophysical Objects}
\label{sec:probes}

A number of astrophysical objects offer the possibility of detecting
directly the observable consequences of two strong-field predictions
of General Relativity that have no weak-field or Newtonian
counterparts: the presence of a horizon around a collapsed object and
the lack of stable circular orbits in the vicinity of a neutron star
or black hole. As in most other areas of astrophysics research, we
have to rely on imaging, spectral, or timing observations in order to
reveal the information of the strong-field effects that is encoded in
the detected photons. The construction of gravitational wave
observatories will offer, for the first time in the near future, a
wealth of additional probes into the inner workings of gravitational
fields in the vicinities of compact objects.

In the following, I review a number of recent attempts to probe
strong-field phenomena that have used a variety of techniques and were
applied to different astrophysical objects. I will only discuss
phenomena that are observable in the electromagnetic spectrum and
refer to a number of excellent reviews on the gravitational phenomena
that are anticipated to be detected by gravitational wave
observatories~\cite{Hugh00, Flanagan05}.

\subsection{Black hole images}
\label{subsec:image}

To paraphrase the common proverb, {\em a picture is worth a thousand
spectra.} Directly imaging the vicinity of a black hole promises to
provide a direct evidence for the existence of a horizon. However,
black holes are notoriously small, and the resolution required for
imaging their horizons is, for most cases, beyond current
capabilities. For a stellar-mass black hole in the galaxy, the opening
angle of the horizon, as viewed by an observer on Earth, is only
\begin{equation}
\theta=2\times 10^{-4}\left(\frac{M}{10\,M_\odot}\right)
\left(\frac{1~\mbox{kpc}}{D}\right)~\mu\mbox{arcsec}\;,
\end{equation}
where $M$ is the mass of the black hole and $D$ is its distance.

For a supermassive black hole in a distant galaxy, the opening angle is
\begin{equation}
\theta=20\left(\frac{M}{10^9\,M_\odot}\right)
\left(\frac{1~\mbox{Mpc}}{D}\right)~\mu\mbox{arcsec}\;.
\end{equation}
This is shown in Figure~\ref{fig:images} for a number of supermassive
black holes with secure mass determinations. The angular size of the
horizons of some of the sources are barely resolvable today with
interferometric observations in the sub-mm/infrared wavelengths and
will be resolvable in the X-rays in near future with the Black Hole
Imager~\cite{Ozel01}.

\epubtkImage{images.png}{
\begin{figure}[htbp]
\centerline{\includegraphics[width=12.0cm]{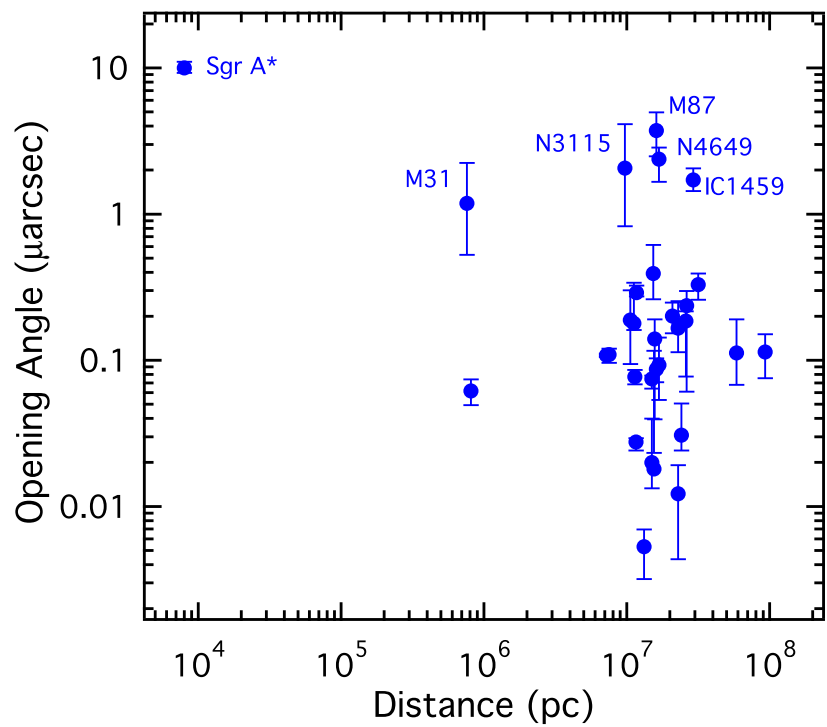}}
\caption{The opening angles, as viewed by an observer on
Earth, of the horizons of a number of supermassive black holes in
distant galaxies with a secure dynamical mass measurement (sample
of~\cite{Tremaine02}). The opening angle of the black hole horizon in
the center of the Milky Way (Sgr~A$^*$) is also shown for comparison.}
\label{fig:images}
\end{figure}}

The black hole that combines the highest brightness with the largest
angular size of the horizon is the one that powers the source
Sgr~A$^*$, in the center of the Milky Way. Since the first
measurements of the size of the source at 7~mm~\cite{Lo98} and at
1.4~mm~\cite{Krichbaum98} demonstrated that the emitting region is only
a few times larger than the radius of the horizon (see
Figure~\ref{fig:sgrsize}), a number of observational and theoretical
investigations have aimed to probe deeper into the gravitational field
of the black hole and constrain its properties.

\epubtkImage{size.png}{
\begin{figure}[htbp]
\centerline{\includegraphics[width=12.0cm]{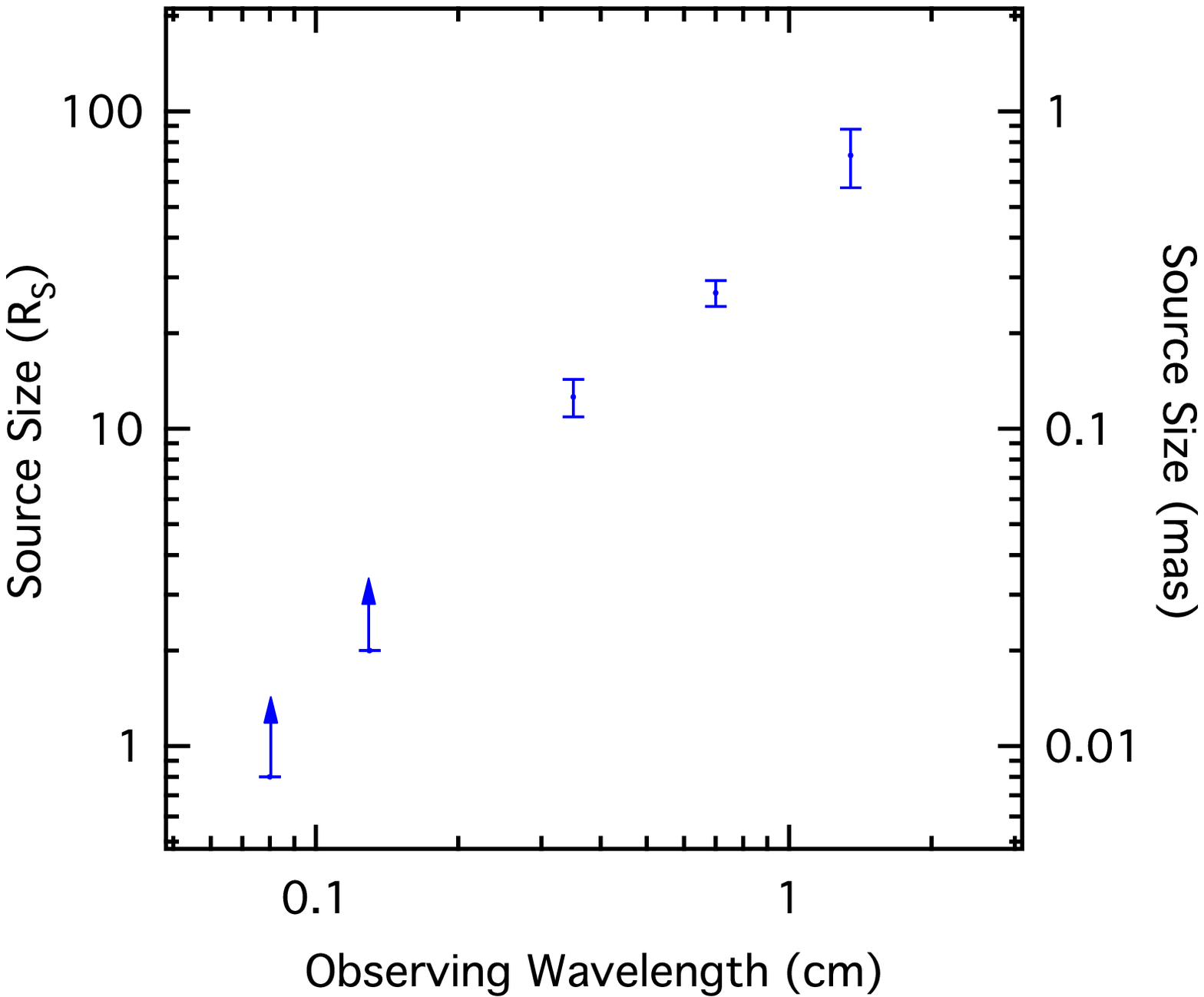}}
\caption{The major axis of the accretion flow around
the black hole in the center of the Milky Way, as measured at
different wavelengths, in units of the Schwarzschild radius (left axis) and
in milliarcsec (right axis; adapted from~\cite{Shen05}). Even with current
technology, the innermost radii of the accretion flow can be readily
observed.}
\label{fig:sgrsize}
\end{figure}}

The long-wavelength spectrum of Sgr~A$^*$ peaks at a frequency of
$\simeq 10^{12}$~Hz, suggesting that the emission changes from
optically thick (probably synchrotron emission) to optically thin at a
comparable frequency (see, e.g., \cite{Narayan95}). As a
result, observations at frequencies comparable to or higher than the
transition frequency can, in principle, probe the accretion flow at
regions very close to the horizon of the black hole.

Even though the exact shape and size of the image of Sgr~A$^*$ at long
wavelengths depends on the detailed structure of the underlying
accretion flow (cf.~\cite{Ozel00} and \cite{Yuan03}), there
exist two generic observable signatures of its strong gravitational
field. First, the horizon leaves a `shadow' on the image of the
source, which is equal to $\simeq\sqrt{27}GM/c^2$ and roughly
independent of the spin of the black hole~\cite{Bardeen73, Falcke00,
Takahashi04, Broderick06,Noble07}. Second, the brightness of the image of the
accretion flow is highly non-uniform because of the high velocity of
the accreting plasma and the effects of the strong gravitational
lensing. Simultaenously fitting the size, shape, polarization map, and
centroid of the image observed at different wavelengths with future
telescopes, will offer the unique possibility of removing the
complications introduced by the unknown nature of the accretion flow,
imaging directly the black hole shadow, and measuring the spin of the
black hole~\cite{Broderick06b}.

\subsection{Continuum spectroscopy of accreting black holes}
\label{subsec:contspec}

There have been at least three different efforts published in the
literature that use the luminosities and the continuum spectra of
accreting black holes to look for evidence of strong-field phenomena.

\epubtkImage{bh_vs_ns.png}{
\begin{figure}[htbp]
\centerline{\includegraphics[width=10.0cm]{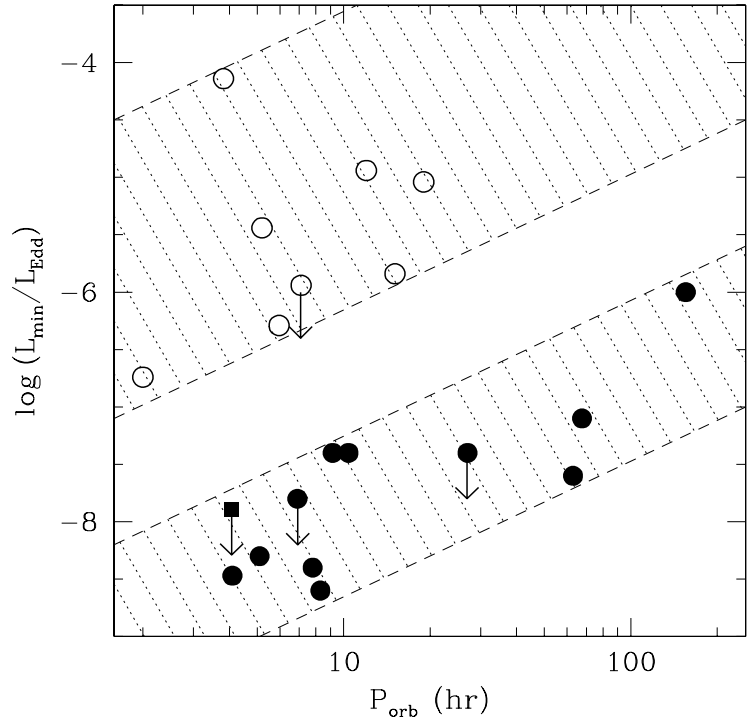}}
\caption{The 2--20~keV quiescent luminosities of black hole
candidates (filled circles) and neutron stars (open circles) in units
of the Eddington luminosity for different galactic binary systems, as
a function of their orbital periods, which are thought to determine
the mass transfer rate between the two stars. The systematically lower
luminisoties of the black hole systems have been attributed to the
presence of the event horizon~\cite{Narayan97, McClintock04}.}
\label{fig:bhvsns}
\end{figure}}

\subsubsection{Luminosities of black holes in quiescence and the absence 
of a hard surface}
\label{subsub:horizon}

More low-mass X-ray binaries are stellar systems in which the primary
star is a compact object and the secondary star is filling its Roche
lobe. Matter is transferred from the companion star to the compact
object and releases its gravitational potential energy mostly as
high-energy radiation, making these systems the brightest sources in
the X-ray sky~\cite{Psaltis06, McClintock06}.

The rate with which mass is transfered from the companion star to the
compact object is determined by the ratio of masses of the two stars,
the evolutionary state of the companion star, and the orbital
separation~\cite{Verbunt93}. On the other hand, the
rate with which energy is released in the form of high-energy
radiation depends on the rate of mass transfer, the state of the
accretion flow (i.e., whether it is via a geometrically thin disk or a
geometrically thick but radiatively inefficient flow), and on whether the
compact object has a hard surface or an event horizon. Indeed, for a
neutron-star system in steady state, most of the released
gravitational potential energy has to be radiated away (only a small
fraction heats the stellar core~\cite{Brown98}), whereas
for a black hole system, a significant amount of the potential energy
may be advected inwards past the event horizon, and hence may be forever
lost from the observable universe. For similar systems, in the same
accretion state, one would therefore expect black holes to be
systematically underluminous than neutron stars~\cite{Narayan97}.

The luminosities of transient black holes and neutron stars in their
quiescent states most clearly show this trend. When plotted against
the orbital periods of the binary systems, which are used here as
observable proxies to the mass transfer rates, sources that are
believed to be black holes, based on their large masses, are
systematically underluminous (Figure~\ref{fig:bhvsns}
and~\cite{Narayan97, Garcia01, McClintock04}). Although the physical
mechanism behind the difference in luminosities is still a matter of
debate~\cite{Narayan97, Bildsten00, Lasota00}, the trend shown in
Figure~\ref{fig:bhvsns} appears to be a strong, albeit indirect,
evidence for the presence of an event horizon in compact objects with
masses larger than the highest possible mass of a neutron star.

\subsubsection{Hard X-ray spectra of luminous black holes 
and the presence of an event horizon}
\label{subsub:hard}

Galactic black holes in some of their most luminous states (the
so-called very high states) have mostly thermal spectra in the soft
X-rays with power-law tails that extend well into the soft
$\gamma$-rays~\cite{Grove98}.  It has been hypothesized that these
power-law tails are the result of Compton upscattering of soft X-ray
photons off the relativistic electrons that flow into the black hole
event horizon with speeds that approach the speed of light and,
therefore, constitute an observational signature of the presence of an
event horizon (e.g., see~\cite{Titarchuk98, Laurent99})

A relativistic converging flow has indeed the potential of producing
power-law spectral tails (e.g., see~\cite{Payne81, Titarchuk97,
Psaltis01}).  However, this mechanism is identical to a second order
Fermi acceleration and hence the power-law tail is a result of
multiple scatterings away from the horizon with small energy exchange
per scattering rather than the result of very few scatterings of
photons with ultrarelativistic electrons near the black hole
horizon~\cite{Psaltis97, Papathanassiou00}. Moreover, the model
spectra always cut-off at energies smaller than the electron rest
mass~\cite{Laurent99, Niedzwiecki06} whereas the observed spectra
extend into the MeV range~\cite{Grove98}. Successful theoretical
models of the power-law spectra of black holes that are based on
Comptonization of soft photons by non-thermal
electrons~\cite{Gierlinski99} as well as the discovery of similar
power-law tails in the spectra of accreting neutron stars that extend
to $\sim 100-200$~keV~\cite{DiSalvo01, DiSalvo06} have shown
conclusively that the observed power-law tails do not constitute
evidence for black hole event horizons.

\subsubsection{Measuring the radii of the innermost stable 
circular orbits of black holes using continuum spectra}
\label{subsub:isco}

The thermal spectrum of a black hole source in some of its most
luminous states is believed to originate in a geometrically thin
accretion disk. The temperature profile of such an accretion disk away
from the black hole is determined entirely by energy conservation and
is independent of the magnitude and properties of the mechanism that
transports angular momentum and allows for matter to accrete (as long
as this mechanism is local; see~\cite{Shakura73, Balbus99}). The
situation is very different, however, near the radius of the innermost
stable circular orbit (hereafter ISCO).

Inside the ISCO, fluid elements cannot stay in circular orbits but
instead quickly loose centrifugal support and rapidly fall into the
black hole. The density of the accretion disk inside the ISCO is very
small and the viscous heating is believed to be strongly
diminished. It is, therefore, expected that only material outside the
ISCO contributes to the observed thermal spectrum. The temperature
profile of the accretion flow just outside the ISCO depends rather
strongly on the mechanism that transports angular momentum outwards
and in particular on the magnitude of the torque at the
ISCO~\cite{Krolik99, Gammie99, Agol00}. To lowest order, however, if
the entire accretion disk spectrum can be decomposed as a sum of
blackbodies, each at the local temperature of every radial annulus,
then the highest temperature will be that of the plasma near the ISCO
and the corresponding flux of radiation will be directly proportional
to the square of the ISCO radius.

\epubtkImage{spin.png}{
\begin{figure}[htbp]
\centerline{\includegraphics[width=13.0cm]{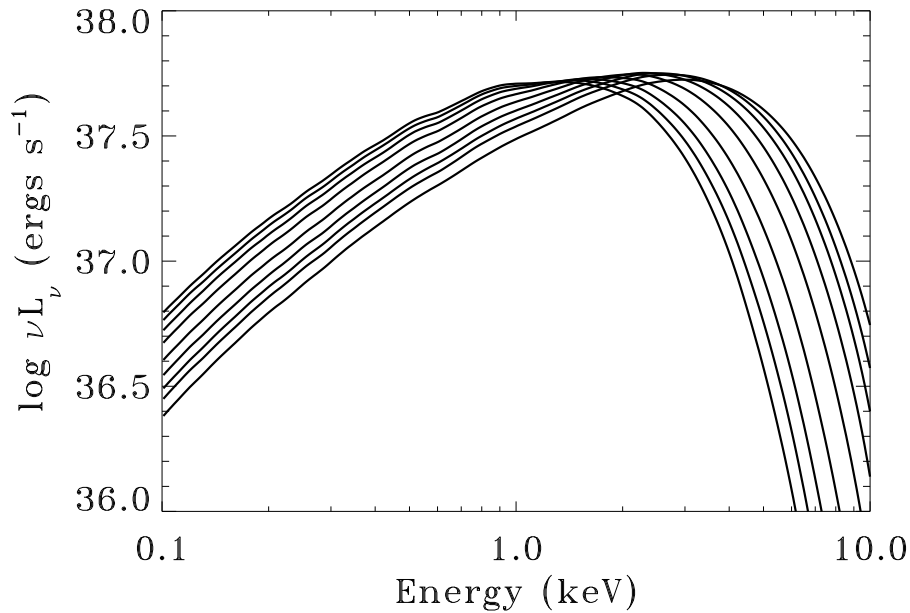}}
\caption{The spectra emerging from geometrically thin
accretion disks around black holes with different spins, but with the
same accretion luminosity~\cite{Davis05}.From left to right, the
curves correspond to spins ($a/M$) of 0, 0.2, 0.4, 0.6, 0.78, 0.881,
0.936, 0.966, and 0.99.  The spin values were chosen to give roughly
equal variation in the position of the spectral peak for spins $>0.8$.
The other parameters which determine the model are the viscosity
parameter, $\alpha=0.01$, the inclination of the observer, $\cos i=0.5$,
the mass of the black hole, $M=10\,M_\odot$, and the accretion
luminosity, $L=0.1 L_{\rm Edd}$. The peak energies of the spectra increase
with increasing spin, as a consequence of the fact that the ISCO
radius decreases with spin.}
\label{fig:bhspin}
\end{figure}}

Phenomenological fits of multi-temperature blackbody models to the
observed spectra of black holes provide strong support to the above
interpretation. When model spectra are fit to observations of any
given black hole in luminosity states that differ by more than one
order of magnitude, the inferred ISCO radius remains approximately
constant~\cite{Tanaka95}. For systems with a dynamically measured mass
and with a known distance, such an observation can lead to a
measurement of the physical size of the ISCO and hence of the spin of
the black hole~\cite{Zhang97, Gierlinski01} (see
Figure~\ref{fig:bhspin}).

There are a number of complications associated with producing the
model spectra of multitemperature blackbody disks that are required in
measuring spectroscopically the ISCO radius around a black
hole. First, as discussed above, the temperature profile of an
accretion disk at the region around the ISCO depends very strongly on
the details of the mechanism of angular momentum transport, which are
poorly understood~\cite{Krolik99, Gammie99, Agol00}. Second, the
vertical structure of the disk at each annulus, which determines the
emerging spectrum, may or may not be in hydrostatic equilibrium near
the ISCO, as it is often assumed, and its structure depends strongly
on the external irradiation of the disk plasma by photons that
originate in other parts of the disk. Finally, material in the inner
accretion disk is highly ionized and often far from local
thermodynamic equilibrium, generating spectra that can be
significantly different from blackbodies~\cite{Hubeny97}.

There have been a number of approximate models of multi-temperature
accretion disks that take into account some of these effects, in a
phenomenological or in an {\em ab initio} way. The models of Li et
al.~\cite{Li05} are based on the alpha-model for angular momentum
transport, assume that the local emission from each annulus is a
blackbody at the local temperature, and take into account the strong
lensing of the emitted photons by the central black hole. On the other
hand, the models of Davis et al.~\cite{Davis05} are the result of
ionization-equilibrium and radiative transfer calculations at each
annulus, they are based on the alpha model for angular momentum but
allow for non-zero torques at the ISCO, and take into account the
strong lensing of photons by the black hole. 

Given the flux $F$ of the accretion disk measured by an observer on
Earth, the color temperature $T_{\rm col}$ that corresponds to the
innermost region in the disk that is emitting (which presumably is
near the ISCO), the distance $D$ to the source, and the mass $M$ of
the black hole, the spin $\alpha$ of the black hole can be
inferred~\cite{Zhang97} by equating the radius of the ISCO, i.e.,
\begin{equation}
r_{\rm ISCO}=
\sqrt{\frac{2GM}{c^2}}\left\{3+A_2\pm [(3-A_1)(3+A_1+2A_2)]^{1/2}\right\}
\label{eq:fisco}
\end{equation}
to the one inferred spectroscopically (since $F\sim T^4 R^2$) by
\begin{equation}
r_{\rm spec}=
D \left[\frac{F}{2\sigma g(\theta, \alpha)}\right]^{1/2}
\left[\frac{f_{\rm col} f_{\rm GR}(\theta, \alpha)}{T_{\rm col}}\right]^2\;.
\label{eq:rspec}
\end{equation}
Here $A_1=1+(1-a^2)^{1/3}[(1+a)^{1/3}+(1-a)^{1/3}]$,
$A_2=(3a^2+A_1^2)^{1/2}$, $a$ is the specific angular momentum per
unit mass for the black hole, and the positive (negative) sign is
taken for prograde (retrograde) disks. In these equations, $\sigma$ is
the Stefan--Boltzmann constant and $\theta$ is the inclination of the
observer with respect to the symmetry axis of the accretion disk. The
functions $g(\theta, \alpha$) and $f_{\rm GR}(\theta, \alpha)$ are
correction factors for the flux and the temperature, respectively,
that need to be calculated when going from an accretion disk annulus
to a distant observer and incorporate the combined effects of
gravitational lensing, gravitational redshift, and Doppler boosting of
the disk photons. Given a thickness of the accretion disk, both these
transfer functions can be computed to any desired degree of
accuracy. Finally, the factor $f_{\rm col}$ measures the ratio of the
color temperature of the spectrum (as measured by fitting a blackbody
to the observed spectrum) to the effective temperature in that annulus
in the accretion disk (which is a measure of the total radiation flux
emerging from that annulus). Computing the value of the factor $f_{\rm
col}$ is the goal of the recent calculations of the ionization
equilibrium and radiative transfer in accretion disks~\cite{Davis05}.

Fitting these spectral models to a number of observations of black hole
candidates with dynamically measured masses has resulted in
approximate measurements of their spins: $a>0.7$ for
GRS~1915$+$105~\cite{Midleton05, McClintock06b}; $a=0.75-0.85$ for
4U~1543-44~\cite{Shafee06}; $a=0.65-0.75$ for
GRO~J1655$-$40~\cite{Shafee06}. It is remarkable that all inferred
values of the black hole spins are high, comparable to the maximum
allowed by the Kerr solution.

Equations~(\ref{eq:fisco}) and (\ref{eq:rspec}) demonstrate the strong
dependence of the inferred values of black hole spins on various
observable quantities (the mass of, distance to, and inclination of
the black hole, as well as the flux, and temperature of its disk
spectrum) and on a model parameter (the color correction factor
$f_{\rm col}$). Numerical simulations of magnetohydrodynamic flows
onto black holes are finely tuned to resolve the length- and
timescales of phenomena that occur in the vicinity of the horizon of a
black hole (see, e.g.,~\cite{Gammie03, deVilliers03}). When such
models incorporate accurate multi-dimensional radiative transfer, they
will provide the best theoretical spectra to be compared directly to
observations (see, e.g.,~\cite{Blaes06}). Moreover, monitoring of
the same sources at long wavelengths will improve the measurements of
their masses and distances.  Finally, combination of this with other
methods based on line spectra and the rapid variability properties of
accreting black holes will enable us to tighten the uncertainties in
the various model parameters and observed quantities that enter
Equation~(\ref{eq:fisco}) and measure with high precision the spins of
galactic black holes.

\subsection{Line spectroscopy of accreting compact objects}
\label{subsec:lines}

Heavy elements on the surface layers of neutron stars or in the
accretion flows around black holes that are not fully ionized generate
atomic emission and absorption lines that can be detected by a distant
observer with a large gravitational redshift. The value of the
gravitational redshift can be used to uniquely identify the region in
the spacetime of the compact object in which the observed photons are
produced.

\subsubsection{Atomic lines from the surfaces of neutron 
stars}
\label{subsub:linesns}

The gravitational redshift of an atomic line from the surface layer of
a neutron star leads to a unique determination of the relation between
its mass and radius. The detection of a rotationally broadened atomic
line from a rapidly spinning neutron star offers the additional
possibility of measuring directly the stellar radius~\cite{Ozel03,
Chang06} and, therefore, of determining its mass, as well.  The
profile of a rotationally broadened atomic line can be used to study
frame-dragging effects in the strong-field
regime~\cite{Bhattacharya05}. Moreover, detecting a gravitationally
redshifted and rotationally broadened atomic line can lead to a
measurement of the oblateness of the spinning star~\cite{Cadeau07},
which is determined by the strong-field coupling of matter with the
gravitationally field. Unfortunately, this is one of the very few
astrophysical settings discussed in this review in which observations
significantly trail behind theoretical investigations.

Despite many optimistic expectations and early claims (see,
e.g.,~\cite{Lewin93}), the observed spectra of almost all
weakly-magnetic neutron stars are remarkably featureless. The best
studied case is that of the nearby isolated neutron star
RX~J1856$-$3754, which was observed for 450~ks with the Chandra X-ray
Observatory and showed no evidence for any atomic lines from heavy
elements~\cite{Braje02}. This is in fact not surprising, given that
heavy elements drift inwards of the photosphere in timescales of
minutes~\cite{Bildsten92} and it takes only $\simeq 10^{-7}\,M_\odot$
of light elements to blanket a heavy element surface.

There are two types of neutron stars, however, in the atmospheres of
which heavy elements may abound: young cooling neutron stars and
accreting X-ray bursters~\cite{Ozel03}. On the one hand, the escaping
latent heat of the supernova explosion makes young neutron stars
relatively bright sources of X-rays. Their strong magnetic fields can
inhibit accretion of light elements either from the supernova fallback
or from the interstellar medium, leaving the surface heavy elements
exposed. On the other hand, in the atmospheres of accreting, weakly
magnetic neutron stars, heavy elements are continuously replenished.
Moreover, large radiation fluxes pass through their atmospheres during
thermonuclear bursts~\cite{Strohmayer06} making them very bright and
easily detectable.

The most promising detection to date of gravitationally redshifted
lines from the surface of a neutron star came from an observation with
XMM-Newton of the source EXO~0748$-$676, which showed redshifted
atomic lines during thermonuclear flashes~\cite{Cottam02}. This is a
slowly spinning neutron star (47~Hz~\cite{Villarreal04}) and hence its
external spacetime can be accurately described by the Schwarzschild
metric. In this case, the measurement of a gravitational redshift of
$z=0.35$ leads to a unique determination of the relation between the
mass and the radius of the neutron star, i.e., $M\simeq 1.4
(R/10$~km$)M_\odot$. Combination of this result with the spectral
properties of thermonuclear bursts during periods of photospheric
radius expansion and in the cooling tales also allowed for an
independent determination of the mass and radius of the neutron
star~\cite{Ozel06}.

Future observations of bursting or young neutron stars with upcoming
X-ray missions such as Constellation-X~\cite{ConX} and
XEUS~\cite{XEUS} have the potential of detecting many gravitationally
redshifted atomic lines and, hence, of probing the coupling of matter
to the strong gravitational fields found in the interiors of neutron
stars.

\subsubsection{Relativistically broadened iron lines in accreting
black holes}
\label{subsub:ironbh}

Astrophysical black holes in active galactic nuclei accreting at
moderate rates offer another possibility of probing strong
gravitational fields using atomic spectroscopy (for an extensive
review on the subject see~\cite{Reynolds03}; see also
\cite{Miller07} for a review of iron line observations from 
stellar-mass black holes). The relatively cool accretion disks in
these systems act as large mirrors, reflecting the high-energy
radiation that is believed to be produced in the disk coronae by
magnetic flaring~\cite{Guilbert88}. The spectrum of reflected
radiation in hard X-rays is determined by electron scattering, whereas
the spectrum in the soft X-rays is characterized by a large number of
fluorescent lines caused by bound-bound transitions of the partially
ionized material. The combination of the high yield and relatively
high abundance of iron atoms in the accreting material make the iron
K$\alpha$ line, with a rest energy of 6.4~keV for a neutral atom, the
most prominent feature of the spectrum.

\epubtkImage{iron_lines_a.png}{
\begin{figure}[htbp]
\centerline{\includegraphics[width=7.5cm]{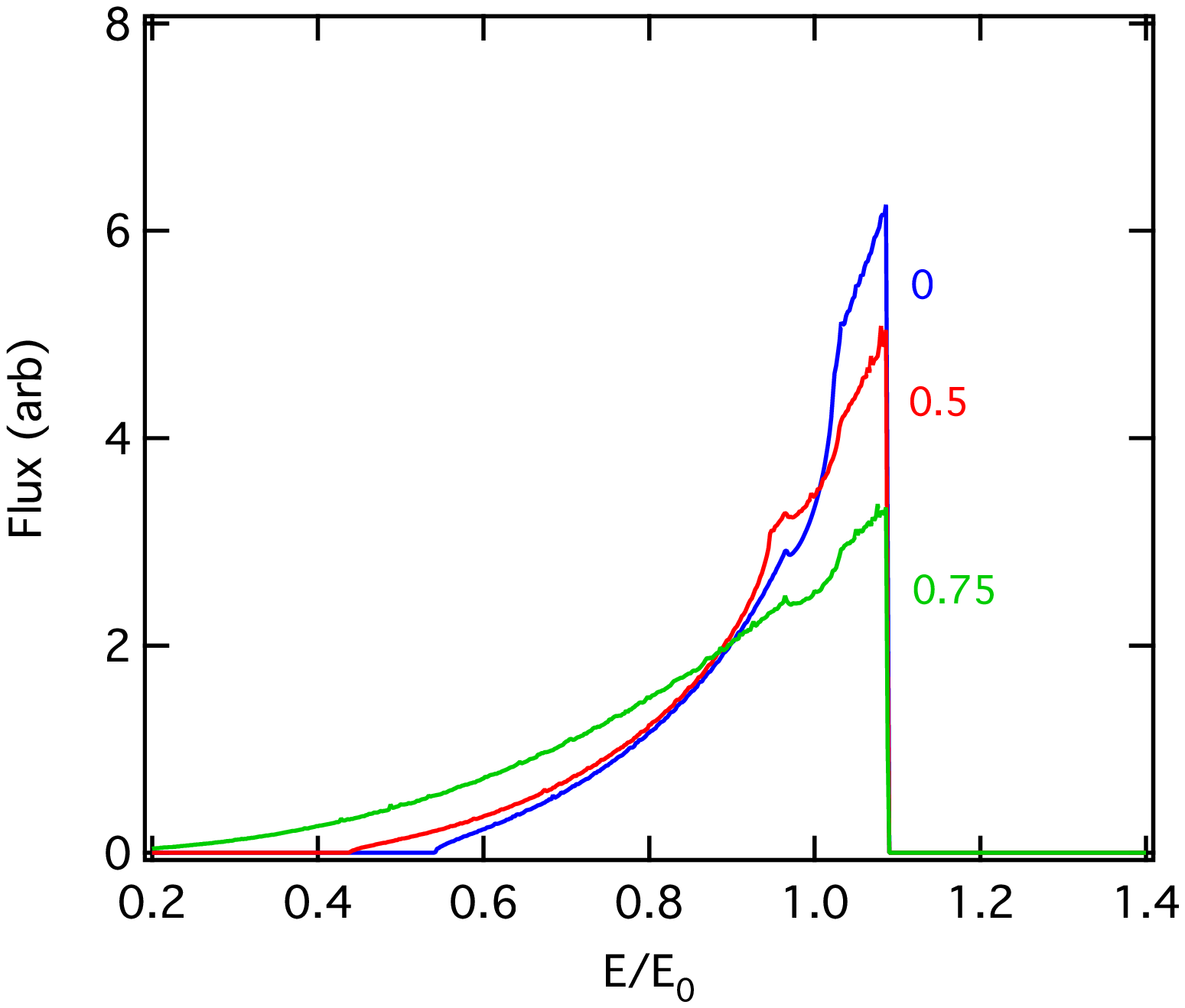}
\includegraphics[width=7.5cm]{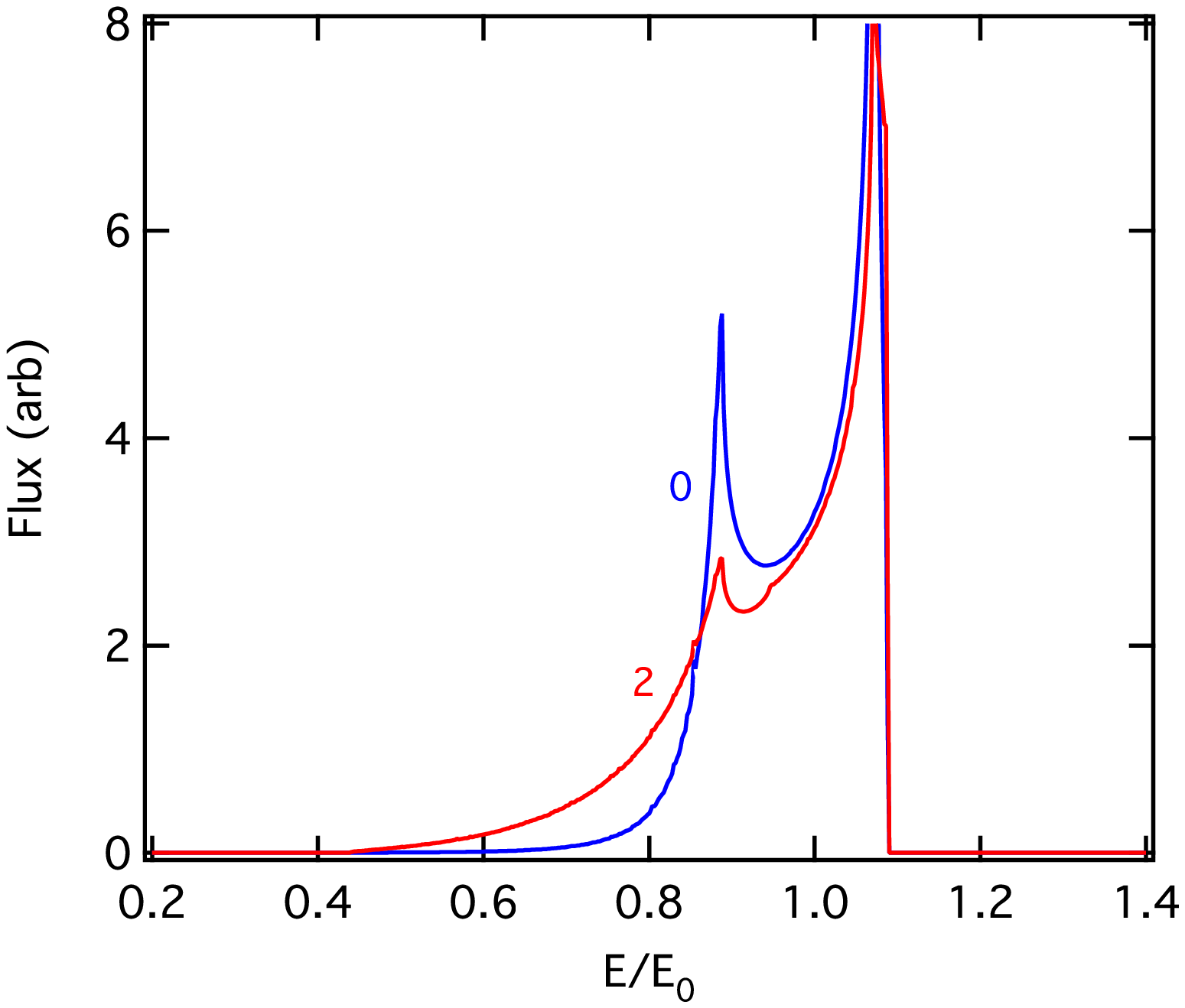}}
\caption{Theoretical models of relativistically broadened
iron line profiles from accretion flows around black holes. The left
panel shows the dependence of the line profile on the spin parameter
of the black hole, whereas the right panel shows its dependence on the
emissivity index (see text). All calculations were performed for an
inclination angle of 40$^\circ$~\cite{Brenneman06}.}
\label{fig:ironlines_th}
\end{figure}}

The profile of the fluorescent iron line as observed at infinity is
determined mainly by general and special relativistic effects that
influence the propagation of photons from the point of reflection to the
observer~\cite{Laor91}. Dividing an accretion disk into a series of
concentric rings orbiting at the local Keplerian frequency, special
relativistic effects produce a rotational splitting of the line
emerging from each ring, whereas general relativistic effects generate
an overall redshift~\cite{Fabian89}. The combination of these effects
integrated over the entire surface of the accretion disk leads to a
characteristic profile for the iron reflection line, which is broad
with a shallow and extended red wing (Figure~\ref{fig:ironlines_th}).

The magnitude of the relativistic effects depends on the specifics of
the spacetime of the black hole, on the position and orientation of
the observer, on the position and properties of the source of X-rays
above the accretion disk, and on the dependence of fluorescence yield
on position on the accretion disk through its dependence on the 
ionization states of the elements~\cite{George91}. Given a model for
the source of X-rays and the accretion disk, fitting the profile of an
iron line from an accreting black hole can lead, in principle, to a
direct mapping of its spacetime. Unfortunately, the source of X-ray
illumination and the physical properties of the accretion flows
themselves are poorly understood.

If we make assumptions regarding these astrophysical complications
that are largely model independent, a general property of the
spacetime, such as the spin of the black hole, can be measured. The
accretion disk is typically modeled as a geometrically thin reflecting
surface at the rotational equator of the black hole that is extending
inwards until the radius of the innermost stable circular orbit. Even
though the density of the material inside this radius is significant
and might reflect the illuminating X-rays, its ionization state
changes rapidly, leading to small changes in the resulting iron line
profile~\cite{Reynolds97, Brenneman06}.  The extent of the iron line
towards lower energies is a measure of the innermost radius of the
accretion disk. By assumption, this radius is set as the radius of the
innermost stable circular orbit, which depends on the spin of the
black hole. Fitting theoretical models to observations can, therefore,
lead to a measurement of the black hole spin.

The uncertainties in the position of the illuminating source and in
the disk structure are often modeled by a single function for the
``emissivity'' of the iron line, which measures the flux in the iron
line that emerges locally from each patch on the accretion disk. This
is typically taken to be axisymmetric and to have a power-law
dependence on radius, i.e., $r^{-a}$. Increasing the emissivity index
$a$ results in iron line profiles with more extended red wings, which
is degenerate with increasing the spin of the black hole (see
Figure~\ref{fig:ironlines_th} and~\cite{Beckwith04}). This uncertainty
can introduce significant systematics in modeling iron-line profiles
from slowly spinning black holes. For rapidly spinning black holes,
however, masking the effect of the black hole spin by steepening the
emissivity function requires an unphysically high value for the
emissivity index~\cite{Brenneman06}.

Since the original observation of broadened iron lines from the
supermasive black hole MCG-6-15-30 with ASCA~\cite{Tanaka95b},
observations of other active galactic nuclei with
ASCA~\cite{Nandra97}, XMM-Newton~\cite{Nandra06}, and more recently
with Suzaku~\cite{Reeves06} as well as of stellar mass black
holes~\cite{Miller06} have revealed many more examples of such
redshifted atomic lines. The best studied case remains MCG-6-15-30
(see Figure~\ref{fig:ironlines_obs}), in which the extended red wing of
the line has been discussed as evidence for a rapidly spinning black
hole ($\alpha\ge 0.98$~\cite{Brenneman06}).

\epubtkImage{iron_line_obs.pbg}{
\begin{figure}[htbp]
\centerline{\includegraphics[width=10.0cm]{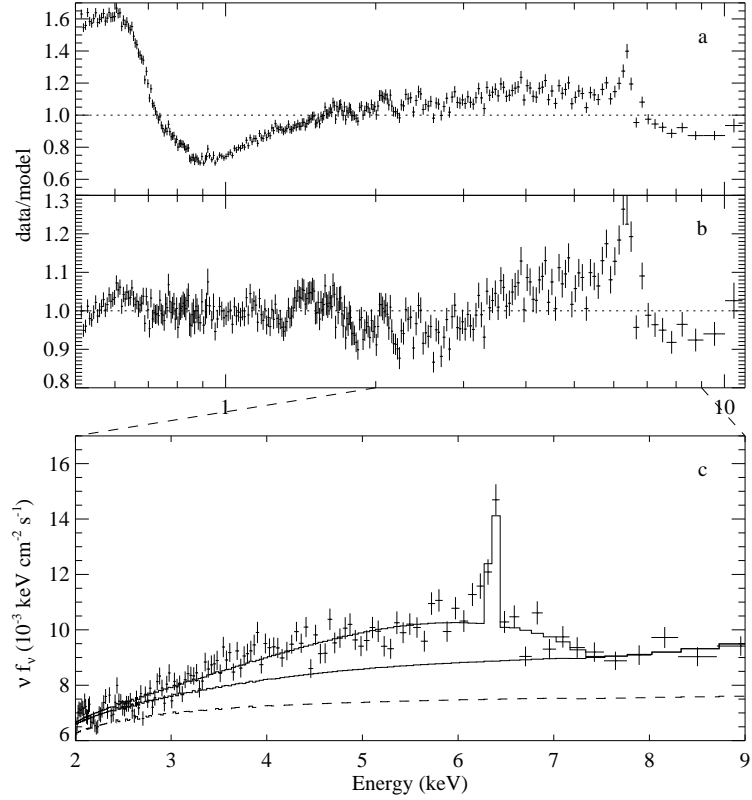}}
\caption{The 0.5-10~keV spectrum of the supermassive black hole 
in the center of the galaxy MCG-6-15-30 as observed with
XMM-EPIC. Panel (a) shows the ratio of the observed spectrum to a
power-law model and reveals the complicated structure of the
residuals. Panel (b) shows the ratio of the observed spectrum to a
model of the warm absorber, which accounts for the low-energy
residuals.  Panel (c) shows the 2-9~keV spectrum of the source
together with a model of the relativistically broadened iron
line~\cite{Wilms01}}
\label{fig:ironlines_obs}
\end{figure}}

Perhaps the most challenging, although most rewarding to understand,
property of iron lines is their time variability. Current observations
of iron lines from accreting black holes (e.g., the one shown in
Figure~\ref{fig:ironlines_obs}) are integrated over a time that is
equal to many hundred times the dynamical timescale in the
accretion-disk region where the lines are formed. As a result, an
observed line profile is not the result of reflection from an
accretion disk of a single flaring event, but rather the convolution
of many such events that occurred over the duration of the
observation. Moreover, the continuum spectrum of the black hole, which
is presumably reflected off the accretion disk to produce the
fluorescent iron line, changes over longer timescales, implying a
correlated variability of the line itself.

Observations with current instruments can only investigate the
correlated variability of the iron line with the continuum spectrum
(see, however,~\cite{Iwasawa04}). They have shown that the flux in the
line remains remarkably constant even though the continuum flux
changes by almost an order of magnitude~\cite{Fabian03}. General
relativistic light bending, which leads to focusing of the
photon rays towards the innermost regions of the accretion disk may
be responsible for this puzzling effect~\cite{Miniutti04}.

Future observations with upcoming X-ray missions such as
Constellation~X~\cite{ConX} and XEUS~\cite{XEUS} will resolve the time
evolution of the reflected iron line from a single magnetic
flare~\cite{Reynolds99}. Because density inhomogeneities in the
turbulent accretion flow move, roughly, in test-particle
orbits~\cite{Armitage03}, the time evolution of the redshift of the
iron line from a single flare reflected mainly off a localized density
inhomogeneity will allow for a direct mapping of the spacetime around 
the black hole.

\subsection{The fast variability of accreting compact objects}
\label{subsec:var}

The strongest gravitational fields in astrophysics can be probed only
with rapidly variable phenomena around neutron stars and galactic
black holes (see Figure~\ref{fig:tests_future}). Such phenomena have
been discovered in almost all known accreting compact objects in the
galaxy. They are quasi-periodic oscillations (QPOs) with frequencies
in the range $\sim 1$~Hz$-1$~kHz that remain coherent for tens to
hundreds of cycles and follow a rich and often complicated
phenomenology (for an extensive review of the observations
see~\cite{vanderKlis06}).

\epubtkImage{qpo1820.pbg}{
\begin{figure}[htbp]
\centerline{\includegraphics[width=10.0cm,angle=-90]{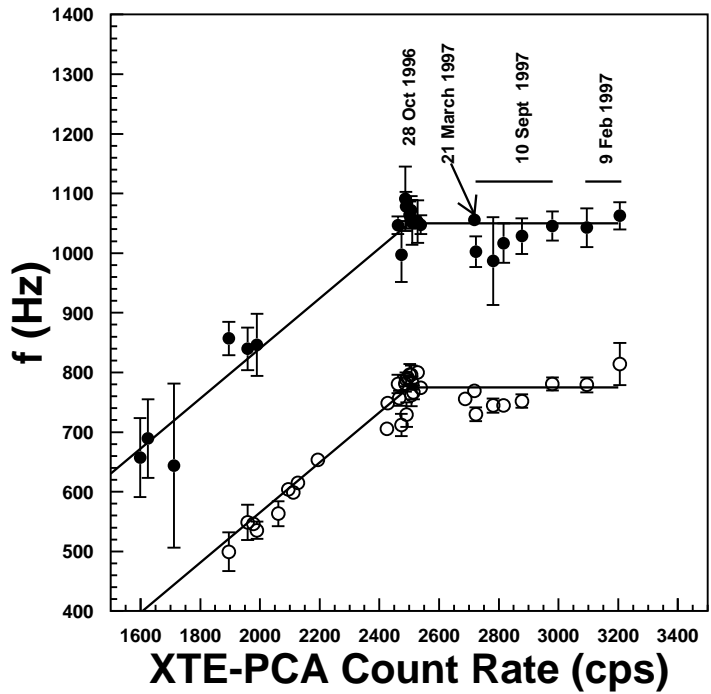}}
\caption{The dependence of the twin QPO frequencies
on the X-ray countrate observed by the PCA instrument onboard RXTE,
for the neutron-star source 4U~1820$-$30~\cite{Zhang98}.  The
flattening of the correlation at high frequencies has been discussed
as a signature of the innermost stable circular orbit.}
\label{fig:qpo_isco}
\end{figure}}

\subsubsection{Quasi-periodic oscillations in neutron stars}

The fastest oscillations detected from accreting, weakly magnetic
neutron stars are pairs of QPOs with variable frequencies that reach
up to $\sim 1300$~Hz and with frequency separations of order $\sim
300$~Hz~\cite{vanderKlis06}. The origin of these oscillations is still
a matter of debate. However, all current models associate at least one
of the oscillation frequencies with a characteristic dynamical
frequency in a geometrically thin accretion disk (see discussion
in~\cite{Psaltis01a} and~\cite{Miller98, Stella99, Psaltis00b}).

The highest dynamical frequency excited at any radius in an equatorial
accretion disk around a compact object is the one associated with the
circular orbit of a test particle at that radius~\cite{Bardeen73b};
this is often referred to as the azimuthal, orbital, or Keplerian
frequency. A mode in the accretion disk associated with this frequency
can give rise to a long-lived quasi-periodic oscillation only if it
lives outside the innermost stable circular orbit. The azimuthal
frequency at this radius provides, therefore, an upper limit on the
frequency of any observed oscillation~\cite{Kluzniak85, Miller98}. As
a result, detecting such rapid oscillations offers the possibility of
measuring the location and of understanding the properties of the
region near the innermost stable circular orbit around a neutron star.

\begin{figure}[t]
\centerline{
\includegraphics[width=10.0cm, height=10.5cm]{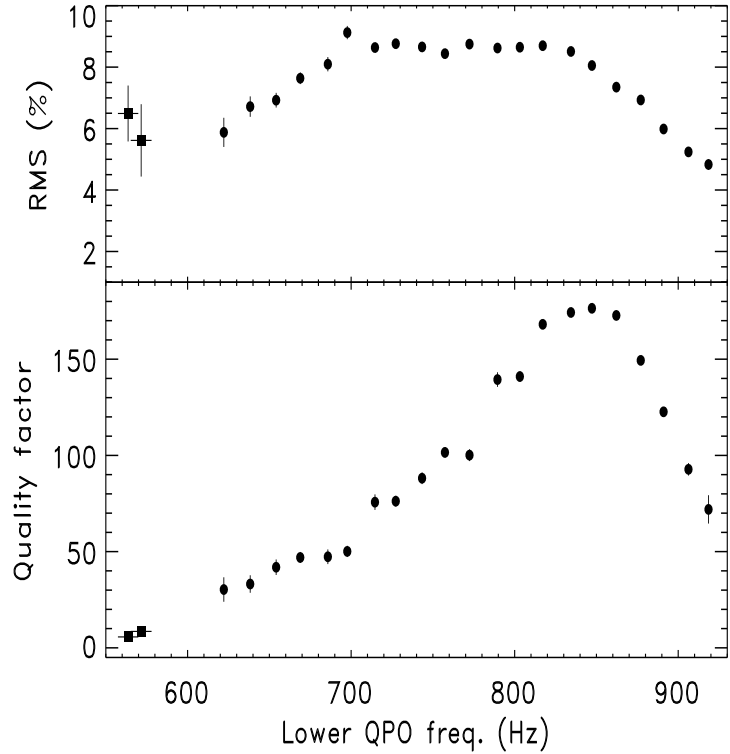}}
\caption{
The dependence of the amplitude and quality factor of the lower kHz QPO on
its frequency for the neutron-star source
4U~1636$-$56~\cite{Barret06}. The drop of the QPO amplitude and
coherence at high frequencies have been discussed as signatures of
the innermost stable circular orbit.}
\label{fig:qpo_isco2}
\end{figure}

The signature of the ISCO on the amplitudes and characteristics of the
observed oscillations is hard to predict without a firm model for the
generation of the oscillations in the X-ray flux. Two potential
signatures have been discussed, however, based on the phenomenology of
the oscillations. The first one is associated with the fact that the
frequencies of the oscillations appear to increase roughly with
accretion rate. When an oscillation frequency reaches that of the
innermost stable circular orbit, one would expect its frequency to
remain constant over a wide range of accretion
rates~\cite{Miller98}. Such a trend has been observed in the
quasi-periodic oscillations of the globular cluster source
4U~1820$-$30~\cite{Zhang98, Kaaret99}; Figure~\ref{fig:qpo_isco}). When
observations of the source obtained over different epochs are
combined, the dependence of the frequency of the fastest oscillation
on the observed accretion rate appears to flatten at a value $\simeq
1050$~Hz.  This is comparable to the azimuthal frequency at the
innermost stable circular orbit for a $\simeq 2.1\,M_\odot$ neutron
star~\cite{Zhang98}. 

Albeit suggestive, the interpretation of the 4U~1820$-$30 data relies
on the assumption that the oscillatory frequencies in an accretion
disk depend monotonically on the accretion rate and, furthermore, that
the X-ray countrate is a good measure of the accretion rate.  This
assumption is probably justified for short timescales (of order one
day) but is known to break down on longer timescales, such as those
used in Figure~\ref{fig:qpo_isco}~\cite{vanderKlis01}. Indeed, in a
given source, the same oscillation frequencies have been observed over
a wide range of X-ray countrates and vice
versa~\cite{vanderKlis01}. The hard X-ray color of a source and not
the countrate appears to be a more unique measure of the accretion rate,
which is presumably the physical parameter that determines the
oscillation frequencies~\cite{Mendez99}. When the data of 4U~1820$-$30
are plotted against hard color, the characteristic flattening seen in
Figure~\ref{fig:qpo_isco} disappears~\cite{Mendez99}.

A second signature of the innermost stable circular orbit is a
potential decrease in the amplitude and coherence of the oscillations
when the region where they are excited approaches the ISCO. Such a
trend has been observed in a number of accreting neutron
stars~(Figure~\ref{fig:qpo_isco2} and~\cite{Barret06, Barret07})
and has been questioned on similar grounds as the study of
4U~1820$-$30~\cite{Mendez06}. The most significant criticism comes from
the fact that the drop in amplitude and coherence is rather gradual
and occurs over a $\sim 150$~Hz range of frequencies. Even assuming
that this drop is a signature of the ISCO, measuring its location will
be possible only within a detailed model of the frequencies of
quasi-periodic oscillations.

Among more model-dependent ideas, perhaps the most exciting prospect
of probing strong-field gravity effects in neutron stars with
quasi-periodic oscillations comes from applying the relativistic model
of QPOs~\cite{Stella99} to the observed correlations between various
pairs of QPO frequencies~\cite{Psaltis99}. In the relativistic model,
the highest-frequency QPO is identified with the azimuthal frequency
of a test particle in orbit at a given radius.  The peak separation of
this QPO with the second higher frequency QPO is identified as the
radial epicyclic frequency of the test particle in the same orbit. A
variant of this model can account for the observed correlations
between oscillations frequencies, when hydrodynamic effects are taken
into account~\cite{Psaltis00b}. Because the two observed frequencies
are directly related to the azimuthal and radial frequencies at
various radii in the accretion flow, interpretation of the data with
this model can provide a direct map of the exterior spacetime of the
neutron stars, to within the $\simeq 10$\% uncertainty introduced by
the hydrodynamic corrections to the oscillation frequencies.

\subsubsection{Quasi-periodic oscillations in black holes}
\label{subsub:bhqpo}

Pairs of rapid quasi-periodic oscillations have also been detected
from a number of accreting systems that harbor black hole
candidates~\cite{McClintock06}. The phenomenology of these
oscillations is very different from the one discussed above for
accreting neutron stars. The frequencies of the rapid oscillations
observed in each source vary at most by a percent over a wide range of
luminosities and their ratios are practically equal to ratios of small
integers (2:3 for XTE~J1550$-$564 and GRO~J1655$-$40, 3:5 for
GRS~1915$+$105, etc.).

The high frequencies of the oscillations observed from black hole
sources with dynamically measured masses demonstrate that they
originate in regions very close to the black hole horizons. In fact,
requiring the frequency of the 450~Hz oscillation observed from
GRO~J1655$-$40 to be limited by the azimuthal frequency at the ISCO
necessitates a spining black hole with a Kerr spin parameter $a/M\ge
0.25$~\cite{Strohmayer01}. Moreover, the frequencies of the observed
oscillations are roughly inversely proportional to the black holes
masses, as one would expect if they were associated to dynamical
frequencies near the innermost stable circular
orbit~\cite{Abramowicz04}.

\epubtkImage{bhqpo_lin.png}{
\begin{figure}[htbp]
\centerline
{\includegraphics[width=7.0cm]{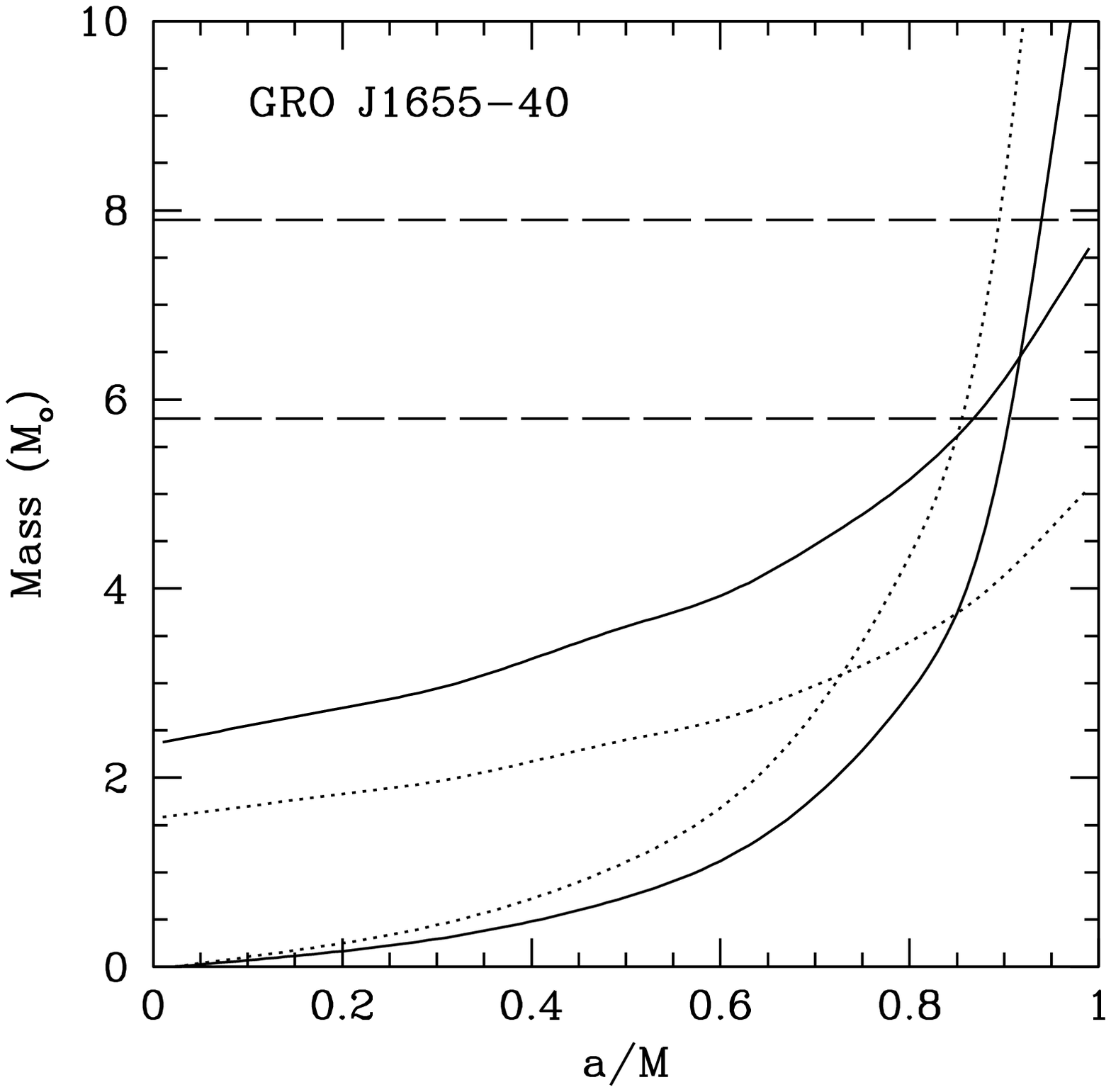}
\includegraphics[width=7.0cm]{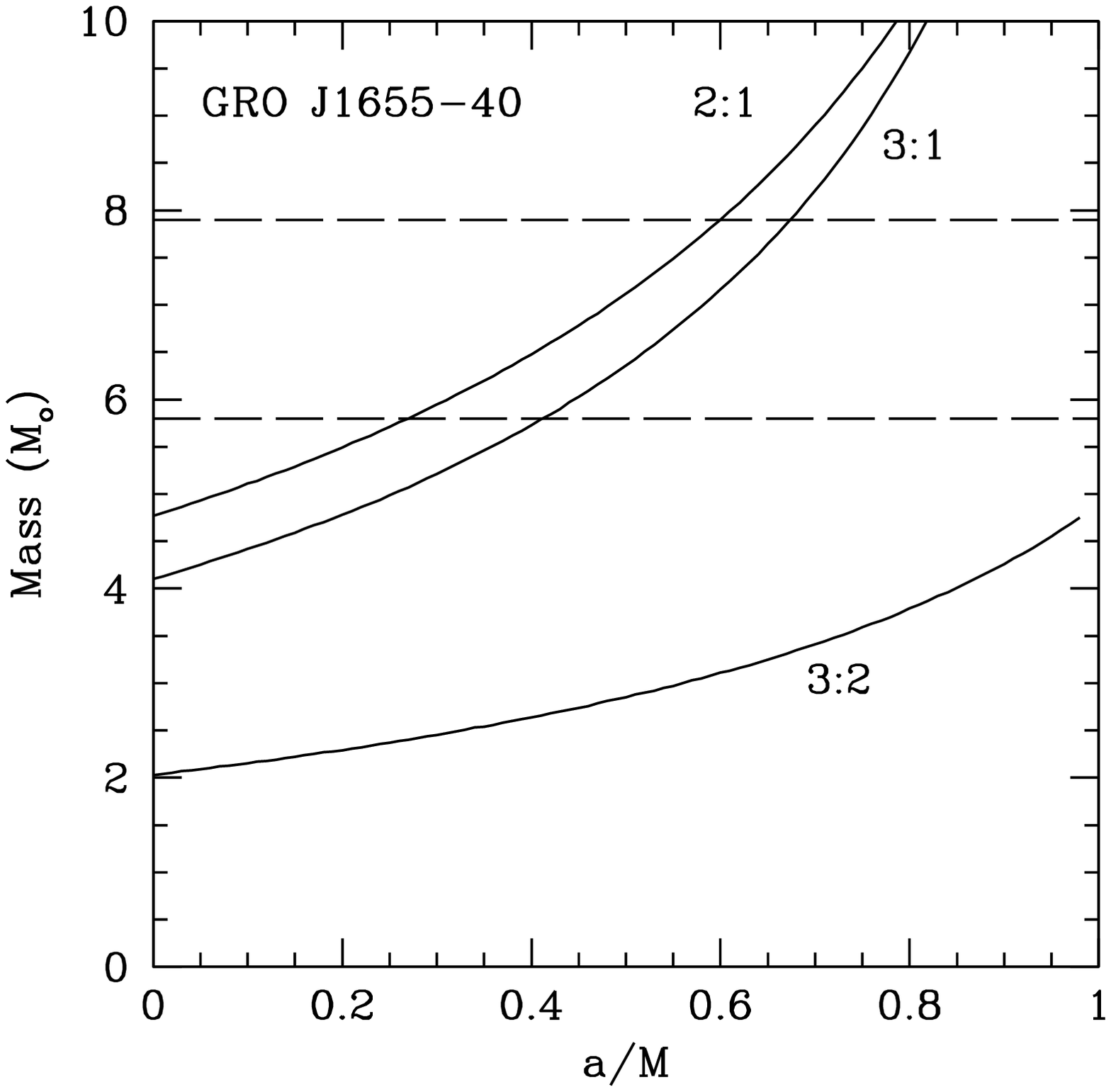}}
\caption{
(Left Panel) The intersection of the two solid lines shows the
black hole mass and spin for the source GRO~J1655$-$40 for which the
observed 300~Hz and the 450~Hz oscillations can be explained as the
lowest order $c$- and $g$-modes, respectively. The intersection of the
dotted lines makes the opposite identification of disk modes to the
observed oscillatory frequencies (after~\cite{Wagoner01}). (Right
Panel) Each solid line traces pairs of black hole mass and spin for
which the observed frequencies correspond to different resonances
between the Keplerian and periastron precession frequencies
(after~\cite{Abramowicz02}). In both panels, the horizontal dashed
lines show the uncertainty in the dynamically measured mass of the
black hole.}
\label{fig:bhqpo}
\end{figure}}

As in the case of neutron stars, using black hole quasi-periodic
oscillations to probe directly strong gravitational fields is hampered
by the lack of a firm understanding of the physical mechanism that is
producing them. In one interpretation, they are associated with linear
oscillatory modes that are trapped just outside the radius of the
innermost stable circular orbit (for reviews see~\cite{Wagoner99,
Kato01, Nowak01}). The frequencies of these modes depend primarily on
the mass and spin of the black hole. Identifying the two observed
oscillations with the lowest order linear modes, therefore, leads to
two pairs of values for the mass and spin of the black hole (depending
on which oscillation is identified with which mode). For example, for
the case of the black hole GRO~J1655$-$40, one of the inferred pairs
of values agrees with the dynamically measured mass of the black hole
of $6.9\pm 1.0\,M_\odot$ and results in an estimated value of the
black hole spin of $a/M\sim 0.9$ (Figure~\ref{fig:bhqpo}
and~\cite{Wagoner01}). Although compelling, this interpretation leaves
to coincidence the fact that the ratios of the oscillation frequencies
are approximately equal to ratios of small integers.

In an alternate model, the oscillations are assumed to be excited in
regions of the accretion disks where two of the dynamical frequencies
are in parametric resonance, i.e., their ratios are equal to ratios of
small integers~\cite{Abramowicz02}. In this case, the
frequencies of the oscillations depend on the mass and spin of the
black hole as well as on the radius at which the resonance occurs. As
a result, the observation of two oscillations from any given source
does not lead to a unique measurement of its mass and spin, but rather
to a families of solutions. For example, identifying the frequencies of
the two oscillations observed from GRO~J1655$-$40 as a 3:2, a 3:1, or
a 2:1 resonance between the Keplerian and the periastron precession
frequencies at any radius in the accretion disk leads to three family
of solutions, as shown in Figure~\ref{fig:bhqpo}. The dynamically
measured mass of the black hole then picks only two of the possible
families of solutions and leads to a smaller value for the inferred
spin.

Future observations of accreting neutron stars and black holes with
upcoming missions that will have fast timing capabilities, such as
XEUS~\cite{XEUS}, will be able to discover a large spectrum of
quasi-periodic oscillations from each source. Such observations will
constrain significantly the underlying physical model for these
oscillations, which remains the most important source of uncertainty
in using fast variability phenomena in probing strong gravitational
fields.

\newpage

\section{The Need for a Theoretical Framework for Strong-Field Gravity
Tests}
\label{sec:need}

Modern observations of black holes and neutron stars in the galaxy
provide ample opportunities for testing the predictions of general
relativity in the strong field regime, as discussed in the previous
section. In several cases, astrophysical complications make such
studies strongly dependent on model assumptions. This will be remedied
in the near future, with the anticipated advances in the observational
techniques and in the theoretical modeling of the various
astrophysical phenomena. A second difficult hurdle, however, in
performing quantitative tests of gravity with compact objects will be
the lack of a parametric extension to General Relativity, i.e., the
equivalent of the PPN formalism, that is suitable for calculations in
the strong-field regime.

In the past, {\em bona fide\/} tests of strong-field general
relativity have been performed using particular parametric extensions
to the Einstein--Hilbert action. This appears a priori to be a
reasonable approach for a number of reasons. First, deriving the
parametric field equations from a Lagrangian action ensures that
fundamental symmetries and conservation laws are obeyed. Second, the
parametric Lagrangian action can be used over the entire range of
field strengths available to an observer and, therefore, even tests of
General Relativity in the weak-field limit (i.e., with the PPN
formalism) can be translated into constraints on the parameters of the
action. This is often important, when strong-field tests lead to
degenerate constraints between different parameters. Finally,
phenomenological Lagrangian extensions can be motivated by ideas of
quantum gravity and string theory and, potentially, help constrain the
fundamental scales of such theories. There are, however, several
issues that need to be settled before any such parametric extension of
the Einstein--Hilbert action can become a useful theoretical framework
for strong-field gravity tests (see also~\cite{Sotiriou08} and
references therein).

{\em First, gravity is highly non-linear and strong-field phenomena
often show a non-perturbative dependence on small changes to the
theory.~--\/} I will illustrate this with scalar-tensor theories that
result from adding a minimally coupled scalar field to the Ricci
curvature in the action. Such fields have been studied for more than
40 years in the form of Brans--Dicke gravity~\cite{Will93} and have
been recently invoked as alternatives to a cosmological constant for
modeling the acceleration of the universe~\cite{Peebles03}. In the
context of compact-object astrophysics, constraints on the relative
contribution of scalar fields coupled in different ways to the metric
have been obtained from observations of the orbital decay of double
neutron stars~\cite{Will89, Damour93} and compact X-ray
binaries~\cite{Will89, Psaltis07a}. More recently, similar constraints
on scalar extensions to General Relativity have been placed using the
observation of redshifted lines from an X-ray burster~\cite{DeDeo03}
and of quasi-periodic oscillations observed in accreting neutron
stars~\cite{DeDeo07}.  The oscillatory modes of neutron stars in such
theories and the prospect of constraining them using gravitational
wave signatures have also been studied~\cite{Sotani04, Sotani05}.

The general form of the Lagrangian of a scalar-tensor theory is given,
in the appropriate frame, by the Bregmann-Wagoner action
(see~\cite{Will93} for details)
\begin{equation}
S=\frac{1}{16\pi}\int d^4x \sqrt{-g_*}\left[R_* \pm 
g_*^{\mu\nu}\partial_\mu \phi \partial_\nu\phi +
2\lambda(\phi)\right]+S_{\rm m}[\phi_m, A^2(\phi)g_{*\mu\nu}]\;,
\label{eq:st}
\end{equation}
where $A(\phi)$ and $\lambda(\phi)$ are two arbitrary functions, and
$S_m$ is the action for the matter field $\phi_m$.  In the
strong-field regime, the potential term $\lambda(\phi)$ in the
action~(\ref{eq:st}) is typically negligible and is set to zero. On the
other hand, the functional form of the coupling function $A(\phi)$ can
be parametrized to measure deviations from General Relativity.

Damour and Esposito-Farese~\cite{Damour93} considered a second-order
parametric form
\begin{equation}
A(\phi)= \exp\left[\alpha_0(\phi-\phi_0)+
   \frac{1}{2}\beta_0(\phi-\phi_0)^2+...\right]\;,
\label{eq:stcoupling}
\end{equation}
with $\phi_0\rightarrow 0$ a background cosmological value for the
scalar field and $\alpha_0$ and $\beta_0$ the two parameters of the
theory to be constrained by observations. The linear term,
parametrized by $\alpha_0$, can be best constrained with weak-field
tests.  On the other hand, constraining significantly the non-linear
term, parametrized by $\beta_0$, requires strong-field phenomena, such
as those found around neutron stars. Indeed, the two main PPN
parameters for such a scalar-tensor theory are
\begin{eqnarray}
  \gamma^{PPN}-1&=&-2\frac{\alpha_0^2}{1+\alpha_0^2}\nonumber\\
  \beta^{PPN}-1&=&\frac{\beta_0\alpha_0^2}{2(1+\alpha_0^2)^2}\;.
\end{eqnarray}
The deviation of the PPN parameters from the general relativistic
values is of second order in $\alpha_0$ and of third order in the
product $\alpha_0^2\beta_0$. As a result, a very good limit on
the parameter $\alpha_0$ renders the parameter $\beta_0$ practically
unconstrainable by weak-field tests.

The study of Damour and Esposito-Farese~\cite{Damour93} revealed one of
the main reasons that necessitate careful theoretical studies of
possible extensions of General Relativity that are suitable for
strong-field tests. The order of a term added to the Lagrangian action
of the gravitational field is not necessarily a good estimate of the
expected magnitude of the observable effects introduced by this
additional term. For example, because of the non-linear coupling
between the scalar field and matter introduced by the coupling
function~(\ref{eq:stcoupling}), the deviation from general
relativistic predictions is not perturbative. For values of $\beta_0$
less that about $-6$, it becomes energetically favorable for neutron
stars to become ``scalarized'', with properties that differ
significantly from their general relativistic
counterparts~\cite{Damour93}. Such non-perturbative effects make
quantitative tests of strong-field gravity possible even when the
astrophysical complications are only marginally understood.

A similar situation, albeit in the opposite regime, arises in an
extended gravity theory in which a term proportional to the inverse of
the Ricci scalar curvature, $R^{-1}$, is added to the Einstein--Hilbert
action in order to explain the accelerated expansion of the
universe~\cite{Carroll04}. Although one would expect that such an
addition can only affect gravitational fields that are extremely
weak, it turns out that it also alters to zeroth order the
post-Newtonian parameter $\gamma$ and can, therefore, be excluded by
simple solar-system tests~\cite{Chiba03}.

{\em Second, Lagrangian extensions of General Relativity often suffer
from serious problems with instabilities.~--\/} This issue can be
understood by considering a Lagrangian action that includes terms of
second order in the Ricci scalar, i.e., $R^2$, as well as the terms of
similar order that can be constructed with the Ricci and Riemann
tensors.  For the sake of the argument, I will consider here the
parametric Lagrangian
\begin{equation}
{\cal S}= \frac{c^4}{16\pi G}\int d^4x
   \sqrt{-g}\left(R+\alpha_2 R^2 + \beta_2 R_{\sigma \tau}R^{\sigma \tau}
   + \gamma_2 R^{\alpha\beta\gamma\delta} R_{\alpha\beta\gamma\delta}\right)\;,
   \label{eq:PL}
\end{equation}
with $\alpha_2$, $\beta_2$, and $\gamma_2$ the three parameters of the
theory. Such terms arise naturally as high-order corrections in
quantum gravity and string theory and their relative importance
increases with the curvature of the metric~\cite{Donoghue94,
Burgess04}.  They have also been invoked as alternatives to the
inflation paradigm for the early expansion of the
universe~\cite{Starobinski80}. The predictions for astrophysical
objects of extended gravity theories that incorporate high-order terms
have been reported only for a few limited cases in the literature. The
dependence of the stellar properties on $R^2$ terms in the action has
been studied by Parker and Simon~\cite{Parker93}, who simply derived
the generalized Tolmann--Oppenheimer--Volkoff equation without solving
it, and by Barraco and Hamity~\cite{Barraco98} who attempted to solve
the problem using a perturbation analysis (unfortunately, this last
study suffers from a large number of errors).

This second-order gravity theory has a number of unappealing
properties (see discussion in~\cite{Simon90, Simon91}). Classically, a
high-order gravity theory requires more than two boundary conditions,
which is a fact that appears to be incompatible with all other
physical theories. Quantum mechanically, second-order gravity theories
lead to unstable vacuum solutions. Both these phenomena could be
artifacts of the possibility that the action~(\ref{eq:PL}) may arise
as a low-energy expansion of a non-local Lagrangian that is
fundamentally of second order~\cite{Simon90, Simon91}.
Phenomenologically speaking, these problems can be overcome by
requiring the field equations to be of second order, when extremizing
the action. This procedure leads to a generalized, high-order gravity
theory that remains consistent with classical expectations and is
stable quantum-mechanically (according to the procedure outlined
in~\cite{Simon90, Simon91}), but requires a different than usual
derivation of the field equations~\cite{DeDeo08}.

Even when these issues are being taken into account, the terms
proportional to $\beta_2$ and $\gamma_2$ lead to field equations with
solutions that suffer from the Ostrogradski
instability~\cite{Woodard06}. And even if these terms are dropped and
only actions that are generic function of only the Ricci scalar are
considered, then the resulting solutions for the expansion of the
universe~\cite{Dolgov03} and for spherically symmetric
stars~\cite{Seifert07} can be violently unstable, depending on the sign
of the second-order term.

A potential resolution of several of these problems in theories with
high-order terms in the action appears to be offered by the Palatini
formalism. In this approach, the field equations are derived by
extremizing the action under variations in the metric and the
connection, which is considered as an independent
field~\cite{Sotiriou07}. For the simple Einstein--Hilbert action, both
approaches are equivalent and give rise to the equations of general
relativity; when the action has non-linear terms in $R$, the two
approaches diverge. Unfortunately, the Palatini formalism leads to
equations that cannot handle in general the transition across the
surface layer of a star to the matter-free space outside it, and is
therefore not a viable alternative~\cite{Barausse07}.

{\em Finally, it is crucial that we identify the astrophysical
phenomena that can be used in testing particular aspects of
strong-field gravity.\/} For example, in the case of the
classical tests of General Relativity, it is easy to show that the
deflection of light during a solar eclipse and the Shapiro time delay
depend on one (and the same) component of the metric of the Sun (i.e.,
on the PPN parameter $\gamma$). Therefore, they do not provide
independent tests of General Relativity (as long as we accept the
validity of the equivalence principle). On the other hand, the
perihelion precession of Mercury and the gravitational redshift depend
on the other component of the metric (i.e., on the PPN parameter
$\beta$) and, therefore, provide complementary tests of the
theory. Understanding such degeneracies is an important component of
performing tests of gravity theories.

In the case of strong gravitational fields, this issue can be
illustrated again by studying the high-order Lagrangian
action~(\ref{eq:PL}) in the metric formalism (see
also~\cite{Burgess04}). In principle, as the strength of the
gravitational field increases, the terms that are of second-order in
the Ricci scalar become more important and, therefore, affect the
observable properties of neutron stars and black holes. However,
because of the Gauss-Bonnet identity,
\begin{equation}
\frac{\delta}{\delta g_{\mu\nu}}\int d^4x \left(
   R^2 -4 R_{\sigma \tau}R^{\sigma \tau}
   +  R^{\alpha\beta\gamma\delta} R_{\alpha\beta\gamma\delta}\right)
   =0\;,
\end{equation}
variations, with respect to the metric, of the term proportional to
$\gamma_2$ in Equation~(\ref{eq:PL}) can be expressed as variations of
the terms proportional to $\alpha_2$ and $\beta_2$. Therefore, for all
non-quantum gravity tests, the predictions of the theory described by
the Lagrangian action~(\ref{eq:PL}) are identical to those of the
Lagrangian
\begin{equation}
{\cal S}= \frac{c^4}{16\pi G}\int d^4x
   \sqrt{-g}\left[R+(\alpha_2-\gamma_2) R^2 + (\beta_2+4\gamma_2) 
    R_{\sigma \tau}R^{\sigma \tau}\right]\;.
   \label{eq:r2_clas}
\end{equation}
As a result, {\em astrophysical tests that do not invoke
quantum-gravity effects can only constrain a particular combination of
the parameters, i.e., $\alpha_2-\gamma_2$ and $\beta_2+4\gamma_2$.} It
is only through phenomena related to quantum gravity, such as the
evaporation of black holes, that the parameter $\gamma$ may
be constrained.

When the spacetime is isotropic and homogeneous, as in the case of
tests using the cosmic evolution of the scale factor, an additional
identity is satisfied, i.e.,
\begin{equation}
\frac{\delta}{\delta g_{\mu\nu}}\int d^4x \left(
   R^2 -3 R_{\sigma \tau}R^{\sigma \tau}\right)=0\;.
\end{equation}
This implies that, for cosmological tests, the predictions of the theory
described by the Lagrangian action~(\ref{eq:PL}) are identical to those 
of the Lagrangian
\begin{equation}
{\cal S}= \frac{c^4}{16\pi G}\int d^4x
   \sqrt{-g}\left[R+(\alpha_2+\frac{1}{3}\beta_2
        +\frac{1}{3}\gamma_2) R^2\right]\;.
   \label{eq:r2_cosm}
\end{equation}
As a result, {\em such cosmological tests of gravity can only constrain a
particular combination of the parameters, i.e.,
$\alpha_2+\beta_2/3+\gamma_2/3$.}

The parameters $\alpha_2$ and $\beta_2$ can be independently
constrained using observations of spacetimes that are strongly curved
but are not isotropic and homogeneous, such as those found in the
vicinities of black holes and neutron stars.  Measuring the properties
of neutron stars, such as their radii, maximum masses and maximum
spins, which require the solution of the field equations in the
presence of matter, will provide independent constraints on the
combination of parameters $\alpha_2-\gamma_2$ and $\beta_2+4\gamma_2$.
However, one can show that in absence of matter, the external
spacetime of a black hole, as given by the solution to Einstein's
field equation, is also one (but not necessarily the only) solution of
the parametric field equation that arises from the Lagrangian
action~(\ref{eq:PL}). As a result, tests that involve black holes will
probably be inadequate in distinguishing between the particular theory
described by Equation~(\ref{eq:PL}) and general
relativity~\cite{Psaltis07b}.

This is, in fact, a general problem of using astrophysical
observations of black holes to test General Relativity in the
strong-field regime. The Kerr solution is not unique to general
relativity~\cite{Psaltis07b}.  For example, there is strong
analytical~\cite{Thorne71, Bekenstein72, Hawking72} and numerical
evidence~\cite{Scheel95} that, in Brans--Dicke scalar-tensor gravity
theories, the end product of the collapse of a stellar configuration
is a black hole described by the same Kerr solution as in Einstein's
theory. The same appears to be true in several other theories
generated by adding additional degrees of freedom to Einstein's
gravity; the only vacuum solutions that are astrophysically relevant
are those described by the Kerr metric~\cite{Psaltis07b}. Until a
counter-example is discovered, studies of the strong gravitational
fields found in the vicinities of black holes can be performed only
within phenomenological frameworks, such as those involving multipole
expansions of the Schwarzschild and Kerr metrics~\cite{Ryan95,
Collins04, Glampedakis06}.

To date, it has only been possible to test quantitatively the
predictions of General Relativity in the strong-field regime using
observations of neutron stars, as I will discuss in the following
section. In all cases, the general relativistic predictions were
contrasted to those of scalar-tensor gravity, with Einstein's theory
passing all the tests with flying colors.

\newpage

\section{Current Tests of Strong-Field Gravity with Neutron Stars}
\label{sec:tests}

Performing tests of strong-field gravity with neutron stars requires
knowledge of the equation of state of neutron-star matter to a degree
better than the required precision of the gravitational test. This
appears {\em a priori\/} to be a serious hurdle given the wide range
of predictions of equally plausible theories of neutron-star matter
(see~\cite{Lattimer01} for a recent compilation). It is easy to
show, however, that current uncertainties in our modeling of the
properties of ultra-dense matter do not preclude significant
constraints on the strong-field behavior of gravity~\cite{DeDeo03}.

\epubtkImage{ns.png}{
\begin{figure}[htbp]
\centerline{\includegraphics[height=10.0cm]{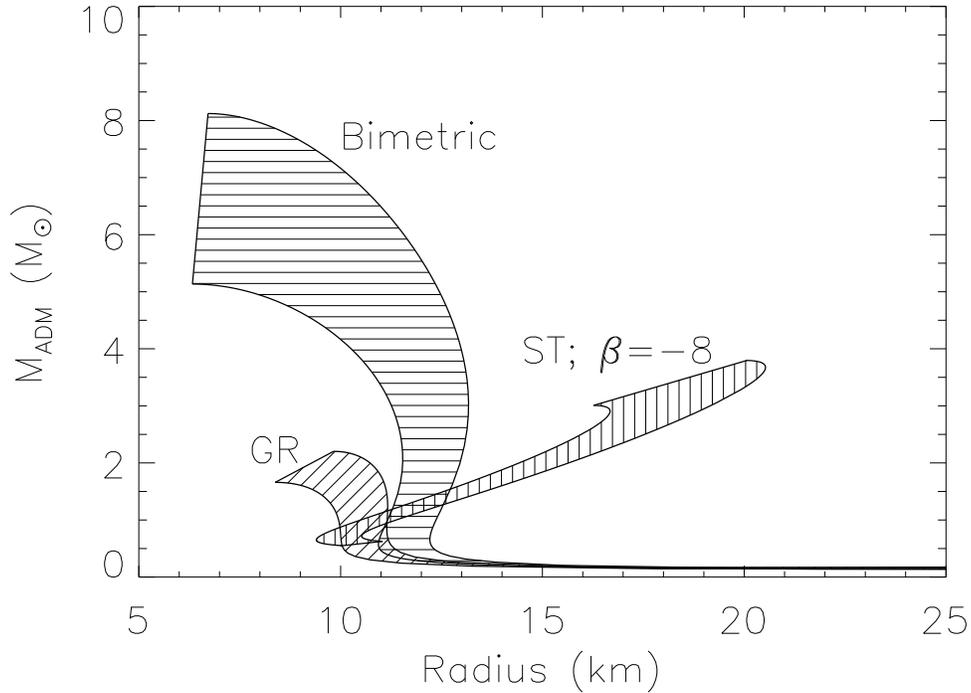}}
\caption{Mass-radius relations of neutron stars 
in General Relativity (GR), scalar-tensor (ST), and Rosen's bimetric
theory of gravity~\cite{DeDeo03}. The shaded areas represent the range
of mass-radius relations predicted in each case by neutron-star
equations of state without unconfined quarks or condensates. All
gravity theories shown in the figure are consistent with solar-system
tests but introduce variations in the predicted sizes of neutron stars
that are significantly larger than the uncertainty caused by the
unknown equation of state.}
\label{fig:mr}
\end{figure}}

During the last three decades, neutron-star models have been
calculated for a variety of gravity theories (see~\cite{Will93} and
references therein) and were invariably different, both in size and in
allowed mass, than their general relativistic counterparts. As an
example, Figure~\ref{fig:mr} shows neutron-star models calculated in
three representative theories that cannot be excluded by current tests
that do not involve neutron stars.  In the figure, the shaded areas
represent the uncertainty introduced by the unknown equation of state
of neutron-star matter (not including quark stars or large neutron
stars with condensates). Clearly, the deviations in neutron-star
properties from the predictions of General Relativity for these
theories (that are still consistent with weak-field tests) are larger
than the uncertainty introduced by the unknown equation of state of
neutron-star matter.

This is a direct consequence of the fact that the curvature around a
neutron star is larger by $\sim 13$ orders of magnitude compared to
the curvature probed by solar-system tests, whereas the density inside
the neutron star is larger by only an order of magnitude compared to
the densities probed by nuclear scattering data that are used to
constrain the equation of state. Given that the current values of the
post-Newtonian parameters are known from weak-field tests to within
$\sim 10^{-4}$, it is reasonable that deviations from general
relativity can be hidden in the weak-field limit but may become
dominant as the curvature is increased by more than ten orders of
magnitude. Neutron stars can indeed be used in testing the
strong-field behavior of a gravity theory.

\epubtkImage{xpsr_bd.png}{
\begin{figure}[htbp]
\centerline{\includegraphics[height=10.0cm]{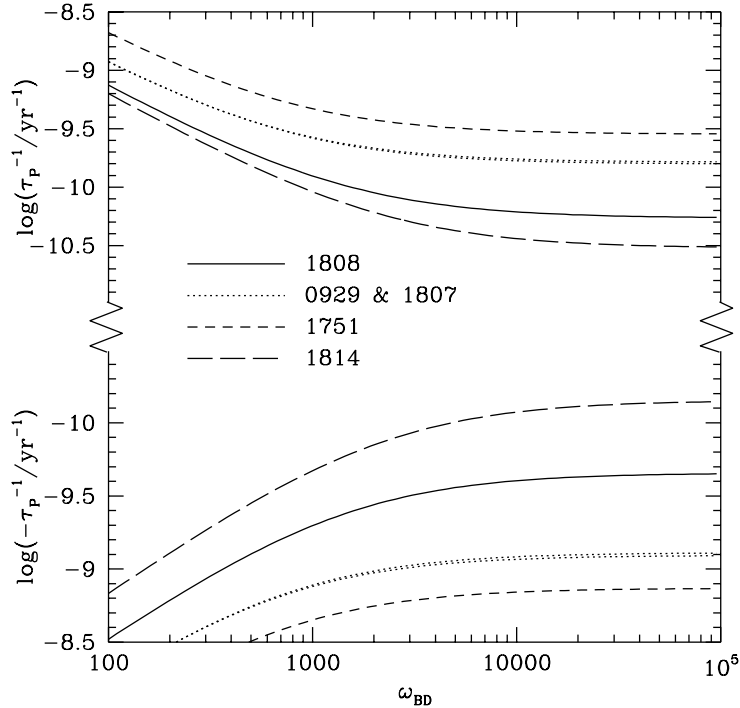}}
\caption{The limiting rate for the evolution of the
orbital periods ($\tau_{\rm P}^{-1}\equiv \dot{P}/P$) of five known
millisecond accreting pulsars as a function of the Brans--Dicke
parameter $\omega_{\rm BD}$.  The lower half of the plot corresponds
to an orbital period that decreases with time ($\dot{P}/P<0$), whereas
the upper half corresponds to an orbital period that increases with
time ($\dot{P}/P>0$). Only the area outside the two curves for each
system is physically allowed~\cite{Psaltis07a}.}
\label{fig:xpsr_bd}
\end{figure}}

\subsection{Brans--Dicke gravity and the orbital decay of binary systems 
with neutron stars}

Binary stellar systems that are currently known to harbor at least one
neutron star have orbital separations that are too large to be used in
probing directly strong gravitational fields. Even at that separation,
however, the orbital evolution of the binary system caused by the
emission of gravitational waves is affected, in a scalar-tensor
theory, by the coupling of matter to the scalar field, which occurs in
a strong gravitational field. This manifests itself as a violation of
the strong equivalence principle, with many observable consequences
such as the rapid decay of the orbit due to emission of dipole
radiation~\cite{Eardley75, Will89}. The various quantitative tests of
strong-field gravity using binary systems with radio pulsars have been
reviewed in detail elsewhere~\cite{Stairs03}. Here, I will focus only
on tests that involve the orbital period evolution of the binary
systems.

The best studied binaries with compact objects are the double neutron
stars, with the Hulse--Taylor pulsar (PSR 1913$+$16) as the
prototypical case. Unfortunately, in all double neutron-star systems,
the masses of the two members of the binary are surprisingly
similar~\cite{Thorsett99} and this severely limits the prospects of
placing strong constraints on the dipole radiation from them.  Indeed,
the magnitude of dipole radiation depends on the difference of the
sensitivities between the two members of the binaries, and for neutron
stars the sensitivities depend primarily on their masses. The
resulting constraint imposed on the Brans--Dicke parameter $\omega$ by
the Hulse--Taylor pulsar is significantly smaller than the limit
$\omega>40000$ set by the Cassini mission~\cite{Bertotti03}.

The constraint is significantly improved when studying binary systems
in which only one of the two stars is a neutron star. There are
several known neutron star-white dwarf binaries that are suitable for
this purpose, in which the neutron stars appear as radio pulsars
(e.g., PSR B0655$+$64~\cite{Damour96};
PSR~J0437$-$4715~\cite{vanStraten01}), as millisecond accreting X-ray
pulsars (e.g., XTE~J1808$-$456~\cite{Psaltis07a}), or as non-pulsing
X-ray sources (e.g., 4U~1820$-$30~\cite{Will89}). In the last two
cases, the evolution of the binary orbit is also affected
significantly by mass transfer from the companion star to the neutron
star. However, for each value of the Brans--Dicke parameter
$\omega_{\rm BD}$, there is a minimum absolute value for the rate of
evolution of the orbital period (see Figure~\ref{fig:xpsr_bd}
and~\cite{Psaltis07a}). An accurate measurement of the orbital period
derivative in any of these systems offers, therefore, the potential of
placing a lower limit on the Brans--Dicke parameter.  Because of the
astrophysical complications introduced by mass transfer, the optimal
constraint on $\omega_{\rm BD}$ is of order $10^4$ in this method,
which is comparable to the Cassini limit.

\epubtkImage{st2_psr.png}{
\begin{figure}[htbp]
\centerline{\includegraphics[height=10.0cm]{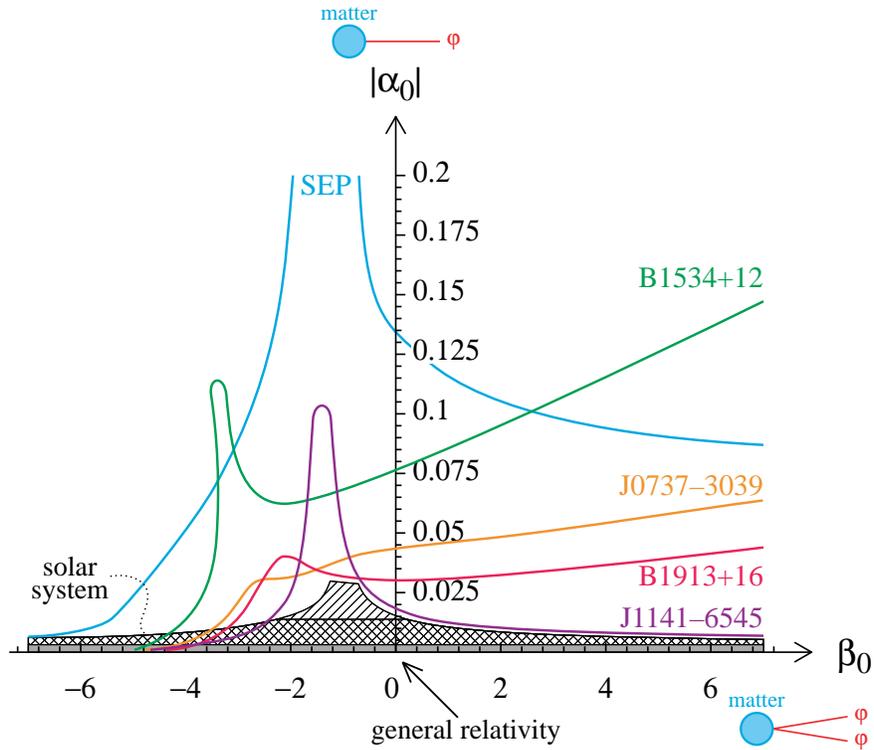}}
\caption{Constraints on the two parameters of a second-order
scalar-tensor theory placed by the timing properties of a number of
binary stellar systems that harbor neutron stars. For two of the
systems, current constraints are contrasted to those expected in the
near future when a measurement of the orbital period derivative is
possible to an accuracy of 1\%. General relativity corresponds to the
origin of the parameter space; the constraint imposed by the Cassini
mission is also shown for comparison~\cite{Damour07}. In all cases,
the allowed part of the parameter space is under the corresponding
curve.}
\label{fig:st_psr}
\end{figure}}

\subsection{Second-order scalar-tensor gravity and radio pulsars}

As discussed in the previous section, observations of strong-field
phenomena provide constraints on Brans--Dicke scalar-tensor gravity,
which are, however, at most comparable to those of solar system tests.
This is true because the fractional deviation of a Brans--Dicke theory
from General Relativity is of order $\omega_{\rm BD}^{-1}$, both for
weak and strong gravitational fields, and the solar-system tests have
superb accuracy. On the other hand, a scalar-tensor theory with a
second-order coupling (e.g., the one arising from the
action~(\ref{eq:st}) with the coupling~(\ref{eq:stcoupling})) allows
for large deviations in the strong-field regime while being consistent
with the weak-field limits~\cite{Damour93, Damour96}.

In the case of neutron stars, the second-order scalar-tensor theory
described by Damour and Esposito-Farese~\cite{Damour93} leads to a
non-perturbative effect known as spontaneous scalarization (similar to
the spontaneous magnetization in ferromagnetism).  For significantly
large negative values of the parameter $\beta_0$, there is a range of
neutron-star masses for which it becomes energetically favorable for
the scalar field to acquire high values inside the neutron star and
affect significantly its structure compared to the general
relativistic predictions. An example of the mass-radius relation for
neutron stars in a second-order scalar-tensor theory with $\beta_0=-8$
is shown in Figure~\ref{fig:mr}.

The properties and stability of scalarized neutron stars have been
studied extensively in the literature~\cite{Damour93, Salgado98,
Harada98}. For the purposes of tests of strong-field gravity, the
coupling of matter with the gravitational field and the external
spacetimes of scalar stars are so different compared to their general
relativistic counterparts that large negative values of $\beta_0$ can
be firmly excluded with current observations of binary stellar systems
that harbor radio pulsars. Figure~{\ref{fig:st_psr}} shows the current
constraints on the two parameters $\alpha_0$ and $\beta_0$ of the
theory imposed by the timing observations of the Hulse--Taylor pulsar
(PSR~J1913$+$16), of a pulsar in an asymmetric binary with a white
dwarf (PSR~J1141$-$6545), and of two other pulsars (PSR~J0737$-$3079
and PSR~B1534$+$12). The best weak-field limits, including those
imposed by the Cassini mission, are also shown for
comparison~\cite{Damour07}.

As expected, weak-field tests bound significantly the value of the
parameter $\alpha_0$, leaving the parameter $\beta_0$ largely
unconstrained. Between the binary systems with radio pulsars, the one
with the white-dwarf companion provides the most stringent
constraints because the large asymmetry between the two compact object
leads to the prediction of strong dipole gravitational radiation that
can be excluded observationally. Finally, for large negative values
of the parameter $\beta_0$, the scalarization of the neutron stars makes
the predictions of the theory incompatible with observations.

\epubtkImage{redshift.png}{
\begin{figure}[htbp]
\centerline{\includegraphics[height=10.0cm]{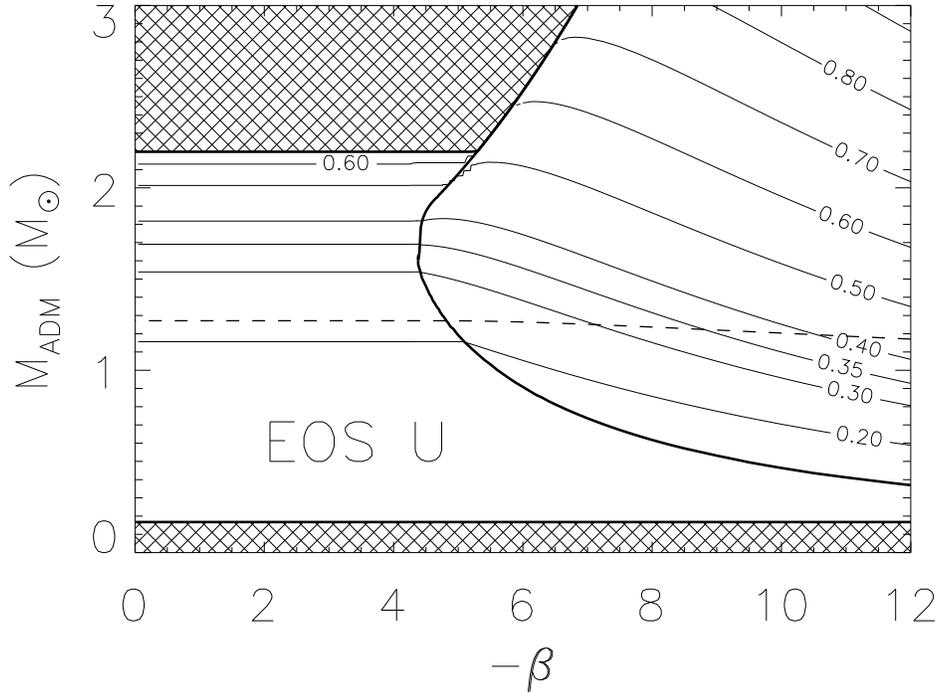}}
\caption{Contours of constant
gravitational redshift measured at infinity for an atomic line
originating at the surface of a neutron star in a scalar-tensor
gravity theory, for different values of the parameter $\beta_0$ that
measures the relative contribution of the scalar field. The thick
curve separates the scalarized stars from the general relativistic
counterparts.  The measurements of a redshift of $z=0.35$ from a
burster~\cite{Cottam02} and the astrophysical constraint of a baryonic
mass of at least $1.4\,M_\odot$ (dashed lines) result in a bound on the
parameter $\beta$ of $-\beta<9$~\cite{DeDeo03}. }
\label{fig:z}
\end{figure}}

\subsection{Second-order scalar-tensor gravity and X-ray observations 
of accreting neutron stars}

The quantitative features of a number of phenomena observed in the
X-rays from accreting neutron stars depend strongly on their masses
and radii, as discussed in~\ref{sec:strong}. The constraints imposed by two of
these phenomena on the parameters of the second-order scalar-tensor
gravity of Damour and Esposito-Farese~\cite{Damour93} have been studied
recently~\cite{DeDeo03, DeDeo07}.

The first phenomenon is the observation of gravitationally redshifted
atomic lines during X-ray bursts from the source
EXO~0748$-$56~\cite{Cottam02}. Figure~\ref{fig:z} shows the values of
the gravitational redshift from the surface of neutron stars with
different masses, in second-order scalar-tensor theories with
different values of the parameter $\beta_0$~\cite{DeDeo03}.  In this
calculation, the parameter $\alpha_0$ was set to zero and the
neutron-star structure was calculated using the equation of state
U~\cite{Cook94}. The hatch-filled area corresponds to neutron-star
masses that are unacceptable for each value of the parameter
$\beta_0$, while the thick curve separates the scalarized stars from
the general relativistic counterparts.

A dynamical measurement of the mass of EXO~0748$-$56 can rule out the
possibility that the neutron star in this source is scalarized,
because scalarized stars have very different surface redshifts
compared to the general relativistic stars of the same mass. The
source EXO~0748$-$56 lies in an eclipsing binary system which makes it
a prime candidate for a dynamical mass measurement. In the absence of
such a measurement, however, a limit on the parameter $\beta_0$ can be
placed under the astrophysical constraint that the baryonic mass of
the neutron stars is larger than $\simeq 1.4\,M_\odot$. This is a
reasonable assumption, given that a progenitor core of a lower mass
would not have collapsed to form a neutron star. Combining this
constraint with the measured redshift of $z=0.35$ leads to a limit on
the parameter $-\beta_0<9$, which depends only weakly on the assumed
equation of state of neutron-star matter~\cite{DeDeo03}.

\epubtkImage{maxorb.png}{
\begin{figure}[htbp]
\centerline{\includegraphics[height=10.0cm]{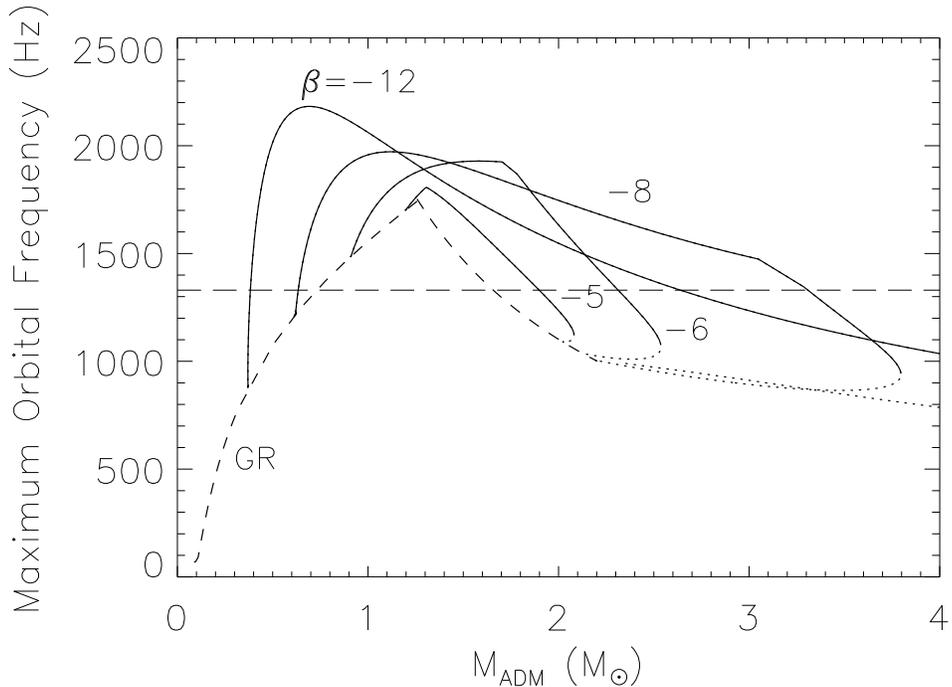}}
\caption{The maximum orbital frequency outside a neutron star
of mass $M_{\rm ADM}$ for different scalar-tensor theories identified
by the parameter $\beta$. The dashed line shows the maximum observed
frequency of a quasi-periodic oscillation from an accreting neutron
star~\cite{DeDeo07}.}
\label{fig:QPO}
\end{figure}}

A second set of phenomena that can lead to strong-field tests of
gravity are the fast quasi-periodic oscillations observed from many
bright accreting neutron stars~\cite{vanderKlis06}. The highest known
frequency of such an oscillations is 1330~Hz, observed from the source
4U~1636$-$53 and corresponds to the Keplerian frequency of the
innermost stable circular orbit of a $1.6\,M_\odot$ slowly spinning
neutron star. Figure~\ref{fig:QPO} shows the maximum Keplerian
frequency outside a neutron star in the second-order scalar tensor
theory, for different values of the parameter $\beta_0$.  For small
stellar masses, the limiting frequency is achieved at the surface of
the star, whereas for large stellar masses, the limiting frequency is
reached at the innermost stable circular orbit. This figure shows that
scalarized stars allow for higher frequencies than their general
relativistic counterparts. Requiring, therefore, the observed
oscillation frequency to be smaller than the highest Keplerian
frequency of a stable orbit outside the compact object cannot be used
to constrain the parameters of this theory. On the other hand, the
correlations between the various dynamical frequencies outside the
compact object depend strongly on the parameter $\beta$ and hence the
gravity theory can be constrained given a particular model for the
oscillations~\cite{DeDeo07}.

\newpage

\section{Going Beyond Einstein}
\label{section:beyond}

\epubtkImage{tests_future.png}{
\begin{figure}[htbp]
\centerline{\includegraphics[width=12.0cm]{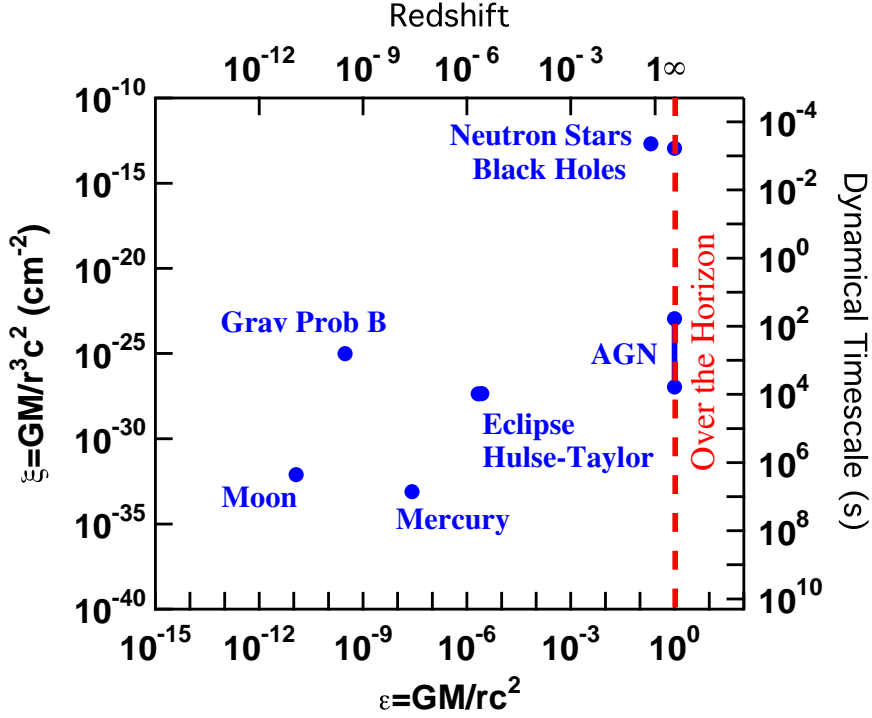}}
\caption{The spectral (redshift) and timing capabilities 
required for an observatory to probe different strengths of 
gravitational fields. Phenomena that occur in the vicinities
of neutron stars and stellar-mass black holes experience large
redshift and occur over sub-millisecond timescales.}
\label{fig:tests_future}
\end{figure}}

Testing General Relativity in the strong-field regime with neutron
stars and black holes will require advanced observatories that will be
able to resolve various phenomena in the characteristic energy and
time-scales in which they occur. The two parameters used to quantify
the strength of a gravitational field in
Section~\ref{subsec:definition} are also useful in discussion the
specifications required by such future observatories.

The potential and the curvature in a gravitational field are related
directly to the characteristic energy- and time-scales, respectively,
that need to be resolved in order for an observation to be able to
probe a particular region of the parameter space.  The potential
$\epsilon$ gives directly the gravitational redshift $z$ according to
\begin{equation}
z=1-(1-2\epsilon)^{-1/2}\;,
\end{equation}
the measurement of which is the goal of spectroscopic observations;
for weak gravitational fields $z\simeq\epsilon$. At the same time, the
curvature $\xi$ is directly related to the dynamical timescale $\tau$
in the same region of a gravitational field by
\begin{equation}
\tau=\frac{2\pi}{c}\xi^{-1/2}\;.
\end{equation}
As shown in Figure~\ref{fig:tests_future}, only observatories with
excellent spectroscopic and millisecond timing capabilities will be
able to resolve phenomena that occur in the strongest gravitational
fields found in astrophysics, i.e., those in the vicinities of neutron
stars and stellar-mass black holes.

\epubtkImage{gwaves.png}{
\begin{figure}[htbp]
\centerline{\includegraphics[width=12.0cm]{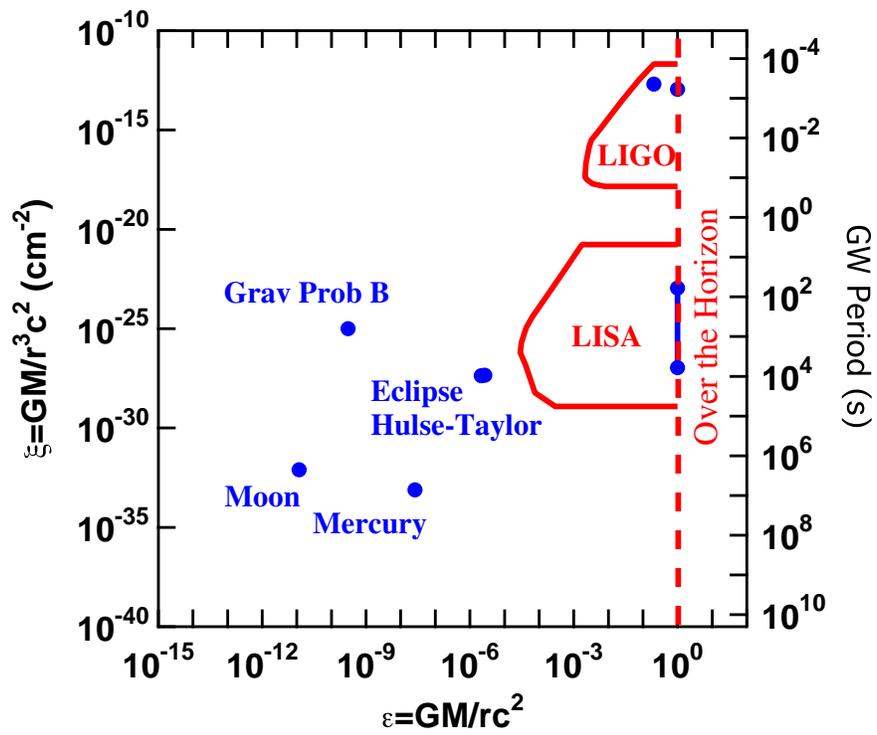}}
\caption{The parameter space that will be probed by 
an experiment based on a gravitational wave detection with LIGO and
LISA, for an assumed source at a distance of 1~Mpc.}
\label{fig:gwaves}
\end{figure}}

One of the most promising avenues towards testing strong-field general
relativity is via the detection of the gravitational waves emitted
during the coalescence of compact objects. In the simple case in which
two compact objects of mass $M$ are orbiting each other in circular
orbits with separation $a$, slowly approaching because of the emission
of gravitational waves, the characteristic period $P$ of the
gravitational wave is half of the orbital period and, therefore, is
related to the spacetime curvature by
\begin{equation}
P=\frac{\pi}{c}\xi^{-1/2}\;.
\label{eq:Pgw}
\end{equation}
At the same time, the strain $h$ detected by an observatory on Earth
for a gravitational wave emitted by such a source placed at a distance
$D$, is~\cite{Flanagan05}
\begin{equation}
h=\left(\frac{GM}{ac^2}\right)\left(\frac{GM}{Dc^2}\right)\;.
\end{equation}
Given the distance to a source and the measurement of a strain, the
curvature of the gravitational field probed is
\begin{equation}
\xi=\frac{\epsilon^5}{h^2 D^2}=10^{-3}\epsilon^5
   \left(\frac{h}{10^{-23}}\right)^{-2}
   \left(\frac{D}{1~\mbox{Mpc}}\right)^{-2}~\mbox{cm}^{-2}\;.
\label{eq:xigw}
\end{equation}
The sensitivity of each detector of gravitational waves depends
strongly on the period of the wave. Using equations~(\ref{eq:Pgw})
and~(\ref{eq:xigw}), the sensitivity curve of a detector can be
converted into a region of the parameter space that can be probed,
given the distance to the source. This is shown in
Figure~\ref{fig:gwaves} for the advanced LIGO and LISA, for an assumed
source distance of 1~Mpc. Gravitational waves detected by LISA will
probe the same curvatures as current tests of General Relativity but
significantly larger potentials.  On the other hand, gravitational
waves detected by the advanced LIGO have the potential of probing
directly the strongest gravitational fields found around astrophysical
objects.

In the near future, a number of observatories will exploit new
techniques and open new horizons in gravitational physics by exploring
the strong-field region of the parameter space shown in
Figure~\ref{fig:tests_future}. Observations with the Square Kilometre
Array~\cite{SKA} may lead to the discovery of the most optimal binary
systems for strong-field gravity tests with pulsar timing, in which a
pulsar is orbiting a black hole~\cite{Kramer05}. High energy
observations of black holes and neutron stars with
Constellation-X~\cite{ConX} and XEUS~\cite{XEUS} will detect highly
redshifted atomic lines and measure their rapid variability
properties. Finally, gravitational wave observatories, either from the
ground (such as LIGO~\cite{LIGO}, GEO600~\cite{GEO},
TAMA300~\cite{TAMA}, and VIRGO~\cite{VIRGO}) or from space (such as
LISA~\cite{LISA}) will detect directly for the first time one of the
most remarkable predictions of General Relativity, the generation of
gravitational waves from orbiting compact objects and black hole
ringing.

\newpage

\section{Acknowledgements}
\label{acknowledgements}

It is my great pleasure to acknowledge the many fruitful discussions
and collaborations with a number of people that have shaped my ideas
on astrophysical tests of strong-field gravity. In particular, I thank
T.\ Belloni, D.\ Chakrabarty, S.\ DeDeo, F.\ Lamb, C.\ Miller, R.\
Narayan, J.\ McClintock, F.\ \"Ozel, and M.\ van der Klis. I am
indebted to S.\ DeDeo and F.\ \"Ozel for helping me settle on and
understand the defition of strong-field gravity. I am also grateful to
G.\ Esposito-Far\'ese, T.\ Johanssen, J.\ McClintock, F.\ \"Ozel, and
C.\ Reynolds for their detailed comments that helped me greatly
improve the presentation of this review.

\newpage

\bibliography{ms}

\begin{thebibliography}{100}

\bibitem{Abramowicz02}
Abramowicz, M.A., and Klu{\'z}niak, W., ``A precise determination of black hole
  spin in GRO J1655-40'', {\em Astron. Astrophys.}, {\bf 374}, L19--L20,
  (2001).  Related online version (cited on 24 July 2007):
  \newline\url{http://arXiv.org/abs/astro-ph/0105077}.

\bibitem{Abramowicz04}
Abramowicz, M.A., Klu{\'z}niak, W., McClintock, J.E., and Remillard, R.A.,
  ``The Importance of Discovering a 3:2 Twin-Peak Quasi-periodic Oscillation in
  an Ultraluminous X-Ray Source, or How to Solve the Puzzle of
  Intermediate-Mass Black Holes'', {\em Astrophys. J.}, {\bf 609}, L63--L65,
  (2004).  Related online version (cited on 24 July 2007):
  \newline\url{http://arXiv.org/abs/astro-ph/0402012}.

\bibitem{Agol00}
Agol, E., and Krolik, J.H., ``Magnetic Stress at the Marginally Stable Orbit:
  Altered Disk Structure, Radiation, and Black Hole Spin Evolution'', {\em
  Astrophys. J.}, {\bf 528}, 161--170, (2000). \newline ADS:
  \url{http://adsabs.harvard.edu/abs/2000ApJ...528..161A}.

\bibitem{Armitage03}
Armitage, P.J., and Reynolds, C.S., ``The variability of accretion on to
  Schwarzschild black holes from turbulent magnetized discs'', {\em Mon. Not.
  R. Astron. Soc.}, {\bf 341}, 1041--1050, (2003). \newline ADS:
  \url{http://adsabs.harvard.edu/abs/2003MNRAS.341.1041A}.

\bibitem{Balbus99}
Balbus, S.A., and Papaloizou, J.C.B., ``On the Dynamical Foundations of alpha
  Disks'', {\em Astrophys. J.}, {\bf 521}, 650--658, (1999). \newline ADS:
  \url{http://adsabs.harvard.edu/abs/1999ApJ...521..650B}.

\bibitem{Barausse07}
Barausse, E., Sotiriou, T.P., and Miller, J.C., ``Curvature singularities,
  tidal forces and the viability of Palatini $f(R)$ gravity'', {\em Class.
  Quantum Grav.}, {\bf 25}, 105008, (2008).  Related online version (cited on
  29 May 2008): \newline\url{http://arXiv.org/abs/0712.1141}.

\bibitem{Bardeen73}
Bardeen, J.M., ``Timelike and Null Geodesics in the Kerr Metric'', in DeWitt,
  C., and DeWitt, B.S., eds., {\em Black Holes}, Lectures delivered at the
  Summer School of Theoretical Physics, Les Houches, France, 23rd session,
  1972,  215--240, (Gordon and Breach, New York, U.S.A., 1973).

\bibitem{Bardeen73b}
Bardeen, J.M., Press, W.H., and Teukolsky, S.A., ``Rotating Black Holes:
  Locally Nonrotating Frames, Energy Extraction, and Scalar Synchrotron
  Radiation'', {\em Astrophys. J.}, {\bf 178}, 347--369, (1972). \newline ADS:
  \url{http://adsabs.harvard.edu/abs/1972ApJ...178..347B}.

\bibitem{Barraco98}
Barraco, D.E., and Hamity, V.H., ``Stellar model in a fourth order theory of
  gravity'', {\em Phys. Rev. D}, {\bf 57}, 954--960,  (1998).

\bibitem{Barret06}
Barret, D., Olive, J.-F., and Miller, M.C., ``The coherence of kilohertz
  quasi-periodic oscillations in the X-rays from accreting neutron stars'',
  {\em Mon. Not. R. Astron. Soc.}, {\bf 370}, 1140--1146, (2006).  Related
  online version (cited on 24 July 2007):
  \newline\url{http://arXiv.org/abs/astro-ph/0605486}.

\bibitem{Barret07}
Barret, D., Olive, J.-F., and Miller, M.C., ``Supporting evidence for the
  signature of the innermost stable circular orbit in Rossi X-ray data from 4U
  1636-536'', {\em Mon. Not. R. Astron. Soc.}, {\bf 376}, 1139--1144,  (2007).

\bibitem{Beckwith04}
Beckwith, K., and Done, C., ``Iron line profiles in strong gravity'', {\em Mon.
  Not. R. Astron. Soc.}, {\bf 352}, 353--362, (2004).  Related online version
  (cited on 24 July 2007): \newline\url{http://arXiv.org/abs/astro-ph/0402199}.

\bibitem{Bekenstein72}
Bekenstein, J.D., ``Nonexistence of Baryon Number for Black Holes. II'', {\em
  Phys. Rev. D}, {\bf 5}, 2403--2412,  (1972).

\bibitem{Bekenstein07}
Bekenstein, J.D., ``The modified Newtonian dynamics-MOND and its implications
  for new physics'', {\em Contemp. Phys.}, {\bf 47}, 387--403, (2007).  Related
  online version (cited on 24 July 2007):
  \newline\url{http://arXiv.org/abs/astro-ph/0701848}.

\bibitem{Bertotti03}
Bertotti, B., Iess, L., and Tortora, P., ``A test of general relativity using
  radio links with the Cassini spacecraft'', {\em Nature}, {\bf 425}, 374--376,
   (2003).

\bibitem{beyondEins}
NASA, ``Beyond Einstein'', project homepage.  URL (cited on 05 July 2007):
  \newline\url{http://universe.nasa.gov/}.

\bibitem{Bhattacharya05}
Bhattacharyya, S., Miller, M.C., and Lamb, F.K., ``The Shapes of Atomic Lines
  from the Surfaces of Weakly Magnetic Rotating Neutron Stars and Their
  Implications'', {\em Astrophys. J.}, {\bf 644}, 1085--1089, (2006).  Related
  online version (cited on 24 July 2007):
  \newline\url{http://arXiv.org/abs/astro-ph/0412107}.

\bibitem{Bildsten00}
Bildsten, L., and Rutledge, R.E., ``Coronal X-Ray Emission from the Stellar
  Companions to Transiently Accreting Black Holes'', {\em Astrophys. J.}, {\bf
  541}, 908--917, (2000). \newline ADS:
  \url{http://adsabs.harvard.edu/abs/2000ApJ...541..908B}.

\bibitem{Bildsten92}
Bildsten, L., Salpeter, E.E., and Wasserman, I., ``The fate of accreted CNO
  elements in neutron star atmospheres: X-ray bursts and gamma-ray lines'',
  {\em Astrophys. J.}, {\bf 384}, 143--176, (1992). \newline ADS:
  \url{http://adsabs.harvard.edu/abs/1992ApJ...384..143B}.

\bibitem{Blaes06}
Blaes, O.M., Davis, S.W., Hirose, S., Krolik, J.H., and Stone, J.M., ``Magnetic
  Pressure Support and Accretion Disk Spectra'', {\em Astrophys. J.}, {\bf
  645}, 1402--1407, (2006).  Related online version (cited on 24 July 2007):
  \newline\url{http://arXiv.org/abs/astro-ph/0601380}.

\bibitem{Braje02}
Braje, T.M., and Romani, R.W., ``RX J1856-3754: Evidence for a Stiff Equation
  of State'', {\em Astrophys. J.}, {\bf 580}, 1043--1047, (2002). \newline ADS:
  \url{http://adsabs.harvard.edu/abs/2002ApJ...580.1043B}.

\bibitem{Brans61}
Brans, C., and Dicke, R.H., ``Mach's Principle and a Relativistic Theory of
  Gravitation'', {\em Phys. Rev.}, {\bf 124}, 925--935,  (1961).

\bibitem{Brenneman06}
Brenneman, L.W., and Reynolds, C.S., ``Constraining Black Hole Spin via X-Ray
  Spectroscopy'', {\em Astrophys. J.}, {\bf 652}, 1028--1043, (2006).  Related
  online version (cited on 24 July 2007):
  \newline\url{http://arXiv.org/abs/astro-ph/0608502}.

\bibitem{Broderick06}
Broderick, A.E., and Loeb, A., ``Frequency-dependent Shift in the Image
  Centroid of the Black Hole at the Galactic Center as a Test of General
  Relativity'', {\em Astrophys. J.}, {\bf 636}, L109--L112, (2006).  Related
  online version (cited on 24 July 2007):
  \newline\url{http://arXiv.org/abs/astro-ph/0508386}.

\bibitem{Broderick06b}
Broderick, A.E., and Loeb, A., ``Testing General Relativity with
  High-Resolution Imaging of Sgr A*'', {\em J. Phys.: Conf. Ser.}, {\bf 54},
  448--455, (2006).  URL (cited on 29 May 2008):
  \newline\url{http://stacks.iop.org/1742-6596/54/448}.

\bibitem{Brown98}
Brown, E.F., Bildsten, L., and Rutledge, R.E., ``Crustal Heating and Quiescent
  Emission from Transiently Accreting Neutron Stars'', {\em Astrophys. J.},
  {\bf 504}, L95--L98, (1998). \newline ADS:
  \url{http://adsabs.harvard.edu/abs/1998ApJ...504L..95B}.

\bibitem{Burgess04}
Burgess, C.P., ``Quantum Gravity in Everyday Life: General Relativity as an
  Effective Field Theory'', {\em Living Rev. Relativity}, {\bf 7}, lrr-2004-5,
  (2004).  URL (cited on 05 July 2007):
  \newline\url{http://www.livingreviews.org/lrr-2004-5}.

\bibitem{Cadeau07}
Cadeau, C., Morsink, S.M., Leahy, D., and Campbell, S.S., ``Light Curves for
  Rapidly Rotating Neutron Stars'', {\em Astrophys. J.}, {\bf 654}, 458--469,
  (2007).  Related online version (cited on 24 July 2007):
  \newline\url{http://arXiv.org/abs/astro-ph/0609325}.

\bibitem{Carroll01}
Carroll, S.M., ``The Cosmological Constant'', {\em Living Rev. Relativity},
  {\bf 4}, lrr-2001-1, (2001).  URL (cited on 05 July 2007):
  \newline\url{http://www.livingreviews.org/lrr-2001-1}.

\bibitem{Carroll04}
Carroll, S.M., Duvvuri, V., Trodden, M., and Turner, M.S., ``Is cosmic speed-up
  due to new gravitational physics?'', {\em Phys. Rev. D}, {\bf 70}, (2004).
  Related online version (cited on 24 July 2007):
  \newline\url{http://arXiv.org/abs/astro-ph/0306438}.

\bibitem{Carroll02}
Carroll, S.M., and Kaplinghat, M., ``Testing the Friedmann equation: The
  expansion of the universe during big-bang nucleosynthesis'', {\em Phys. Rev.
  D}, {\bf 65}, 063507, (2002).  Related online version (cited on 24 July
  2007): \newline\url{http://arXiv.org/abs/astro-ph/0108002}.

\bibitem{Chang06}
Chang, P., Morsink, S., Bildsten, L., and Wasserman, I., ``Rotational
  Broadening of Atomic Spectral Features from Neutron Stars'', {\em Astrophys.
  J.}, {\bf 636}, L117--L120, (2006).  Related online version (cited on 24 July
  2007): \newline\url{http://arXiv.org/abs/astro-ph/0511246}.

\bibitem{Chiba03}
Chiba, T., ``1/R gravity and scalar-tensor gravity'', {\em Phys. Lett. B}, {\bf
  575}, 1--3, (2003).  Related online version (cited on 24 July 2007):
  \newline\url{http://arXiv.org/abs/astro-ph/0307338}.

\bibitem{Clayton83}
Clayton, D.D., {\em Principles of stellar evolution and nucleosynthesis},
  (University of Chicago Press, Chicago, U.S.A., 1983).

\bibitem{Collins04}
Collins, N.A., and Hughes, S.A., ``Towards a formalism for mapping the
  spacetimes of massive compact objects: Bumpy black holes and their orbits'',
  {\em Phys. Rev. D}, {\bf 69}, 124022, (2004).  Related online version (cited
  on 24 July 2007): \newline\url{http://arXiv.org/abs/gr-qc/0402063}.

\bibitem{ConX}
NASA GSFC / SAO, ``Constellation-X: The Constellation X-Ray Mission'', project
  homepage.  URL (cited on 05 July 2007):
  \newline\url{http://constellation.gsfc.nasa.gov/}.

\bibitem{Cook94}
Cook, G.B., Shapiro, S.L., and Teukolsky, S.A., ``Rapidly rotating neutron
  stars in general relativity: Realistic equations of state'', {\em Astrophys.
  J.}, {\bf 424}, 823--845, (1994). \newline ADS:
  \url{http://adsabs.harvard.edu/abs/1994ApJ...424..823C}.

\bibitem{Cottam02}
Cottam, J., Paerels, F., and Mendez, M., ``Gravitationally redshifted
  absorption lines in the X-ray burst spectra of a neutron star'', {\em
  Nature}, {\bf 420}, 51--54, (2002).  Related online version (cited on 24 July
  2007): \newline\url{http://arXiv.org/abs/astro-ph/0211126}.

\bibitem{Damour07}
{Damour}, T., ``{Binary Systems as Test-beds of Gravity Theories}'', {\em ArXiv
  e-prints}, {\bf 704}, (April, 2007). \newline ADS:
  \url{http://adsabs.harvard.edu/abs/2007arXiv0704.0749D}.

\bibitem{Damour93}
Damour, T., and Esposito-Farese, G., ``Nonperturbative strong-field effects in
  tensor-scalar theories of gravitation'', {\em Phys. Rev. Lett.}, {\bf 70},
  2220--2223,  (1993).

\bibitem{Damour96}
Damour, T., and Esposito-Far{\`e}se, G., ``Tensor-scalar gravity and
  binary-pulsar experiments'', {\em Phys. Rev. D}, {\bf 54}, 1474--1491,
  (1996).  Related online version (cited on 24 July 2007):
  \newline\url{http://arXiv.org/abs/gr-qc/9602056}.

\bibitem{Davis05}
Davis, S.W., Blaes, O.M., Hubeny, I., and Turner, N.J., ``Relativistic
  Accretion Disk Models of High-State Black Hole X-Ray Binary Spectra'', {\em
  Astrophys. J.}, {\bf 621}, 372--387, (2005).  Related online version (cited
  on 24 July 2007): \newline\url{http://arXiv.org/abs/astro-ph/0408590}.

\bibitem{deVilliers03}
De~Villiers, J.-P., and Hawley, J.F., ``A Numerical Method for General
  Relativistic Magnetohydrodynamics'', {\em Astrophys. J.}, {\bf 589},
  458--480, (2003). \newline ADS:
  \url{http://adsabs.harvard.edu/abs/2003ApJ...589..458D}.

\bibitem{DeDeo03}
Dedeo, S., and Psaltis, D., ``Towards New Tests of Strong-Field Gravity with
  Measurements of Surface Atomic Line Redshifts from Neutron Stars'', {\em
  Phys. Rev. Lett.}, {\bf 90}, 141101, (2003).  Related online version (cited
  on 24 July 2007): \newline\url{http://arXiv.org/abs/astro-ph/0302095}.

\bibitem{DeDeo08}
DeDeo, S., and Psaltis, D., ``Stable, Accelerating Universes in Modified
  Gravity'', arxiv, (December, 2007).

\bibitem{DeDeo07}
DeDeo, S., and Psaltis, D, ``Testing Strong-field Gravity with Quasi-Periodic
  Oscillations'', {\em Phys. Rev. D}, submitted, (2007).  Related online
  version (cited on 05 July 2007):
  \newline\url{http://arXiv.org/abs/astro-ph/0405067}.

\bibitem{Dehnen98}
Dehnen, W., and Binney, J., ``Mass models of the Milky Way'', {\em Mon. Not. R.
  Astron. Soc.}, {\bf 294}, 429--438, (1998). \newline ADS:
  \url{http://adsabs.harvard.edu/abs/1998MNRAS.294..429D}.

\bibitem{DiSalvo06}
Di~Salvo, T., Goldoni, P., Stella, L., van~der Klis, M., Bazzano, A., Burderi,
  L., Farinelli, R., Frontera, F., Israel, G.L., M\'endez, M., Mirabel, I.F.,
  Robba, N.R., Sizun, P., Ubertini, P., and Lewin, W.H.G., ``A Hard X-Ray View
  of Scorpius X-1 with INTEGRAL: Nonthermal Emission?'', {\em Astrophys. J.},
  {\bf 649}, L91--L94, (2006).  Related online version (cited on 24 July 2007):
  \newline\url{http://arXiv.org/abs/astro-ph/0608335}.

\bibitem{DiSalvo01}
Di~Salvo, T., Robba, N.R., Iaria, R., Stella, L., Burderi, L., and Israel,
  G.L., ``Detection of a Hard Tail in the X-Ray Spectrum of the Z Source GX
  349+2'', {\em Astrophys. J.}, {\bf 554}, 49--55, (2001). \newline ADS:
  \url{http://adsabs.harvard.edu/abs/2001ApJ...554...49D}.

\bibitem{Dolgov03}
Dolgov, A.D., and Kawasaki, M., ``Can modified gravity explain accelerated
  cosmic expansion?'', {\em Phys. Lett. B}, {\bf 573}, 1--4,  (2003).

\bibitem{Donoghue94}
Donoghue, J.F., ``General relativity as an effective field theory: The leading
  quantum corrections'', {\em Phys. Rev. D}, {\bf 50}, 3874--3888, (1994).
  Related online version (cited on 24 July 2007):
  \newline\url{http://arXiv.org/abs/gr-qc/9405057}.

\bibitem{DGP00a}
Dvali, G., Gabadadze, G., and Porrati, M., ``4D gravity on a brane in 5D
  Minkowski space'', {\em Phys. Lett. B}, {\bf 485}, 208--214, (2000). \newline
  ADS: \url{http://adsabs.harvard.edu/abs/2000PhLB..485..208D}.

\bibitem{DGP00b}
Dvali, G., Gabadadze, G., and Porrati, M., ``A comment on brane bending and
  ghosts in theories with infinite extra dimensions'', {\em Phys. Lett. B},
  {\bf 484}, 129--132, (2000). \newline ADS:
  \url{http://adsabs.harvard.edu/abs/2000PhLB..484..129D}.

\bibitem{DGP00c}
Dvali, G., Gabadadze, G., and Porrati, M., ``Metastable gravitons and infinite
  volume extra dimensions'', {\em Phys. Lett. B}, {\bf 484}, 112--118, (2000).
  \newline ADS: \url{http://adsabs.harvard.edu/abs/2000PhLB..484..112D}.

\bibitem{Eardley75}
Eardley, D.M., ``Observable effects of a scalar gravitational field in a binary
  pulsar'', {\em Astrophys. J.}, {\bf 196}, L59--L62, (1975). \newline ADS:
  \url{http://adsabs.harvard.edu/abs/1975ApJ...196L..59E}.

\bibitem{Fabian89}
Fabian, A.C., Rees, M.J., Stella, L., and White, N.E., ``X-ray fluorescence
  from the inner disc in Cygnus X-1'', {\em Mon. Not. R. Astron. Soc.}, {\bf
  238}, 729--736, (1989). \newline ADS:
  \url{http://adsabs.harvard.edu/abs/1989MNRAS.238..729F}.

\bibitem{Fabian03}
Fabian, A.C., and Vaughan, S., ``The iron line in MCG-6-30-15 from XMM-Newton:
  evidence for gravitational light bending?'', {\em Mon. Not. R. Astron. Soc.},
  {\bf 340}, L28--L32, (2003). \newline ADS:
  \url{http://adsabs.harvard.edu/abs/2003MNRAS.340L..28F}.

\bibitem{Falcke00}
Falcke, H., Melia, F., and Agol, E., ``Viewing the Shadow of the Black Hole at
  the Galactic Center'', {\em Astrophys. J.}, {\bf 528}, L13--L16, (2000).
  \newline ADS: \url{http://adsabs.harvard.edu/abs/2000ApJ...528L..13F}.

\bibitem{Flanagan05}
Flanagan, {\'E}.{\'E}., and Hughes, S.A., ``The basics of gravitational wave
  theory'', {\em New J. Phys.}, {\bf 7}, 204, (2005).  Related online version
  (cited on 24 July 2007): \newline\url{http://arXiv.org/abs/gr-qc/0501041}.

\bibitem{Gammie99}
Gammie, C.F., ``Efficiency of Magnetized Thin Accretion Disks in the Kerr
  Metric'', {\em Astrophys. J.}, {\bf 522}, L57--L60, (1999). \newline ADS:
  \url{http://adsabs.harvard.edu/abs/1999ApJ...522L..57G}.

\bibitem{Gammie03}
Gammie, C.F., McKinney, J.C., and T{\'o}th, G., ``HARM: A Numerical Scheme for
  General Relativistic Magnetohydrodynamics'', {\em Astrophys. J.}, {\bf 589},
  444--457, (2003). \newline ADS:
  \url{http://adsabs.harvard.edu/abs/2003ApJ...589..444G}.

\bibitem{Garcia01}
Garcia, M.R., McClintock, J.E., Narayan, R., Callanan, P., Barret, D., and
  Murray, S.S., ``New Evidence for Black Hole Event Horizons from Chandra'',
  {\em Astrophys. J.}, {\bf 553}, L47--L50, (2001). \newline ADS:
  \url{http://adsabs.harvard.edu/abs/2001ApJ...553L..47G}.

\bibitem{GEO}
MPI for Gravitational Physics (Albert Einstein Institute), ``GEO 600: The
  German-British Gravitational Wave Detector'', project homepage.  URL (cited
  on 05 July 2007): \newline\url{http://geo600.aei.mpg.de}.

\bibitem{George91}
George, I.M., and Fabian, A.C., ``X-ray reflection from cold matter in active
  galactic nuclei and X-ray binaries'', {\em Mon. Not. R. Astron. Soc.}, {\bf
  249}, 352--367, (1991). \newline ADS:
  \url{http://adsabs.harvard.edu/abs/1991MNRAS.249..352G}.

\bibitem{Gierlinski01}
Gierli{\'n}ski, M., Macio{\l}ek-Nied{\'z}wiecki, A., and Ebisawa, K.,
  ``Application of a relativistic accretion disc model to X-ray spectra of LMC
  X-1 and GRO J1655-40'', {\em Mon. Not. R. Astron. Soc.}, {\bf 325},
  1253--1265, (2001). \newline ADS:
  \url{http://adsabs.harvard.edu/abs/2001MNRAS.325.1253G}.

\bibitem{Gierlinski99}
Gierli{\'n}ski, M., Zdziarski, A.A., Poutanen, J., Coppi, P.S., Ebisawa, K.,
  and Johnson, W.N., ``Radiation mechanisms and geometry of Cygnus X-1 in the
  soft state'', {\em Mon. Not. R. Astron. Soc.}, {\bf 309}, 496--512, (1999).
  \newline ADS: \url{http://adsabs.harvard.edu/abs/1999MNRAS.309..496G}.

\bibitem{Glampedakis06}
Glampedakis, K., and Babak, S., ``Mapping spacetimes with LISA: inspiral of a
  test body in a 'quasi-Kerr' field'', {\em Class. Quantum Grav.}, {\bf 23},
  4167--4188, (2006).  Related online version (cited on 24 July 2007):
  \newline\url{http://arXiv.org/abs/gr-qc/0510057}.

\bibitem{Green88}
Green, M.B., Schwarz, J.H., and Witten, E., {\em Superstring Theory},
  (Cambridge University Press, Cambridge, U.K.; New York, U.S.A., 1988).

\bibitem{Grove98}
Grove, J.E., Johnson, W.N., Kroeger, R.A., McNaron-Brown, K., Skibo, J.G., and
  Phlips, B.F., ``Gamma-Ray Spectral States of Galactic Black Hole
  Candidates'', {\em Astrophys. J.}, {\bf 500}, 899--908, (1998). \newline ADS:
  \url{http://adsabs.harvard.edu/abs/1998ApJ...500..899G}.

\bibitem{Guilbert88}
Guilbert, P.W., and Rees, M.J., ``\,`Cold' material in non-thermal sources'',
  {\em Mon. Not. R. Astron. Soc.}, {\bf 233}, 475--484, (1988). \newline ADS:
  \url{http://adsabs.harvard.edu/abs/1988MNRAS.233..475G}.

\bibitem{Harada98}
Harada, T., ``Neutron stars in scalar-tensor theories of gravity and
  catastrophe theory'', {\em Phys. Rev. D}, {\bf 57}, 4802--4811, (1998).
  Related online version (cited on 24 July 2007):
  \newline\url{http://arXiv.org/abs/gr-qc/9801049}.

\bibitem{Hawking72}
Hawking, S.W., ``Black Holes in the Brans--Dicke: Theory of Gravitation'', {\em
  Commun. Math. Phys.}, {\bf 25}, 167--171,  (1972).

\bibitem{Hubeny97}
Hubeny, I., and Hubeny, V., ``Non-LTE Models and Theoretical Spectra of
  Accretion Disks in Active Galactic Nuclei'', {\em Astrophys. J.}, {\bf 484},
  L37--L40, (1997). \newline ADS:
  \url{http://adsabs.harvard.edu/abs/1997ApJ...484L..37H}.

\bibitem{Iwasawa04}
Iwasawa, K., Miniutti, G., and Fabian, A.C., ``Flux and energy modulation of
  redshifted iron emission in NGC 3516: implications for the black hole mass'',
  {\em Mon. Not. R. Astron. Soc.}, {\bf 355}, 1073--1079, (2004).  Related
  online version (cited on 24 July 2007):
  \newline\url{http://arXiv.org/abs/astro-ph/0409293}.

\bibitem{Kaaret99}
Kaaret, P., Piraino, S., Bloser, P.F., Ford, E.C., Grindlay, J.E., Santangelo,
  A., Smale, A.P., and Zhang, W., ``Strong-Field Gravity and X-Ray Observations
  of 4U 1820-30'', {\em Astrophys. J.}, {\bf 520}, L37--L40, (1999). \newline
  ADS: \url{http://adsabs.harvard.edu/abs/1999ApJ...520L..37K}.

\bibitem{Kato01}
Kato, S., ``Basic Properties of Thin-Disk Oscillations'', {\em Publ. Astron.
  Soc. Japan}, {\bf 53}, 1--24,  (2001).

\bibitem{Kluzniak85}
Kluzniak, W., and Wagoner, R.V., ``Evolution of the innermost stable orbits
  around accreting neutron stars'', {\em Astrophys. J.}, {\bf 297}, 548--554,
  (1985). \newline ADS:
  \url{http://adsabs.harvard.edu/abs/1985ApJ...297..548K}.

\bibitem{Kramer05}
Kramer, M., Backer, D.C., Cordes, J.M., Lazio, T.J.W., Stappers, B.W., and
  Johnston, S., ``Strong-field tests of gravity using pulsars and black
  holes'', {\em New Astron. Rev.}, {\bf 48}, 993--1002,  (2004).

\bibitem{Krichbaum98}
Krichbaum, T.P., Graham, D.A., Witzel, A., Greve, A., Wink, J.E., Grewing, M.,
  Colomer, F., de~Vicente, P., Gomez-Gonzalez, J., Baudry, A., and Zensus,
  J.A., ``VLBI observations of the galactic center source SGR A* at 86 GHz and
  215 GHz'', {\em Astron. Astrophys.}, {\bf 335}, L106--L110, (1998). \newline
  ADS: \url{http://adsabs.harvard.edu/abs/1998A&A...335L.106K}.

\bibitem{Krolik99}
Krolik, J.H., ``Magnetized Accretion inside the Marginally Stable Orbit around
  a Black Hole'', {\em Astrophys. J.}, {\bf 515}, L73--L76, (1999). \newline
  ADS: \url{http://adsabs.harvard.edu/abs/1999ApJ...515L..73K}.

\bibitem{Laor91}
Laor, A., ``Line profiles from a disk around a rotating black hole'', {\em
  Astrophys. J.}, {\bf 376}, 90--94, (1991). \newline ADS:
  \url{http://adsabs.harvard.edu/abs/1991ApJ...376...90L}.

\bibitem{Lasota00}
Lasota, J.-P., ``X-rays from quiescent low-mass X-ray binary transients'', {\em
  Astron. Astrophys.}, {\bf 360}, 575--582, (2000). \newline ADS:
  \url{http://adsabs.harvard.edu/abs/2000A&A...360..575L}.

\bibitem{Lattimer01}
Lattimer, J.M., and Prakash, M., ``Neutron Star Structure and the Equation of
  State'', {\em Astrophys. J.}, {\bf 550}, 426--442, (2001). \newline ADS:
  \url{http://adsabs.harvard.edu/abs/2001ApJ...550..426L}.

\bibitem{Laurent99}
Laurent, P., and Titarchuk, L., ``The Converging Inflow Spectrum Is an
  Intrinsic Signature for a Black Hole: Monte Carlo Simulations of
  Comptonization on Free-falling Electrons'', {\em Astrophys. J.}, {\bf 511},
  289--297, (1999). \newline ADS:
  \url{http://adsabs.harvard.edu/abs/1999ApJ...511..289L}.

\bibitem{Lewin93}
Lewin, W.H.G., van Paradijs, J., and Taam, R.E., ``X-Ray Bursts'', {\em Space
  Sci. Rev.}, {\bf 62}, 223--389,  (1993).

\bibitem{Tanaka95}
Lewin, W.H.G., van Paradijs, J., and Taam, R.E., ``Black-Hole Candidates'', in
  Lewin, W.H.G., van Paradijs, J., and van~den Heuvel, E.P.J., eds., {\em X-ray
  Binaries}, vol. 126 of Cambridge Astrophysics Series,  126,  (Cambridge
  University Press, Cambridge, U.K.; New York, U.S.A., 1995).

\bibitem{Li05}
Li, L.-X., Zimmerman, E.R., Narayan, R., and McClintock, J.E.,
  ``Multitemperature Blackbody Spectrum of a Thin Accretion Disk around a Kerr
  Black Hole: Model Computations and Comparison with Observations'', {\em
  Astrophys. J. Suppl. Ser.}, {\bf 157}, 335--370, (2005). \newline ADS:
  \url{http://adsabs.harvard.edu/abs/2005ApJS..157..335L}.

\bibitem{LIGO}
California Institute of Technology, ``LIGO Scientific Collaboration'', project
  homepage.  URL (cited on 05 July 2007): \newline\url{http://www.ligo.org}.

\bibitem{Lo98}
Lo, K.Y., Shen, Z.-Q., Zhao, J.-H., and Ho, P.T.P., ``Intrinsic Size of
  Sagittarius A*: 72 Schwarzschild Radii'', {\em Astrophys. J.}, {\bf 508},
  L61--L64, (1998). \newline ADS:
  \url{http://adsabs.harvard.edu/abs/1998ApJ...508L..61L}.

\bibitem{Maartens04}
Maartens, R., ``Brane-World Gravity'', {\em Living Rev. Relativity}, {\bf 7},
  lrr-2004-7, (2004).  URL (cited on 05 July 2007):
  \newline\url{http://www.livingreviews.org/lrr-2004-7}.

\bibitem{McClintock04}
McClintock, J.E., Narayan, R., and Rybicki, G.B., ``On the Lack of Thermal
  Emission from the Quiescent Black Hole XTE J1118+480: Evidence for the Event
  Horizon'', {\em Astrophys. J.}, {\bf 615}, 402--415, (2004).  Related online
  version (cited on 24 July 2007):
  \newline\url{http://arXiv.org/abs/astro-ph/0403251}.

\bibitem{McClintock06}
McClintock, J.E., and Remillard, R.A., ``Black hole binaries'', in Lewin,
  W.H.G., and van~der Klis, M., eds., {\em Compact Stellar X-Ray Sources},
  vol.~39 of Cambridge Astrophysics Series,  157--213,  (Cambridge University
  Press, Cambridge, U.K.; New York, U.S.A., 2006).

\bibitem{McClintock06b}
McClintock, J.E., Shafee, R., Narayan, R., Remillard, R.A., Davis, S.W., and
  Li, L.-X., ``The Spin of the Near-Extreme Kerr Black Hole GRS 1915+105'',
  {\em Astrophys. J.}, {\bf 652}, 518--539, (2006).  Related online version
  (cited on 24 July 2007): \newline\url{http://arXiv.org/abs/astro-ph/0606076}.

\bibitem{Mendez06}
Mendez, M., ``The elusive Innermost Stable Circular Orbit: Now you see it, now
  you don't'', Proceedings of the conference `The Multicoloured Landscape of
  Compact Objects and their Explosive Origins', held in Cefalu, Sicily, June
  11\,--\,24, 2006, submitted, (2006). Related online version (cited on 05 July
  2007): \newline\url{http://arXiv.org/abs/astro-ph/0611469}. to be published
  by AIP.

\bibitem{Mendez99}
M{\'e}ndez, M., van~der Klis, M., Ford, E.C., Wijnands, R., and van Paradijs,
  J., ``Dependence of the Frequency of the Kilohertz Quasi-periodic
  Oscillationson X-Ray Count Rate and Colors in 4U 1608-52'', {\em Astrophys.
  J.}, {\bf 511}, L49--L52, (1999). \newline ADS:
  \url{http://adsabs.harvard.edu/abs/1999ApJ...511L..49M}.

\bibitem{Midleton05}
Middleton, M., Done, C., Gierli{\'n}ski, M., and Davis, S.W., ``Black hole spin
  in GRS 1915+105'', {\em Mon. Not. R. Astron. Soc.}, {\bf 373}, 1004--1012,
  (2006).  Related online version (cited on 24 July 2007):
  \newline\url{http://arXiv.org/abs/astro-ph/0601540}.

\bibitem{Milgrom83}
Milgrom, M., ``A modification of the Newtonian dynamics as a possible
  alternative to the hidden mass hypothesis'', {\em Astrophys. J.}, {\bf 270},
  365--370, (1983). \newline ADS:
  \url{http://adsabs.harvard.edu/abs/1983ApJ...270..365M}.

\bibitem{Miller06}
Miller, J.M., ``A short review of relativistic iron lines from stellar-mass
  black holes'', {\em Astron. Nachr.}, {\bf 327}, 997--1003, (2006).  Related
  online version (cited on 24 July 2007):
  \newline\url{http://arXiv.org/abs/astro-ph/0609447}.

\bibitem{Miller07}
Miller, J.M., ``Relativistic X-ray Lines from the Inner Accretion Disks Around
  Black Holes'', {\em Annu. Rev. Astron. Astrophys.}, submitted, (2007).
  Related online version (cited on 05 July 2007):
  \newline\url{http://arXiv.org/abs/0705.0540}.

\bibitem{Miller98}
Miller, M.C., Lamb, F.K., and Psaltis, D., ``Sonic-Point Model of Kilohertz
  Quasi-periodic Brightness Oscillations in Low-Mass X-Ray Binaries'', {\em
  Astrophys. J.}, {\bf 508}, 791--830, (1998). \newline ADS:
  \url{http://adsabs.harvard.edu/abs/1998ApJ...508..791M}.

\bibitem{Miniutti04}
Miniutti, G., and Fabian, A.C., ``A light bending model for the X-ray temporal
  and spectral properties of accreting black holes'', {\em Mon. Not. R. Astron.
  Soc.}, {\bf 349}, 1435--1448, (2004). \newline ADS:
  \url{http://adsabs.harvard.edu/abs/2004MNRAS.349.1435M}.

\bibitem{Misner73}
Misner, C.W., Thorne, K.S., and Wheeler, J.A., {\em Gravitation},  (W.H.
  Freeman, San Francisco, U.S.A., 1973).

\bibitem{Nandra97}
Nandra, K., George, I.M., Mushotzky, R.F., Turner, T.J., and Yaqoob, T., ``ASCA
  Observations of Seyfert 1 Galaxies. II. Relativistic Iron K alpha Emission'',
  {\em Astrophys. J.}, {\bf 477}, 602--622, (1997). \newline ADS:
  \url{http://adsabs.harvard.edu/abs/1997ApJ...477..602N}.

\bibitem{Nandra06}
Nandra, K., O'Neill, P.M., George, I.M., Reeves, J.N., and Turner, T.J., ``An
  XMM-Newton survey of broad iron lines in AGN'', {\em Astron. Nachr.}, {\bf
  327}, 1039, (2006).  Related online version (cited on 24 July 2007):
  \newline\url{http://arXiv.org/abs/astro-ph/0610585}.

\bibitem{Narayan97}
Narayan, R., Garcia, M.R., and McClintock, J.E., ``Advection-dominated
  Accretion and Black Hole Event Horizons'', {\em Astrophys. J.}, {\bf 478},
  L79--L82, (1997). \newline ADS:
  \url{http://adsabs.harvard.edu/abs/1997ApJ...478L..79N}.

\bibitem{Narayan95}
Narayan, R., Yi, I., and Mahadevan, R., ``Explaining the Spectrum of
  Sagittarius A* with a Model of an Accreting Black-Hole'', {\em Nature}, {\bf
  374}, 623--625,  (1995).

\bibitem{LISA}
NASA, ``LISA: Laser Interferometer Space Antenna'', project homepage.  URL
  (cited on 05 July 2007): \newline\url{http://lisa.nasa.gov}.

\bibitem{Niedzwiecki06}
Nied{\'z}wiecki, A., and Zdziarski, A.A., ``Bulk motion Comptonization in black
  hole accretion flows'', {\em Mon. Not. R. Astron. Soc.}, {\bf 365}, 606--614,
  (2006).  Related online version (cited on 24 July 2007):
  \newline\url{http://arXiv.org/abs/astro-ph/0507579}.

\bibitem{Noble07}
Noble, S.C., Leung, P.K., Gammie, C.F., and Book, L.G., ``Simulating the
  emission and outflows of accretion disks'', {\em Class. Quantum Grav.}, {\bf
  24}, 259--274, (2007).  Related online version (cited on 18 May 2008):
  \newline\url{http://arXiv.org/abs/astro-ph/0507579}.

\bibitem{Nowak01}
Nowak, M., and Lehr, D., ``Stable oscillations of black hole accretion discs'',
  in Abramowicz, M.A., Bj{\"o}rnsson, G., and Pringle, J.E., eds., {\em Theory
  of Black Hole Accretion Discs}, Cambridge Contemporary Astrophysics,
  233--253, (Cambridge University Press, Cambridge, U.K.; New York, U.S.A.,
  1998).  Related online version (cited on 24 July 2007):
  \newline\url{http://arXiv.org/abs/astro-ph/9812004}.

\bibitem{Oppenheimer39}
Oppenheimer, J.R., and Snyder, H., ``On Continued Gravitational Contraction'',
  {\em Phys. Rev.}, {\bf 56}, 455--459,  (1939).

\bibitem{Ozel06}
{\"O}zel, F., ``Soft equations of state for neutron-star matter ruled out by
  EXO 0748-676'', {\em Nature}, {\bf 441}, 1115--1117,  (2006).

\bibitem{Ozel01}
{\"O}zel, F., and Di~Matteo, T., ``X-Ray Images of Hot Accretion Flows'', {\em
  Astrophys. J.}, {\bf 548}, 213--218, (2001). \newline ADS:
  \url{http://adsabs.harvard.edu/abs/2001ApJ...548..213O}.

\bibitem{Ozel03}
{\"O}zel, F., and Psaltis, D., ``Spectral Lines from Rotating Neutron Stars'',
  {\em Astrophys. J.}, {\bf 582}, L31--L34, (2003). \newline ADS:
  \url{http://adsabs.harvard.edu/abs/2003ApJ...582L..31O}.

\bibitem{Ozel00}
{\"O}zel, F., Psaltis, D., and Narayan, R., ``Hybrid Thermal-Nonthermal
  Synchrotron Emission from Hot Accretion Flows'', {\em Astrophys. J.}, {\bf
  541}, 234--249, (2000). \newline ADS:
  \url{http://adsabs.harvard.edu/abs/2000ApJ...541..234O}.

\bibitem{Pais82}
Pais, A., {\em \,`Subtle is the Lord': The Science and Life of Albert
  Einstein},  (Oxford University Press, Oxford, U.K., 1982).

\bibitem{Papathanassiou00}
Papathanassiou, H., and Psaltis, D., ``Photon Scattering by Relativistic Flows
  in Schwarzschild Spacetimes. I. The Generation of Power-Law Spectra'',
  (2000).  URL (cited on 05 July 2007):
  \newline\url{http://arXiv.org/abs/astro-ph/0011447}.

\bibitem{Parker93}
Parker, L., and Simon, J.Z., ``Einstein equation with quantum corrections
  reduced to second order'', {\em Phys. Rev. D}, {\bf 47}, 1339--1355, (1993).
  Related online version (cited on 24 July 2007):
  \newline\url{http://arXiv.org/abs/gr-qc/9211002}.

\bibitem{Payne81}
Payne, D.G., and Blandford, R.D., ``Compton scattering in a converging fluid
  flow. III - Spherical supercritical accretion'', {\em Mon. Not. R. Astron.
  Soc.}, {\bf 196}, 781--795, (1981). \newline ADS:
  \url{http://adsabs.harvard.edu/abs/1981MNRAS.196..781P}.

\bibitem{Peebles03}
Peebles, P.J., and Ratra, B., ``The cosmological constant and dark energy'',
  {\em Rev. Mod. Phys.}, {\bf 75}, 559--606,  (2003).

\bibitem{Perl97}
Perlmutter, S., Gabi, S., Goldhaber, G., Goobar, A., Groom, D.E., Hook, I.M.,
  Kim, A.G., Kim, M.Y., Lee, J.C., Pain, R., Pennypacker, C.R., Small, I.A.,
  Ellis, R.S., McMahon, R.G., Boyle, B.J., Bunclark, P.S., Carter, D., Irwin,
  M.J., Glazebrook, K., Newberg, H.J.M., Filippenko, A.V., Matheson, T.,
  Dopita, M., and Couch, W.J. (The Supernova Cosmology~Project), ``Measurements
  of the Cosmological Parameters Omega and Lambda from the First Seven
  Supernovae at $z \ge 0.35$'', {\em Astrophys. J.}, {\bf 483}, 565--581,
  (1997). \newline ADS:
  \url{http://adsabs.harvard.edu/abs/1997ApJ...483..565P}.

\bibitem{Psaltis01}
Psaltis, D., ``Compton Scattering in Static and Moving Media. II. System-Frame
  Solutions for Spherically Symmetric Flows'', {\em Astrophys. J.}, {\bf 555},
  786--800, (2001). \newline ADS:
  \url{http://adsabs.harvard.edu/abs/2001ApJ...555..786P}.

\bibitem{Psaltis01a}
Psaltis, D., ``Models of quasi-periodic variability in neutron stars and black
  holes'', {\em Adv. Space Res.}, {\bf 28}, 481--491,  (2001).

\bibitem{Psaltis07a}
Psaltis, D., ``Constraining Brans--Dicke Gravity with Millisecond Pulsars in
  Ultracompact Binaries'', (2005).  URL (cited on 05 July 2007):
  \newline\url{http://arXiv.org/abs/astro-ph/0501234}.

\bibitem{Psaltis06}
Psaltis, D., ``Accreting neutron stars and black holes: a decade of
  discoveries'', in Lewin, W.H.G., and van~der Klis, M., eds., {\em Compact
  Stellar X-Ray Sources}, vol.~39 of Cambridge Astrophysics Series,  1--34,
  (Cambridge University Press, Cambridge, U.K.; New York, U.S.A., 2006).

\bibitem{Psaltis99}
Psaltis, D., Belloni, T., and van~der Klis, M., ``Correlations in
  Quasi-periodic Oscillation and Noise Frequencies among Neutron Star and Black
  Hole X-Ray Binaries'', {\em Astrophys. J.}, {\bf 520}, 262--270, (1999).
  \newline ADS: \url{http://adsabs.harvard.edu/abs/1999ApJ...520..262P}.

\bibitem{Psaltis97}
Psaltis, D., and Lamb, F.K., ``Compton Scattering by Static and Moving Media.
  I. The Transfer Equation and Its Moments'', {\em Astrophys. J.}, {\bf 488},
  881--894, (1997). \newline ADS:
  \url{http://adsabs.harvard.edu/abs/1997ApJ...488..881P}.

\bibitem{Psaltis00b}
Psaltis, D., and Norman, C., ``On the Origin of Quasi-Periodic Oscillations and
  Broad-band Noise in Accreting Neutron Stars and Black Holes'', (2000).  URL
  (cited on 05 July 2007): \newline\url{http://arXiv.org/abs/astro-ph/0001391}.

\bibitem{Psaltis07b}
Psaltis, D., Perrodin, D., Dienes, K., and Mocioiu, I., ``Kerr Black Holes are
  not Unique to General Relativity'', {\em Phys. Rev. Lett.}, {\bf 100}, 1101,
  (2008).

\bibitem{Reeves06}
Reeves, J.N., Fabian, A.C., Kataoka, J., Kunieda, H., Markowitz, A., Miniutti,
  G., Okajima, T., Serlemitsos, P., Takahashi, T., Terashima, Y., and Yaqoob,
  T., ``Suzaku observations of iron lines and reflection in AGN'', {\em Astron.
  Nachr.}, {\bf 327}, 1079, (2006).  Related online version (cited on 24 July
  2007): \newline\url{http://arXiv.org/abs/astro-ph/0610436}.

\bibitem{Reynolds97}
Reynolds, C.S., and Begelman, M.C., ``Iron Fluorescence from within the
  Innermost Stable Orbit of Black Hole Accretion Disks'', {\em Astrophys. J.},
  {\bf 488}, 109--118, (1997). \newline ADS:
  \url{http://adsabs.harvard.edu/abs/1997ApJ...488..109R}.

\bibitem{Reynolds03}
Reynolds, C.S., and Nowak, M.A., ``Fluorescent iron lines as a probe of
  astrophysical black hole systems'', {\em Phys. Rep.}, {\bf 377}, 389--466,
  (2003).

\bibitem{Reynolds99}
Reynolds, C.S., Young, A.J., Begelman, M.C., and Fabian, A.C., ``X-Ray Iron
  Line Reverberation from Black Hole Accretion Disks'', {\em Astrophys. J.},
  {\bf 514}, 164--179, (1999). \newline ADS:
  \url{http://adsabs.harvard.edu/abs/1999ApJ...514..164R}.

\bibitem{Riess98}
Riess, A.G., Filippenko, A.V., Challis, P., Clocchiatti, A., Diercks, A.,
  Garnavich, P.M., Gilliland, R.L., Hogan, C.J., Jha, S., Kirshner, R.P.,
  Leibundgut, B., Phillips, M.M., Reiss, D., Schmidt, B.P., Schommer, R.A.,
  Smith, R.C., Spyromilio, J., Stubbs, C., Suntzeff, N.B., and Tonry, J.,
  ``Observational Evidence from Supernovae for an Accelerating Universe and a
  Cosmological Constant'', {\em Astron. J.}, {\bf 116}, 1009--1038, (1998).
  \newline ADS: \url{http://adsabs.harvard.edu/abs/1998AJ....116.1009R}.

\bibitem{Hugh00}
Rowan, S., and Hough, J., ``Gravitational Wave Detection by Interferometry
  (Ground and Space)'', {\em Living Rev. Relativity}, {\bf 3}, lrr-2000-3,
  (2000).  URL (cited on 05 July 2007):
  \newline\url{http://www.livingreviews.org/lrr-2000-3}.

\bibitem{Ryan95}
Ryan, F.D., ``Gravitational waves from the inspiral of a compact object into a
  massive, axisymmetric body with arbitrary multipole moments'', {\em Phys.
  Rev. D}, {\bf 52}, 5707--5718,  (1995).

\bibitem{Salgado98}
Salgado, M., Sudarsky, D., and Nucamendi, U., ``Spontaneous scalarization'',
  {\em Phys. Rev. D}, {\bf 58}, 124003, (1998).  Related online version (cited
  on 24 July 2007): \newline\url{http://arXiv.org/abs/gr-qc/9806070}.

\bibitem{Sanders02}
Sanders, R.H., and McGaugh, S.S., ``Modified Newtonian Dynamics as an
  Alternative to Dark Matter'', {\em Annu. Rev. Astron. Astrophys.}, {\bf 40},
  263--317,  (2002).

\bibitem{Santiago97}
Santiago, D.I., Kalligas, D., and Wagoner, R.V., ``Nucleosynthesis constraints
  on scalar-tensor theories of gravity'', {\em Phys. Rev. D}, {\bf 56},
  7627--7637, (1997).  Related online version (cited on 24 July 2007):
  \newline\url{http://arXiv.org/abs/gr-qc/9706017}.

\bibitem{Scheel95}
Scheel, M.A., Shapiro, S.L., and Teukolsky, S.A., ``Collapse to black holes in
  Brans--Dicke theory. II. Comparison with general relativity'', {\em Phys.
  Rev. D}, {\bf 51}, 4236--4249, (1995).  Related online version (cited on 24
  July 2007): \newline\url{http://arXiv.org/abs/gr-qc/9411026}.

\bibitem{Schodel02}
Sch{\"o}del, R., Ott, T., Genzel, R., Hofmann, R., Lehnert, M., Eckart, A.,
  Mouawad, N., Alexander, T., Reid, M.J., Lenzen, R., Hartung, M., Lacombe, F.,
  Rouan, D., Gendron, E., Rousset, G., Lagrange, A.-M., Brandner, W., Ageorges,
  N., Lidman, C., Moorwood, A.F.M., Spyromilio, J., Hubin, N., and Menten,
  K.M., ``A star in a 15.2-year orbit around the supermassive black hole at the
  centre of the Milky Way'', {\em Nature}, {\bf 419}, 694--696, (2002).
  Related online version (cited on 24 July 2007):
  \newline\url{http://arXiv.org/abs/astro-ph/0210426}.

\bibitem{Seifert07}
Seifert, M.D., ``Stability of spherically symmetric solutions in modified
  theories of gravity'', {\em Phys. Rev. D}, submitted, (2007).  Related online
  version (cited on 05 July 2007):
  \newline\url{http://arXiv.org/abs/gr-qc/0703060}.

\bibitem{Shafee06}
Shafee, R., McClintock, J.E., Narayan, R., Davis, S.W., Li, L.-X., and
  Remillard, R.A., ``Estimating the Spin of Stellar-Mass Black Holes by
  Spectral Fitting of the X-Ray Continuum'', {\em Astrophys. J.}, {\bf 636},
  L113--L116, (2006).  Related online version (cited on 24 July 2007):
  \newline\url{http://arXiv.org/abs/astro-ph/0508302}.

\bibitem{Shakura73}
Shakura, N.I., and Sunyaev, R.A., ``Black Holes in Binary Systems.
  Observational Appearance'', {\em Astron. Astrophys.}, {\bf 24}, 337--355,
  (1973). \newline ADS:
  \url{http://adsabs.harvard.edu/abs/1973A&A....24..337S}.

\bibitem{Shapiro84}
Shapiro, S.L., and Teukolsky, S.A., {\em Black Holes, White Dwarfs and Neutron
  Stars: The Physics of Compact Objects},  (John Wiley \& Sons, Hoboken,
  U.S.A., 1983).

\bibitem{Shen05}
Shen, Z.-Q., Lo, K.Y., Liang, M.-C., Ho, P.T.P., and Zhao, J.-H., ``A size of
  $\sim$1AU for the radio source Sgr A* at the centre of the Milky Way'', {\em
  Nature}, {\bf 438}, 62--64, (2005).  Related online version (cited on 24 July
  2007): \newline\url{http://arXiv.org/abs/astro-ph/0512515}.

\bibitem{Simon90}
Simon, J.Z., ``Higher-derivative Lagrangians, nonlocality, problems, and
  solutions'', {\em Phys. Rev. D}, {\bf 41}, 3720--3733,  (1990).

\bibitem{Simon91}
Simon, J.Z., ``Stability of flat space, semiclassical gravity, and higher
  derivatives'', {\em Phys. Rev. D}, {\bf 43}, 3308--3316,  (1991).

\bibitem{SKA}
International SKA Project Office (ISPO), ``SKA: Square Kilometre Array, the
  international radiotelescope for the 21st century'', project homepage.  URL
  (cited on 05 July 2007): \newline\url{http://www.skatelescope.org/}.

\bibitem{Sotani04}
Sotani, H., and Kokkotas, K.D., ``Probing strong-field scalar-tensor gravity
  with gravitational wave asteroseismology'', {\em Phys. Rev. D}, {\bf 70},
  084026, (2004).  Related online version (cited on 24 July 2007):
  \newline\url{http://arXiv.org/abs/gr-qc/0409066}.

\bibitem{Sotani05}
Sotani, H., and Kokkotas, K.D., ``Stellar oscillations in scalar-tensor theory
  of gravity'', {\em Phys. Rev. D}, {\bf 71}, 124038, (2005).  Related online
  version (cited on 24 July 2007):
  \newline\url{http://arXiv.org/abs/gr-qc/0506060}.

\bibitem{Sotiriou08}
Sotiriou, T.P., and Faraoni, V., ``$f(R)$ Theories of Gravity'', {\em Rev. Mod.
  Phys.}, submitted, (2008).  Related online version (cited on 18 May 2008):
  \newline\url{http://arXiv.org/abs/0805.1726}.

\bibitem{Sotiriou07}
Sotiriou, T.P., and Liberati, S., ``Metric-affine $f(R)$ theories of gravity'',
  {\em Ann. Phys. (N.Y.)}, {\bf 322}, 935--966, (2007).  Related online version
  (cited on 24 July 2007): \newline\url{http://arXiv.org/abs/gr-qc/0604006}.

\bibitem{Spergel03}
Spergel, D.N., Verde, L., Peiris, H.V., Komatsu, E., Nolta, M.R., Bennett,
  C.L., Halpern, M., Hinshaw, G., Jarosik, N., Kogut, A., Limon, M., Meyer,
  S.S., Page, L., Tucker, G.S., Weiland, J.L., Wollack, E., and Wright, E.L.,
  ``First-Year Wilkinson Microwave Anisotropy Probe (WMAP) Observations:
  Determination of Cosmological Parameters'', {\em Astrophys. J. Suppl. Ser.},
  {\bf 148}, 175--194, (2003). \newline ADS:
  \url{http://adsabs.harvard.edu/abs/2003ApJS..148..175S}.

\bibitem{Stairs03}
Stairs, I.H., ``Testing General Relativity with Pulsar Timing'', {\em Living
  Rev. Relativity}, {\bf 6}, lrr-2003-5, (2003).  URL (cited on 05 July 2007):
  \newline\url{http://www.livingreviews.org/lrr-2003-5}.

\bibitem{Starobinski80}
Starobinskij, A.A., ``A new type of isotropic cosmological models without
  singularity'', {\em Phys. Lett. B}, {\bf 91}, 99--102,  (1980).

\bibitem{Stella99}
Stella, L., Vietri, M., and Morsink, S.M., ``Correlations in the Quasi-periodic
  Oscillation Frequencies of Low-Mass X-Ray Binaries and the Relativistic
  Precession Model'', {\em Astrophys. J.}, {\bf 524}, L63--L66, (1999).
  \newline ADS: \url{http://adsabs.harvard.edu/abs/1999ApJ...524L..63S}.

\bibitem{Strohmayer01}
Strohmayer, T.E., ``Discovery of a 450 HZ Quasi-periodic Oscillation from the
  Microquasar GRO J1655-40 with the Rossi X-Ray Timing Explorer'', {\em
  Astrophys. J.}, {\bf 552}, L49--L53, (2001). \newline ADS:
  \url{http://adsabs.harvard.edu/abs/2001ApJ...552L..49S}.

\bibitem{Strohmayer06}
Strohmayer, T.E., and Bildsten, L., ``New views of thermonuclear bursts'', in
  Lewin, W.H.G., and van~der Klis, M., eds., {\em Compact Stellar X-Ray
  Sources}, vol.~39 of Cambridge Astrophysics Series,  113--156, (Cambridge
  University Press, Cambridge, U.K.; New York, U.S.A., 2006).  Related online
  version (cited on 24 July 2007):
  \newline\url{http://arXiv.org/abs/astro-ph/0301544}.

\bibitem{Takahashi04}
Takahashi, R., ``Shapes and Positions of Black Hole Shadows in Accretion Disks
  and Spin Parameters of Black Holes'', {\em Astrophys. J.}, {\bf 611},
  996--1004, (2004).  Related online version (cited on 24 July 2007):
  \newline\url{http://arXiv.org/abs/astro-ph/0405099}.

\bibitem{TAMA}
National Astronomical Observatory of Japan (NAO), ``TAMA: The 300m Laser
  Interferometer Gravitational Wave Antenna'', project homepage.  URL (cited on
  05 July 2007): \newline\url{http://tamago.mtk.nao.ac.jp/}.

\bibitem{Tanaka95b}
Tanaka, Y., Nandra, K., Fabian, A.C., Inoue, H., Otani, C., Dotani, T.,
  Hayashida, K., Iwasawa, K., Kii, T., Kunieda, H., Makino, F., and Matsuoka,
  M., ``Gravitationally redshifted emission implying an accretion disk and
  massive black hole in the active galaxy MCG-6-30-15'', {\em Nature}, {\bf
  375}, 659--661,  (1995).

\bibitem{Thorne71}
Thorne, K.S., and Dykla, J.J., ``Black Holes in the Dicke--Brans Theory of
  Gravity'', {\em Astrophys. J.}, {\bf 166}, L35--L38, (1971). \newline ADS:
  \url{http://adsabs.harvard.edu/abs/1971ApJ...166L..35T}.

\bibitem{Thorsett99}
Thorsett, S.E., and Chakrabarty, D., ``Neutron Star Mass Measurements. I. Radio
  Pulsars'', {\em Astrophys. J.}, {\bf 512}, 288--299, (1999). \newline ADS:
  \url{http://adsabs.harvard.edu/abs/1999ApJ...512..288T}.

\bibitem{Titarchuk97}
Titarchuk, L., Mastichiadis, A., and Kylafis, N.D., ``X-Ray Spectral Formation
  in a Converging Fluid Flow: Spherical Accretion into Black Holes'', {\em
  Astrophys. J.}, {\bf 487}, 834--846, (1997). \newline ADS:
  \url{http://adsabs.harvard.edu/abs/1997ApJ...487..834T}.

\bibitem{Titarchuk98}
Titarchuk, L., and Zannias, T., ``The Extended Power Law as an Intrinsic
  Signature for a Black Hole'', {\em Astrophys. J.}, {\bf 493}, 863--872,
  (1998). \newline ADS:
  \url{http://adsabs.harvard.edu/abs/1998ApJ...493..863T}.

\bibitem{Tremaine02}
Tremaine, S., Gebhardt, K., Bender, R., Bower, G., Dressler, A., Faber, S.M.,
  Filippenko, A.V., Green, R., Grillmair, C., Ho, L.C., Kormendy, J., Lauer,
  T.R., Magorrian, J., Pinkney, J., and Richstone, D., ``The Slope of the Black
  Hole Mass versus Velocity Dispersion Correlation'', {\em Astrophys. J.}, {\bf
  574}, 740--753, (2002). \newline ADS:
  \url{http://adsabs.harvard.edu/abs/2002ApJ...574..740T}.

\bibitem{vanderKlis01}
van~der Klis, M., ``A Possible Explanation for the ``Parallel Tracks''
  Phenomenon in Low-Mass X-Ray Binaries'', {\em Astrophys. J.}, {\bf 561},
  943--949, (2001). \newline ADS:
  \url{http://adsabs.harvard.edu/abs/2001ApJ...561..943V}.

\bibitem{vanderKlis06}
van~der Klis, M., ``Rapid X-ray variability'', in Lewin, W.H.G., and van~der
  Klis, M., eds., {\em Compact Stellar X-Ray Sources}, vol.~39 of Cambridge
  Astrophysics Series,  39--112,  (Cambridge University Press, Cambridge, U.K.;
  New York, U.S.A., 2006).

\bibitem{vanStraten01}
van Straten, W., Bailes, M., Britton, M., Kulkarni, S.R., Anderson, S.B.,
  Manchester, R.N., and Sarkissian, J., ``A test of general relativity from the
  three-dimensional orbital geometry of a binary pulsar'', {\em Nature}, {\bf
  412}, 158--160, (2001).  Related online version (cited on 24 July 2007):
  \newline\url{http://arXiv.org/abs/astro-ph/0108254}.

\bibitem{Verbunt93}
Verbunt, F., ``Origin and evolution of X-ray binaries and binary radio
  pulsars'', {\em Annu. Rev. Astron. Astrophys.}, {\bf 31}, 93--127,  (1993).

\bibitem{Villarreal04}
Villarreal, A.R., and Strohmayer, T.E., ``Discovery of the Neutron Star Spin
  Frequency in EXO 0748-676'', {\em Astrophys. J.}, {\bf 614}, L121--L124,
  (2004).  Related online version (cited on 24 July 2007):
  \newline\url{http://arXiv.org/abs/astro-ph/0409384}.

\bibitem{VIRGO}
Istituto Nazionale di Fisica Nucleare, ``VIRGO Project Central Web Site'',
  project homepage.  URL (cited on 05 July 2007):
  \newline\url{http://www.virgo.infn.it}.

\bibitem{Wagoner01}
Wagoner, R.V., Silbergleit, A.S., and Ortega-Rodr\'iguez, M., ``\,``Stable''
  Quasi-periodic Oscillations and Black Hole Properties from Diskoseismology'',
  {\em Astrophys. J.}, {\bf 559}, L25--L28, (2001). \newline ADS:
  \url{http://adsabs.harvard.edu/abs/2001ApJ...559L..25W}.

\bibitem{Wagoner99}
Wagoner, R.W., ``Relativistic diskoseismology'', {\em Phys. Rep.}, {\bf 311},
  259--269,  (1999).

\bibitem{Wald84}
Wald, R.M., {\em General Relativity},  (University of Chicago Press, Chicago,
  U.S.A., 1984).

\bibitem{Weinberg89}
{Weinberg}, S., ``{The cosmological constant problem}'', {\em Reviews of Modern
  Physics}, {\bf 61}, 1--23, (January, 1989). \newline ADS:
  \url{http://adsabs.harvard.edu/abs/1989RvMP...61....1W}.

\bibitem{Will93}
Will, C.M., {\em Theory and experiment in gravitational physics}, (Cambridge
  University Press, Cambridge, U.K.; New York, U.S.A., 1993),  2nd edition.

\bibitem{Will06}
Will, C.M., ``The Confrontation between General Relativity and Experiment'',
  {\em Living Rev. Relativity}, {\bf 9}, lrr-2006-3, (2006).  URL (cited on 05
  July 2007): \newline\url{http://www.livingreviews.org/lrr-2006-3}.

\bibitem{Will89}
Will, C.M., and Zaglauer, H.W., ``Gravitational radiation, close binary
  systems, and the Brans--Dicke theory of gravity'', {\em Astrophys. J.}, {\bf
  346}, 366--377, (1989). \newline ADS:
  \url{http://adsabs.harvard.edu/abs/1989ApJ...346..366W}.

\bibitem{Wilms01}
Wilms, J., Reynolds, C.S., Begelman, M.C., Reeves, J., Molendi, S., Staubert,
  R., and Kendziorra, E., ``XMM-EPIC observation of MCG-6-30-15: direct
  evidence for the extraction of energy from a spinning black hole?'', {\em
  Mon. Not. R. Astron. Soc.}, {\bf 328}, L27--L31, (2001). \newline ADS:
  \url{http://adsabs.harvard.edu/abs/2001MNRAS.328L..27W}.

\bibitem{Woodard06}
Woodard, R.P., ``Avoiding Dark Energy with 1/R Modifications of Gravity'',
  (2006).  URL (cited on 05 July 2007):
  \newline\url{http://arXiv.org/abs/astro-ph/0601672}.

\bibitem{XEUS}
European Space Agency (ESA), ``XEUS: The X-Ray Evolving Universe
  Spectrometer'', project homepage.  URL (cited on 05 July 2007):
  \newline\url{http://www.rssd.esa.int/index.php?project=XEUS}.

\bibitem{Yuan03}
Yuan, F., Quataert, E., and Narayan, R., ``Nonthermal Electrons in Radiatively
  Inefficient Accretion Flow Models of Sagittarius A*'', {\em Astrophys. J.},
  {\bf 598}, 301--312, (2003). \newline ADS:
  \url{http://adsabs.harvard.edu/abs/2003ApJ...598..301Y}.

\bibitem{Zhang97}
Zhang, S.N., Cui, W., and Chen, W., ``Black Hole Spin in X-Ray Binaries:
  Observational Consequences'', {\em Astrophys. J.}, {\bf 482}, L155--L158,
  (1997). \newline ADS:
  \url{http://adsabs.harvard.edu/abs/1997ApJ...482L.155Z}.

\bibitem{Zhang98}
Zhang, W., Smale, A.P., Strohmayer, T.E., and Swank, J.H., ``Correlation
  between Energy Spectral States and Fast Time Variability and Further Evidence
  for the Marginally Stable Orbit in 4U 1820-30'', {\em Astrophys. J.}, {\bf
  500}, L171--L174, (1998). \newline ADS:
  \url{http://adsabs.harvard.edu/abs/1998ApJ...500L.171Z}.

\end{thebibliography}

\end{document}